\documentclass[aps,twocolumn,prb,reprint,amsmath,amssymb,showpacs,superscriptaddress,floatfix]{revtex4-2}
\usepackage{graphicx}
\usepackage{dcolumn}
\usepackage{bm}
\usepackage{amsmath,amsfonts,amsthm}
\usepackage{color}
\usepackage{mathrsfs}
\usepackage{float}
\usepackage[normalem]{ulem}

\usepackage[caption=false]{subfig}
\usepackage{txfonts}
\usepackage{wasysym}

\usepackage{cancel}

\makeatletter

\newcommand{\Rmnum}[1]{\expandafter\@slowromancap\romannumeral #1@}
\makeatother
\extrafloats{100}

\begin{document}
\newcommand{\LG}[1]{\textcolor{red}{#1}}

\title{Transport features  of a topological superconducting nanowire with a quantum dot: conductance and noise}

\author{Leonel Gru\~neiro}
\thanks{These two authors contributed equally.  }
\affiliation{Escuela de Ciencia y Tecnolog\'{\i}a  and ICIFI, Universidad Nacional de San Mart\'{\i}n-UNSAM, Av 25 de Mayo y Francia, 1650 Buenos Aires, Argentina}
\author{Miguel Alvarado}
\thanks{These two authors contributed equally.  }
\affiliation{Departamento de F\'{\i}sica Te\'orica de la Materia Condensada, Condensed Matter Physics Center (IFIMAC) and Instituto Nicol\'as Cabrera, Universidad Aut\'onoma de Madrid, 28049 Madrid, Spain}
\author{Alfredo Levy Yeyati}
\affiliation{Departamento de F\'{\i}sica Te\'orica de la Materia Condensada, Condensed Matter Physics Center (IFIMAC) and Instituto Nicol\'as Cabrera, Universidad Aut\'onoma de Madrid, 28049 Madrid, Spain}
\author{Liliana Arrachea}
\affiliation{Escuela de Ciencia y Tecnolog\'{\i}a  and ICIFI, Universidad Nacional de San Mart\'{\i}n-UNSAM, Av 25 de Mayo y Francia, 1650 Buenos Aires, Argentina}

\begin{abstract}
We study  two-terminal configurations in junctions between  a topological superconducting wire with spin-orbit coupling and magnetic field, and an  ordinary conductor with an embedded quantum dot. One of the signatures of the Majorana zero modes in the topological phase is a quantization of  the zero-bias conductance at $G(V=0)=2e^2/h$. However, the finite size of the wires and the presence of the quantum dot in the junction generate
 more complicated features which lead to deviations from this simple picture. Here, we analyze the behavior of the conductance at zero and finite bias, $G(V)$, as a function of a gate voltage applied at the quantum dot in the case of a finite-length wire. We analyze the effect of the angle between the magnetic field and the orientation associated to the spin-orbit coupling. We provide a detailed description of the spectral features of the quantum wire weakly and also strongly coupled to the quantum dot and describe the conditions to have zero-energy states 
 in  these two regimes for both the topological and non-topological phases.  We also analyze the concomitant behavior of the noise.
 We identify qualitative features that are useful to distinguish between the topological and non-topological phases. We show that in a strongly coupled quantum dot the  simultaneous hybridization with the topological modes and the supragap states of the wire mask the signatures of the Majorana bound states in both the conductance and the Fano factor. 
\end{abstract}

\date{\today}
\maketitle

\section{Introduction} 
The search for topological superconducting phases in low-dimensional hybrid nanostructures is a very active field of research for some years now.
This is mainly motivated by the fact that this phase is characterized by the existence of  Majorana states localized at the edges. In one-dimensional structures 
these states are zero modes with non-Abelian exchange statistics and are very appealing to realize topologically protected qubits
\cite{kitaev2001unpaired,kitaev2003fault,nayak2008non,alicea2012new}. 

Nanowires with 
spin-orbit coupling (SOC), proximity-induced s-wave superconductivity and 
a magnetic field having a component
perpendicular to the direction of the SOC \cite{wires1,wires2} are one of the most prominent systems to realize localized Majorana modes. 
The topological phase of this model was predicted to take place for perpendicular orientations of the magnetic field $\vec{B}$ and the spin-orbit coupling.
However, when this direction is twisted, so that 
$\vec{B}$ is tilted with respect to the direction of the SOC by an angle $\theta \neq \pi/2$, 
it has been shown that the topological phase survives for a range of parameters  provided that the tilt does not overcome a critical value $\theta_c$ \cite{rex2014tilting,osca2014effects,klinovaja2015fermionic,aligia2020tomography,daroca2021phase}. 

Several experimental works investigated 
 realizations of this platform for topological superconductivity in InAs wires 
\cite{mourik2012signatures,deng2016majorana,chen2017experimental,nichele2017scaling,vaitiekenas2020fullshell}. In most of these works, the signatures of non-trivial topology are searched in the behavior of the conductance, where the Majorana zero modes (MZM) are expected to generate a zero-bias peak
 quantized at $2 G_0$, with $G_0=e^2/h$. Features consistent with 
 such a signature have been reported in these experiments, although several alternative  interpretations based on the formation of non-topological Andreev bound states  have been also proposed in the literature 
 \cite{tanaka95,tanaka04,kells2012near,prada2012transport,roy2013,liu2017andreev,moore2018two,moore2018quantized,fleckenstein2018decaying,prada2020andreev,vuik2019reproducing,zhang2022suppressing}. Furthermore, the success of these experiments to realizing the topological superconducting phase with zero-energy Majorana modes has been put  under debate \cite{frolov2021quantum}. This discussion stimulates theoretical ideas exploring setups with two and three-terminal configurations 
 \cite{tanaka13,roy2019,lai2021quality,lobos2015tunneling,gramich2017andreev,zazunov2017multiterminal,jonckheere2019trijunction,danon2020nonlocal,melo2021conductance,pan2021three,banerjee2023local,hess-2022} 
 and more experimental studies \cite{yu2021non,wang2022observation,wang2022plateau,pikulin2021protocol} to clarify such a fundamental issue. 

 \begin{figure}[htb]
\centering
\includegraphics[width=\linewidth]{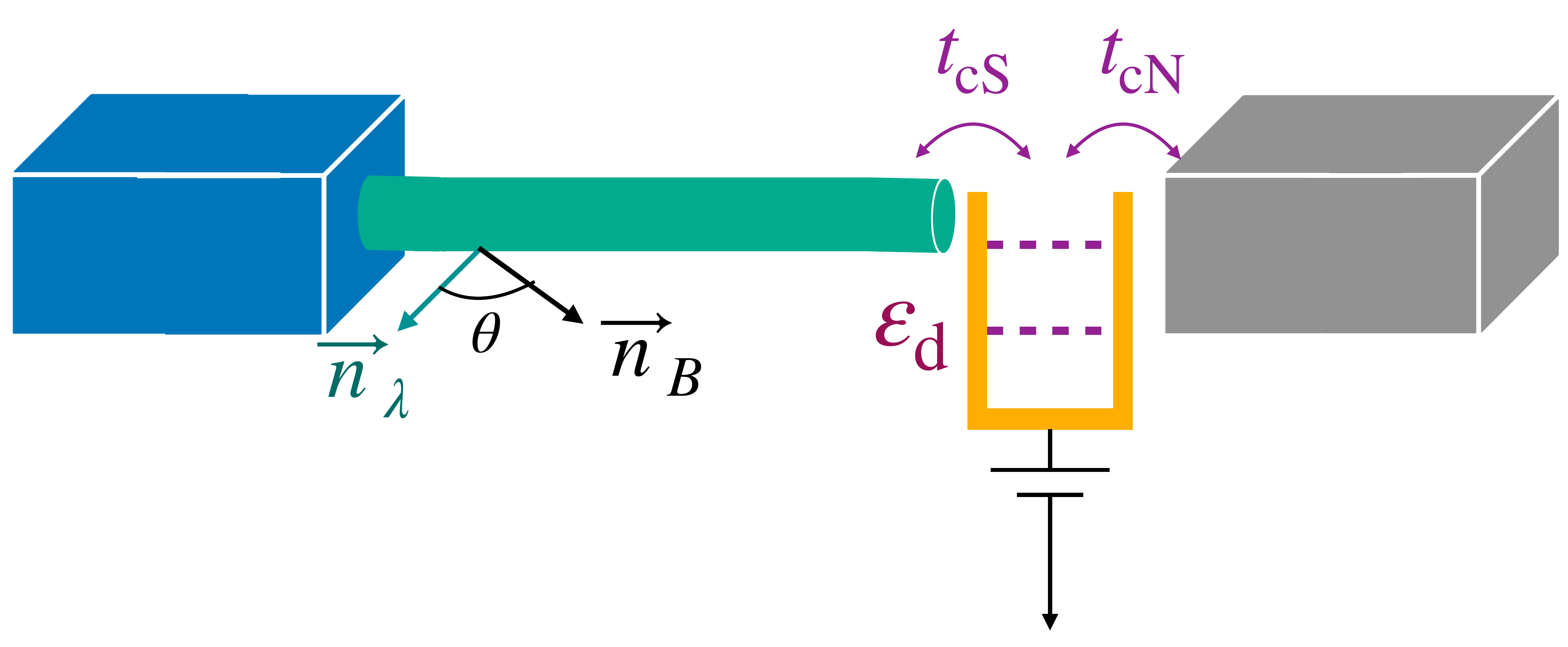}
     \caption{Sketch of the setup. A finite-length superconducting wire with spin-orbit coupling and magnetic field applied in the directions
     $\vec{n}_\lambda$ and $\vec{n}_B$, respectively is contacted to a normal lead through a quantum dot. The energy levels of the quantum dot $\varepsilon_{\rm d}$ can be tuned by means of a gate voltage. The tunneling contacts between the wire and the quantum dot and the normal lead and the quantum dot are, respectively $t_{\rm cS}$ and $t_{\rm cN}$.}
     \label{sketch}
 \end{figure}

 It is important to mention that perfectly localized Majorana modes should exist in ideal infinite-length wires while experiments are  carried out in wires of finite size where some degree of hybridization between these modes  takes place. This results in a pair of  delocalized fermionic particle-hole subgap excitations. 

The impact of the finite size of the wires in the behavior of the local density of states has been studied in Refs. \cite{stanescu2012close,sarma2012splitting,chevallier2013andreev,dmytruk2018suppression} in configurations where  $\vec{B}$ and the SOC are perfectly perpendicular. In Refs. \cite{sarma2012splitting,rainis2013towards,danon2017conductance,ricco2018majorana} the corresponding features in the conductance have been also analyzed. Other realistic ingredients, like the imperfect contact between the wire and the normal contact, as well as the proximity effect to a non-topological superconductor and the hybridization with a quantum dot 
have also been discussed in Refs. \cite{prada2017measuring,deng2018,schuray2020signatures,ricco2021topological}. 
On the other hand, the presence of 
subgap states associated to unintended quantum dots formed in the 
junction between the proximitized region and the normal contacts has 
been reported in several experiments 
\cite{vaitiekenas2020flux,valentini2021nontopological}. 
The tendency to the formation of quantum dot states at the ends of the wires has been discussed in \cite{escribano2018dotformation}.
 The other relevant ingredient, which might play a role in generating peculiar effects in the conductance is disorder \cite{pan2020physical}.

The aim of the present contribution is to further analyze the excitations and transport properties in a heterostructure  consisting of a finite-size superconducting wire with spin orbit coupling and magnetic field contacted to a normal system through a quantum dot. 
In contrast to experiments where 
quantum dots are spontaneously formed in an uncontrolled fashion, we study 
here  the situation depicted in Fig. \ref{sketch}, where the quantum dot 
properties could be ideally controlled by external gates.
 We focus on two aspects of the spectrum of the wire-dot system that have not been previously analyzed: (i) the effect of the supragap states, which plays a relevant role as the coupling between the quantum dot and the topological wire becomes strong (ii) the effect of the tilt angle between the SOC and the magnetic field, away from the perfect perpendicular orientations. Previous literature on the effect of the tilt focused 
mainly on the robustness of the topological phase and the Majorana modes
in infinite wires\cite{rex2014tilting,osca2014effects,klinovaja2015fermionic} or on the behavior of the dc Josephson current \cite{aligia2020tomography}.

We characterize the spectrum of  the wire coupled to the quantum dot in a regime of parameters that is relevant for experiments. We show that zero-energy states may appear in the 
spectrum within the topological as well as in the non-topological phases of the wire, and identify  under which conditions this may happen. In the topological phase, we show that only for a very weakly coupled quantum dot the zero-energy modes and other low-energy features of the spectrum can be unambigously associated with the Majorana modes. As the coupling between the wire and the quantum dot increases, the hybridization with supragap states play a role and introduce
additional features. 

For these systems we also
analyze the two-terminal conductance and identify the features that would allow to distinguish situations where the wire is in the topological
phase from those where the wire is a non-topological superconductor. We 
complement this analysis with the study of the associated current noise. 
This quantity has been analyzed within the topological phase in  previous works without quantum dots  \cite{bolech2007observing,nilsson2008splitting,golub2011shot,zazunov2016low,jonckheere2019trijunction,perrin2021noise}. We  show that this response provides useful information on the topological
nature of the  wire. 

 The paper is organized as follows. The model and the theoretical treatment are presented in Sections \ref{sec:model} and \ref{sec:form}, respectively.  Results are presented in Sec. \ref{sec:results}.
 Sec. \ref{last} is devoted to summary and conclusions. 
 
\section{Model}\label{sec:model}
\subsection{Superconducting wire}
We consider a lattice version of the model for topological superconducting wires introduced in Refs. \cite{wires1,wires2}, with arbitrary orientations of the magnetic field and SOC. 

The corresponding 
Hamiltonian for the bulk system in the Nambu basis
${\bf c}_k=(c_{k \uparrow}, c_{k \downarrow},c^{\dagger}_{-k \downarrow},-
c^{\dagger}_{-k \uparrow})^T$ reads
\begin{equation}
    \label{hs}
H_{\rm S}=\frac{1}{2} \sum_k {\bf c}_k^{\dagger} {\cal H}_k
{\bf c}_k, 
\end{equation}
with the Bogoliubov-de-Gennes (BdG) Hamiltonian matrix given by
\begin{equation}\label{hbdg}
    {\cal H}_k= \tau^z \otimes \left[\xi_k \sigma^0 -  \lambda_k \vec{n}_{\lambda} \cdot \vec{\sigma} \right] -  B_{\rm S} \tau^0 \otimes \; \vec{n}_{\rm B} \cdot
    \vec{\sigma} + \Delta \tau^x \otimes \sigma^0.
\end{equation}
Here, $\vec{\sigma}=\left( \sigma^{x},
\sigma^{y},\sigma^{z}\right) $ and 
$\vec{\tau}=\left( \tau^{x},
\tau^{y},\tau^{z}\right) $
are the Pauli matrices acting, respectively, in the spin and particle-hole degrees of freedom, while 
$\sigma^0, \; \tau^0$ are the 2$\times$2 unitary matrices. 
$\xi_k = - 2 t_{\rm S} \cos (k a) - \mu_{\rm S}$ is the kinetic dispersion relation relative to the chemical potential $\mu_{\rm S}$ being $t_{\rm S}$ the nearest-neighbor hopping parameter in the 1D lattice  along the wire. 
The lattice constant is $a$, while $\lambda_k=  2 \lambda \sin (k a)$
is the amplitude of the SOC oriented in the direction
$\vec{n}_{\lambda}$. Notice that close to the bottom of the band
(for  $\mu_{\rm S} \simeq -2t_{\rm S}$) this Hamiltonian is equivalent to the
continuum Hamiltonian of Refs. \cite{wires1,wires2} with the usual quadratic dispersion relation, $\xi_k \propto k^2$ and the spin-orbit interaction $\lambda_k \propto k$. The other parameters in the Hamiltonian of Eq. (\ref{hbdg}) are the magnetic field 
oriented along  $\vec{n}_{\rm B}$, which 
introduces a Zeeman splitting of amplitude $B_{\rm S}$  and the pairing potential
$\Delta$. 
This model has a
topological phase for some values of the parameters and for a range of orientations of relative orientations of the SOC and the magnetic field.
The evaluation of topological invariants \cite{tewa,budich,aligia2020tomography,daroca2021phase},
leads to the following analytical expressions for the boundaries  
\begin{equation}
|2t_{\rm S}-r| <|\mu_{\rm S} | < |2t_{\rm S}+r|, \;\;\;\;\;\;\;\;\;   |\cos(\theta)| <|\Delta|/B_{\rm S} <1. \label{bound}
\end{equation}
with $r =\sqrt{B_{\rm S}^{2}-\Delta^{2}}$ and $\cos(\theta)=\vec{n}_{\lambda} \cdot \vec{n}_B$. 

\begin{figure}[htb]
\centering
\includegraphics[width=\linewidth]{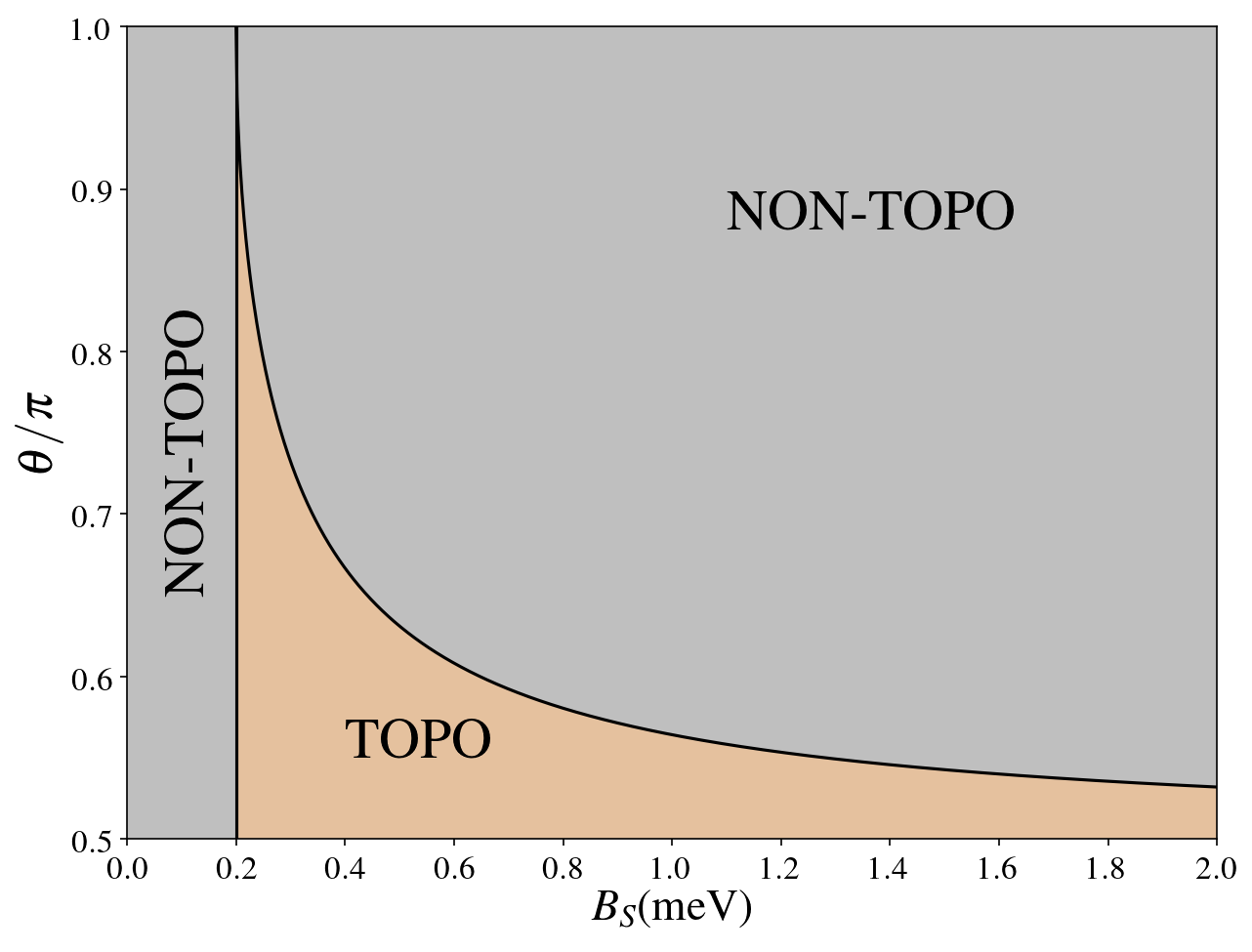}
     \caption{Phase diagram according to Eq. \eqref{bound} for $\Delta=0.2$meV. TOPO and NON-TOPO denotes, respectively, the topological and non-topological phase.}
     \label{PhaseDiagram_BvsTheta}
 \end{figure}

In the present work, we focus on values of $\mu_{\rm S}<0$. Assuming $t_{\rm S},\; \Delta>0$, the  topological phase corresponds to the range
 $-2t_{\rm S} \leq \mu_{\rm S} \leq  \mu_{\rm c}$, with
\begin{equation}\label{muc}
\mu_{\rm c}=-2 t_{\rm S} +  \sqrt{B_{\rm S}^2-\Delta^2}, \;\;\;\;\;\;\;B_{\rm S}>\Delta,
\end{equation}
and relative angles between $\vec{n}_{\rm B}$ and $\vec{n}_\lambda$ satisfying Eq. (\ref{bound}).
The corresponding phase diagram of the topological phase for $\pi/2 \leq \theta \leq \pi $ and a particular value of $\Delta$ is shown in Fig. \ref{PhaseDiagram_BvsTheta}.

\subsection{N-S junction with embedded quantum dot}
\subsubsection{Full Hamiltonian}
We consider a junction between a wire described by Eqs. (\ref{hs}), (\ref{hbdg}) and a normal conductor, with an embedded quantum dot. The full Hamiltonian reads
\begin{equation}\label{hsn}
    H=\frac{1}{2}\left[H_{\rm S}+ H_{\rm d} + H_{\rm N} + H_{\rm cont}\right],
\end{equation}
where the Hamiltonian for the superconducting wire $H_{\rm S}$ is given by Eq. (\ref{hs}) expressed in real space in a system with $N_{\rm w}$ sites and connected to a  semi-infinite BCS Hamiltonian with singlet pairing. The  Hamiltonian reads
\begin{eqnarray}\label{hsreal}
H_{\rm S} & = & - \sum_{j=1}^{\infty} \left[   {\bf c}^{\dagger}_{j} \tau^0 \otimes \left( t_{\rm S} \sigma^0+  i \lambda_j \; \vec{n}_{\lambda} \cdot \vec{\sigma}  \right) {\bf c}_{j+1} + \text{H.c.} \right]  \\
& + & \sum_{j=1}^{\infty} {\bf c}^{\dagger}_{j} \left[\Delta \tau^x \otimes \sigma^0 - B_j \tau^0 \otimes \vec{n}_{B} \cdot \vec{\sigma}  -\mu_{\rm S} \tau^z \otimes \sigma^0  \right]{\bf c}_{j},\nonumber
\end{eqnarray}
with ${\bf c}_{j} =\left( c_{j,\uparrow}, c_{j,\downarrow},c^{\dagger}_{j,\downarrow}, -c^{\dagger}_{j,\uparrow} \right)$ and 
\begin{eqnarray}
\lambda_j&=& \lambda, \;\;\;\;\; 1 \leq j \le N_{\rm w}, \;\;\;\;\;\;\;\; \lambda_j= 0,\;\;\; j > N_{\rm w}, \nonumber \\
B_j&=& B_{\rm S}, \;\;\;1 \leq j \le N_{\rm w}, \;\;\;\;\;\;\;\; B_j= 0,\;\;\; j > N_{\rm w}.
\end{eqnarray}
We work in the coordinate system where the $z$-axis is oriented along the wire and the $x$-axis is oriented along the SOC. We consider $\vec{n}_B$ in the $(x,z)$ plane.
The Hamiltonian for the normal contact is a 1D tight binding Hamiltonian with hopping $t_{\rm N}$,
\begin{eqnarray}\label{hn}
    H_{\rm N} &=& 
\sum_{j=1}^{\infty} \left[-t_{\rm N} \left({\bf b}_{j}^{\dagger } \tau^z \otimes \sigma^0
{\bf b}_{j+1}+\text{H.c.}\right) \right. \nonumber \\
& &  \left. -  \; {\bf b}_{j}^{\dagger } \left(\mu_{\rm N} \tau^z \otimes \sigma^0+ B_{\rm N} \tau^0 \otimes \vec{n}_{B} \cdot \vec{\sigma} \right)
 {\bf b}_{j} \right],
\end{eqnarray}
where we are using the notation ${\bf b}^{\dagger}_j=\left(b^{\dagger}_{j,\uparrow},
b^{\dagger}_{j,\downarrow},b_{j,\downarrow},- b_{j,\uparrow} \right)$ for the Nambu spinor within the normal lead.
 The quantum dot is modeled by 
\begin{eqnarray}\label{hd}
   H_{\rm qd} &=& - {\bf d}^{\dagger} \left(\varepsilon_{\rm d} \tau^z \otimes \sigma^0 + B_{\rm d} \tau^0 \otimes \vec{n}_{B} \cdot \vec{\sigma} \right)
   {\bf d},
\end{eqnarray}
being ${\bf d}=\left(d_{\uparrow}, d_{\downarrow},
d^{\dagger}_{\downarrow}, -d^{\dagger}_{\uparrow}\right)^T$ the Nambu spinor that describes the corresponding degrees of freedom.
 $\vec{B}_{\rm d}$ is the magnetic field  and
$\varepsilon_{d}$ is the local energy of the quantum dot, which can be controlled by an external gate voltage. Notice that we are not explicitly considering here the effect of the Coulomb interaction at the quantum dot.
This is because we focus on the regime where the Zeeman field introduced by the magnetic field dominates, hence, the main effect of the Coulomb interaction would be to introduce
a renormalization of $B_{\rm d}$ and $\varepsilon_{\rm d}$.

The last term of Eq. (\ref{hsn}) is the tunneling-contact between the quantum dot and the S and N leads. It reads
\begin{equation}\label{hcont}
    H_{\rm cont} = -
\left[\left( t_{\rm cS} {\bf c}^{\dagger}_{1} \; + \;  t_{\rm cN} {\bf b}^{\dagger}_{1} \right) \tau^z \otimes \sigma^0 {\bf d} + \text{H. c.} \right],
\end{equation}
where the label $\ell=1$ denotes the sites of the superconducting and normal chains at the boundary with the quantum dot.

\subsubsection{Effective Hamiltonian for the wire hybridized with the quantum dot}
It is useful to guide the study of the transport properties with an analysis of the spectral properties of the finite-size superconducting wire with SOC and magnetic field connected to the quantum dot.

Within the topological phase, 
the Majorana  modes localized at the left/right ($l/r$) ends of the wire can be generically represented as
  $\eta_{\nu}= \gamma^{\dagger}_{\nu}+ \gamma_{\nu}$, with
 $\nu=l,r$, being $\gamma^{\dagger}_{\nu}, \gamma_{\nu}$ creation and anihilation regular fermionic operators. These are related to
 the fermionic operators defining the Hamiltonian of Eq. (\ref{hsreal}) as follows,
\begin{equation}\label{modes}
    \gamma^{\dagger}_{\nu}=e^{i\delta_\nu} \left[ \cos(\theta_\nu/2) c^{\dagger}_{\nu \uparrow} + e^{i\varphi_{\nu}} 
    \sin(\theta_\nu/2) c^{\dagger}_{\nu \downarrow}  \right].
\end{equation}
$c^{\dagger}_{\nu \sigma}$ denotes a linear combination of the operators entering Eq. (\ref{hsreal}), involving a certain number of sites 
close to the $\nu$-end of the wire with a weight of the form $w_{j,\nu} \propto e^{i k_F j a} e^{-j a/\xi_M}$. Here $j$ counts sites starting at the 
$\nu$-end of the wire,
$k_F$ is the Fermi wave vector  and $\xi_M$ is the localization length of the Majorana modes. The angular parameters
$(\delta_{\nu}, \theta_{\nu}, \varphi_{\nu})$ define generalized Bloch coordinates  \cite{aligia2020tomography} and they describe the
phase and the angular coordinates of the spin of the particle component of the Majorana mode. 
For the present model, and focusing on $\vec{n}_B$ and $\vec{n}_{\lambda}$ in the $(x,z)$ plane, 
they satisfy $\delta_l=-\delta_r=\pi/4$ and $\varphi_l=-\varphi_r$.
In wires shorter than $2 \xi_M$ there is some degree of  hybridization of the Majorana modes, which results in subgap quasiparticles 
with energies $\pm \varepsilon_M$. We introduce the fermionic creation and annihilation operators  $\Gamma^{\dagger}_M,\;\Gamma_M $ to describe these modes, with
\begin{equation}\label{gamam}
    \Gamma^{\dagger}_M= \eta_r+i \eta_l,
\end{equation}
being $\eta_{l,r}$ the Majorana modes defined from Eq. (\ref{modes}). We consider the following effective Hamiltonian for the quantum dot hybridized with 
the finite-length 
topological superconducting wire,
\begin{eqnarray}\label{heff}
    H_{\rm eff} &=& H_{\rm qd}+ \sum_{n}E_{n} \Gamma^{\dagger}_{n} \Gamma_{n} + \varepsilon_M \Gamma_M^{\dagger} \Gamma_M \nonumber \\
    &+& \sum_{\sigma} \left(t_{\sigma} \Gamma_M^{\dagger} d_{\sigma} + \Delta_{\sigma} \Gamma_M d_{\sigma} + \text{H.c.} \right) \nonumber \\
    & + &\sum_{\sigma,n} \left(t^{\prime}_{n,\sigma} \Gamma^{\dagger}_{n} d_{\sigma} +  \Delta^{\prime}_{n,\sigma} \Gamma_n d_{\sigma}+ \text{H.c.} \right).
\end{eqnarray}
The first term is the Hamiltonian of the quantum dot defined in Eq. (\ref{hd}) and the second term represents the set of supra-gap excitations
of the superconductor bulk. 
In an infinite-length system, these define a continuum but in a finite wire, they consist of a set of discrete modes (labeled with $n$)
with energies $E_n$ above the superconducting gap. 
The third term is the effective Hamiltonian for the low-energy fermionic modes defined in Eq. (\ref{gamam})
and the last terms describe the hybridization of quantum dot with the Majorana modes as well as with the supra-gap states of the wire. 
The terms with $\Delta^{\prime}_{n,\sigma}$ and $t^{\prime}_{n,\sigma}$ can be dropped for weakly coupled quantum dots. In such case,
the effective parameters are calculated by projecting the contact Hamiltonian
defined in Eq. (\ref{hcont}) on the fermionic states of Eq. (\ref{gamam}). 
To this end, we substitute Eq. (\ref{modes}) in Eq. (\ref{gamam}). The result is
\begin{equation}
\Gamma^{\dagger}_M= \sum_{\sigma}\left(\alpha_{\sigma} c^{\dagger}_{1,\sigma} + \beta_{\sigma} c_{1,\sigma}\right) + \ldots,
\end{equation}
where we write explicitly only the component related to the  site $j=1$ of the wire since this  is the one contacted with the quantum dot
and we indicate the remaining components with $\ldots$. The coefficients are
\begin{eqnarray}
\alpha_{\uparrow} &=& w_{1r} e^{i\delta_r} \cos(\theta_r/2) \pm i w_{1l} e^{i \delta_l} \cos(\theta_l/2), \nonumber \\
\alpha_{\downarrow} &=& w_{1r} e^{i\delta_r} e^{i\varphi_r} \sin(\theta_r/2) \pm i w_{1l} e^{i \delta_l} e^{i\varphi_l} \sin(\theta_l/2), \nonumber \\
\beta_{\uparrow} &=& w_{1r} e^{-i\delta_r} \cos(\theta_r/2) \pm i w_{1l} e^{-i \delta_l} \cos(\theta_l/2), \nonumber \\
\beta_{\downarrow} &=& w_{1r} e^{-i\delta_r} e^{-i\varphi_r} \sin(\theta_r/2) \pm i w_{1l} e^{-i \delta_l} e^{-i\varphi_l} \sin(\theta_l/2), \nonumber \\
\end{eqnarray}
where $w_{1r}, \; w_{1l}$ are the weights of the  states generated by $\gamma_{r}^{\dagger}$ and $\gamma_{l}^{\dagger}$ on the  site $j=1$. Notice that, as the length of the wire
becomes much larger than $\xi_M$, the weights behave as $w_{1l} \rightarrow 0$.
Here, we see that the projection of $c^{\dagger}_{1,\sigma}$ on the low-energy states of the wires is
\begin{equation}
c^{\dagger}_{1,\sigma} = \alpha^*_{\sigma} \Gamma^{\dagger}_M+\beta^*_{\sigma} \Gamma_M.
\end{equation}
Substituting in Eq. (\ref{hcont}) we get
\begin{equation}\label{param}
t_\sigma=t_{\rm cS} \alpha^*_{\sigma},\;\;\;\;
\Delta_{\sigma}=t_{\rm cS} \beta^*_{\sigma}.
\end{equation}
A similar Hamiltonian with $\Delta^{\prime}_{\sigma,n} = t^{\prime}_{\sigma,n}=0$ was  presented in Ref. \cite{prada2017measuring,schuray2020signatures} where the analysis focused on weakly coupled quantum dots. Here, we also analyze the
effect of strongly coupled quantum dots where the hybridization  with the supra-gap states also plays a role. The effect
of such states is represented by the operators $\Gamma_{n}$. In our study, we shall focus only on the effect of the lowest-energy supragap states. 
The hybridization parameters can be calculated from
\begin{equation}\label{supra}
\Gamma^{\dagger}_n= \tilde{\alpha}_{n,\sigma} c^{\dagger}_{1,\sigma} + \tilde{\beta}_{n,\sigma} c_{1,\sigma} + 
\tilde{\alpha}_{n,\overline{\sigma}} c^{\dagger}_{1,\overline{\sigma}} + \tilde{\beta}_{n,\overline{\sigma}} c_{1,\overline{\sigma}}+
\ldots,
\end{equation}
with $\overline{\uparrow}=\downarrow, \;\overline{\downarrow}=\uparrow$. 
The parameters entering the effective Hamiltonian read
\begin{equation}\label{paramp}
t^{\prime}_{n,\sigma}=t_{\rm cS} \left(\tilde{\alpha}^*_{n,\sigma} + \tilde{\alpha}^*_{n,\overline{\sigma}}\right),\;\;\;\;
\Delta^{\prime}_{n,\sigma}=t_{\rm cS} \left(\tilde{\beta}^*_{n,\sigma} + \tilde{\beta}^*_{n,\overline{\sigma}}\right).
\end{equation}

We shall mainly focus on zero energy features in the structure constituted by the superconducting wire and the quantum dot. These can take place in two conceptually 
different scenarios:  (1) The wire is in the topological phase and host low-energy modes as a result of the hybridization of the Majorana end modes. In turn, these modes hybridize with the quantum dot and zero-energy crossings may take place for selected parameters. (2) In the second scenario the wire is not in the topological phase and its states  (isolated from the quantum dot) are above the superconducting gap. 
 In the framework of the effective Hamiltonian, this case corresponds to eliminating the terms containing $\Gamma_M$  in Eq. (\ref{heff}). 
Interestingly,
 because of the hybridization with the quantum dot, low-energy states with zero energy develop inside the gap for certain parameters. Because of the Zeeman field in the quantum dot, the latter 
 behaves as a magnetic impurity within the range of parameters where it is singly occupied. Hence, the development of bound states crossing zero energy within this second scenario is akin to the case of
 Yu-Shiba-Rusinov bound states of a magnetic impurity coupled to a superconductor \cite{yu2005bound,shiba1968classical,rusinov1969theory,balatsky2006impurity}.
 In this scenario when the angle between $\vec{n}_{\lambda}$
  and $\vec{n}_B$ overcomes the critical value $\theta_{\rm c}$, there are gapless states in the wire because of the peculiar nature of this non-topological superconductor. 
  The states resulting from their hybridization with the quantum dot may also cross zero energy for some parameters. 

  Our aim is to analyze the transport features in these two situations in order to identify signatures of the topological phase.

\section{Electrical current, conductance and noise}\label{sec:form}
We consider an electrical bias $eV$ applied at the normal lead. The generated current reads
$J= e \langle \dot{N}_{\rm N} \rangle= -i 2e/\hbar \langle \left[N_{\rm N}, H \right]\rangle $ and it can be written as follows
\begin{equation}
    J= \frac{e}{\hbar} t_{\rm c N} \mbox{Re}\mbox{Tr}\left\{
    G^<_{Nd}(t,t)\right\}.
\end{equation}
 We have expressed the
mean values of the operators entering the definition of the current in terms of the Green's function matrix
\begin{equation}
    G^<_{Nd}(t,t^{\prime}) = -i \langle {\bf d}^{\dagger}(t^{\prime}) {\bf b}(t) \rangle.
\end{equation}
After operating within the Schwinger-Keldysh Green's function formalism \cite{rammer2011quantum,cuevas1996hamiltonian} we get the following expression 
\begin{eqnarray}
    J&=& \frac{e}{h}\int d \varepsilon \left\{ 
    \left[f(\varepsilon+eV)- f(\varepsilon)\right] {\cal T}_N(\varepsilon)
    \right.\nonumber \\
   & &  \left. \;\;\;\;\;\;\;\;\;\; +\left[f(\varepsilon+eV)- f(\varepsilon-eV)\right] {\cal R}_A(\varepsilon)\right\},
\end{eqnarray}
where $f(\varepsilon)$ is the Fermi-Dirac distribution function. 
As usual \cite{blonder1982transition} we have separated the contributions of the normal transmission and the Andreev reflections
${\cal T}_N(\varepsilon)$ and ${\cal R}_A(\varepsilon)$, respectively. The expressions for these two functions in terms of Green's functions are presented in Appendix \ref{apa}.
The numerical calculations for the results presented in Sec. \ref{sec:results} have been carried out by following the procedures of Refs. 
\cite{lopezsancho} and  \cite{zazunov2016low,alvarado2020boundary}, finding an excellent agreement between them.

The conductance at zero temperature $T=0$ is calculated as
\begin{equation}\label{g}
    G= \frac{dJ}{dV}= G_0 \left[{\cal T}_N(eV)+{\cal R}_A(eV)\right],
\end{equation}
where $G_0=e^2/h$ is the conductance quantum per spin channel. 
The first term of Eq. (\ref{g}) accounts for the normal transport of quasiparticles and is dominant for $eV$ above the gap, while the Andreev reflection contributes within the gap and describes the conversion of particles and holes in the normal side to Cooper pairs in the superconducting one. For a ballistic contact we expect  $G=4 G_0$ in an ordinary superconductor and $G=2 G_0$ in a topological wire
with perfectly decoupled Majorana modes. 

We  also analyze the zero-frequency noise associated to this current, $S(eV)=\int_{-\infty}^{\infty} d\tau S(t,t-\tau)$, with
$S(t,t^{\prime})=\langle \left[\delta J(t) \delta J(t^{\prime})+\delta J(t^{\prime}) \delta J(t) \right]\rangle  $, with
$\delta J(t) = \dot{N}_{\rm N}(t) - J(t)$. This expression can be also written in the present model as \cite{zazunov2016low}
\begin{widetext}
\begin{equation}
    S(eV) = \frac{e^2 t_{\rm c N}^2}{h^2} \int\frac{d \varepsilon}{2\pi} \mbox{Tr}\left\{
    G^<_{NN}(\varepsilon)
   G^>_{dd}(\varepsilon)-
   G^<_{Nd}(\varepsilon)
   G^>_{Nd}(\varepsilon) + N\leftrightarrow d
    \right\}.
\end{equation}
 \end{widetext}
The corresponding the Fano factor reads
\begin{equation}\label{fano}
F=\frac{S(eV)}{2 e J}.
\end{equation}
This quantity has been analyzed within the topological phase in  previous works without quantum dots  \cite{bolech2007observing,nilsson2008splitting,golub2011shot,zazunov2016low} and the outcome 
at zero temperature is
$F=0$ when the conductance is perfectly quantized at ($G=2 G_0$) in a topological N-S junction.

\section{Results}\label{sec:results}
We consider parameters of the Hamiltonian of Eq. (\ref{hs}) that are representative of reported experimental research in InAs wires \cite{chen2017experimental,deng2016majorana,wang2022observation}. We assign $t_{\rm S}=10meV$ in order to ensure a large bandwidth
which fits properly with the quadratic dispersion relation of the continuum model for the wires and
$a=\sqrt{2mt_{\rm S}}/\hbar\simeq 15 nm$. For the normal wire we consider  a large bandwidth, in order to guarantee its hybridization with all the states of the superconducting structure and the quantum dot, $t_{\rm N}=2 t_{\rm S}$.

\subsection{Magnetic field perpendicular to the direction of the SOC}
We start by analyzing the results for the most studied configuration, which corresponds to $\theta=\pi/2$.

Notice that, because of the combination of the s-wave superconductivity with the SOC and the magnetic field,
the effective superconducting pairing has singlet and triplet components when the pairing term of 
Eq. (\ref{hbdg}) is expressed in the basis that diagonalizes this Hamiltonian with $\Delta=0$ (see Appendix \ref{sing-trip}). 
They read, respectively, 
\begin{equation}\label{deltas}
\Delta^{\rm S}_k=\frac{\Delta B}{\sqrt{B^2+\lambda_k^2}}, \;\;\;\;\; \Delta^{\rm T}_k=- \frac{\Delta\lambda_k}{\sqrt{B_{\rm S}^2+\lambda_k^2}}.
\end{equation}
The singlet component acts on particles at the different bands -- see Eq. (\ref{wirekit}) -- while the
triplet component acts intra-band.
In the topological phase the Fermi energy of the system with $\Delta=0$ is in the gap between the two bands. Hence, when
$\Delta$ is switched on,  the triplet pairing
represented by $\Delta^{\rm T}_k$ is dominant.

The estimate for the localization length of the Majorana zero mode is $\xi_M=(\hbar v_F)/\Delta_{\rm eff}$ \cite{klinovaja2012composite}, being $v_F$ the Fermi velocity and 
$\Delta_{\rm eff}$ being the effective pairing for the corresponding values of the rest of the parameters. 
In the topological phase the relevant  pairing is defined by the triplet component at the Fermi point $k_F$, 
$\Delta_{\rm eff} \simeq \Delta^{\rm T}_{k_F}$, which leads to
\begin{equation}\label{xim}
\xi_M \simeq \frac{\hbar v_F \sqrt{B_{\rm S}^2+\lambda_{k_F}^2}}{\Delta \lambda_{k_F a}},
\end{equation}
with $\lambda_{k_F a}= 2 \lambda \sin(k_F a)$,
$-2 t_{\rm S} \cos(k_F a) \simeq \mu$ and $\hbar v_F \simeq -2 t_{\rm S} a \sin (k_F a)$.

 Considering 
$\vec{n}_B$ along the $z$-axis (parallel to the wire) and 
 $\vec{n}_\lambda$ along the $x$-axis (perpendicular to the wire), the parameters entering Eqs. (\ref{gamam}), (\ref{heff}) and (\ref{param}) are
 \begin{equation}
     \delta_r=-\delta_l=-\pi/4, \;\;\; \varphi_r=-\varphi_l=-\pi/2, \;\;\; \theta_r=\theta_l\equiv\theta_M.
 \end{equation}
 Therefore, the parameters of Eq. (\ref{param}) defining the effective Hamiltonian of Eq. (\ref{heff}) read
  \begin{eqnarray}\label{param1}
 t_{\uparrow}&=&t_{\rm cS} e^{i \pi/4}\left(w_{1r} -w_{1l} \right) \cos(\theta_M/2), \nonumber \\
t_{\downarrow}&=&t_{\rm cS} e^{i \pi/4} i \left(w_{1r} + w_{1l} \right) \sin(\theta_M/2) , \nonumber \\
 \Delta_{\uparrow}&=&t_{\rm cS} e^{-i \pi/4}\left(w_{1r} + w_{1l} \right) \cos(\theta_M/2), \nonumber \\
\Delta_{\downarrow}&=&-t_{\rm cS} e^{-i \pi/4} i \left(w_{1r}  - w_{1l}\right) \sin(\theta_M/2),
\end{eqnarray}
 
Along this section, we fix $\Delta =0.2meV$ and $\lambda=0.5 meV$- We also consider $B_{\rm S}=1 meV$, in which case
$\mu_{\rm c} \simeq -19.02 meV$. Within the topological phase, we focus on $\mu_{\rm S}=1.01 \mu_{\rm c}$. For these parameters
$\xi_M \simeq 104 a$. 
 We expect that only wires
with lengths significantly larger than $ 2 \xi_M$
are free from effects related to the hybridization of the Majorana end modes. In this section we analyze in detail a chain with
$L=250a$, which corresponds to a length slightly larger than $ 2 \xi_M$.

\begin{figure}
\subfloat{\includegraphics[width= 3.3in]{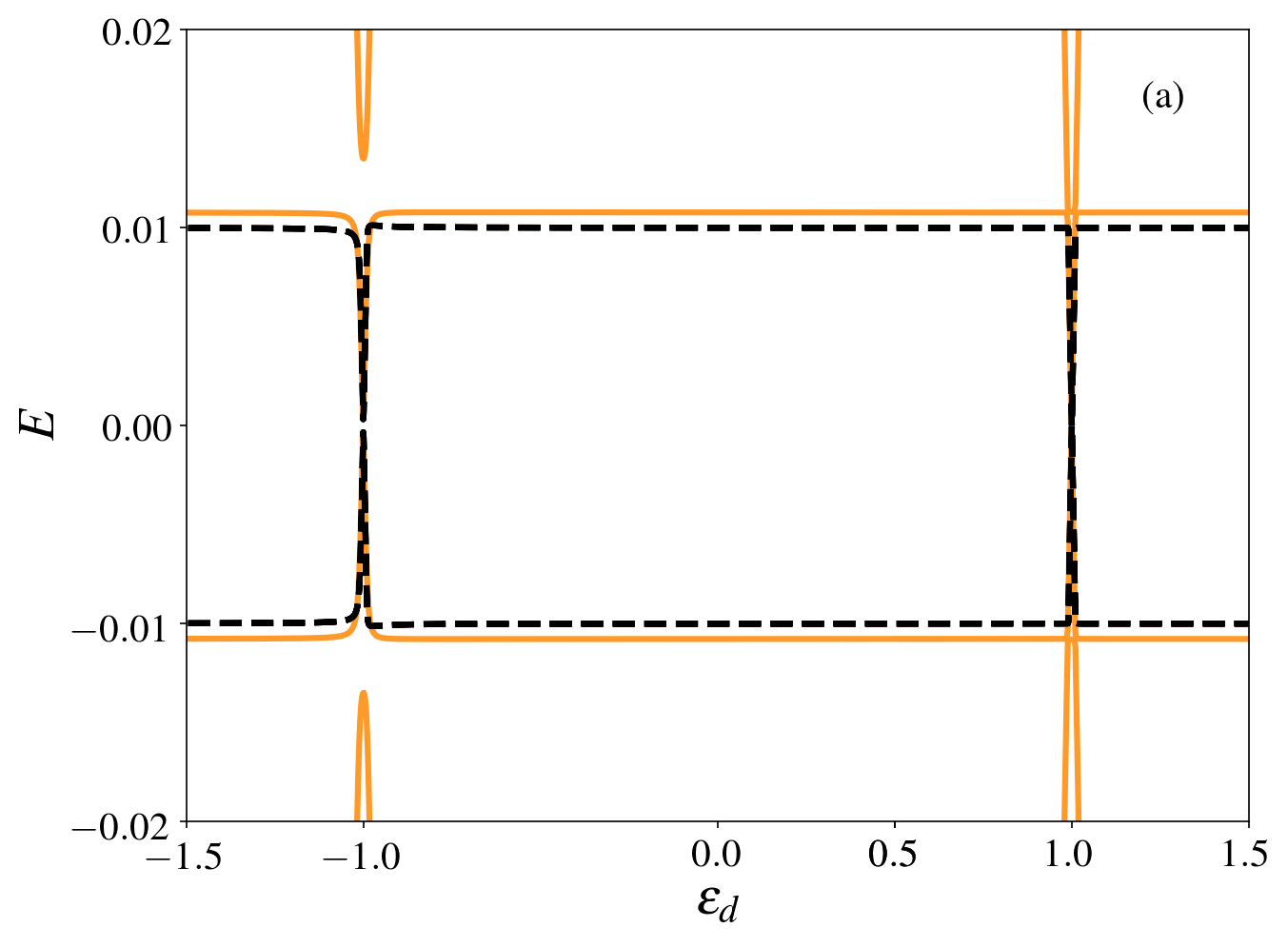}}\vspace{-0.1cm}
\subfloat{\includegraphics[width= 3.3in]{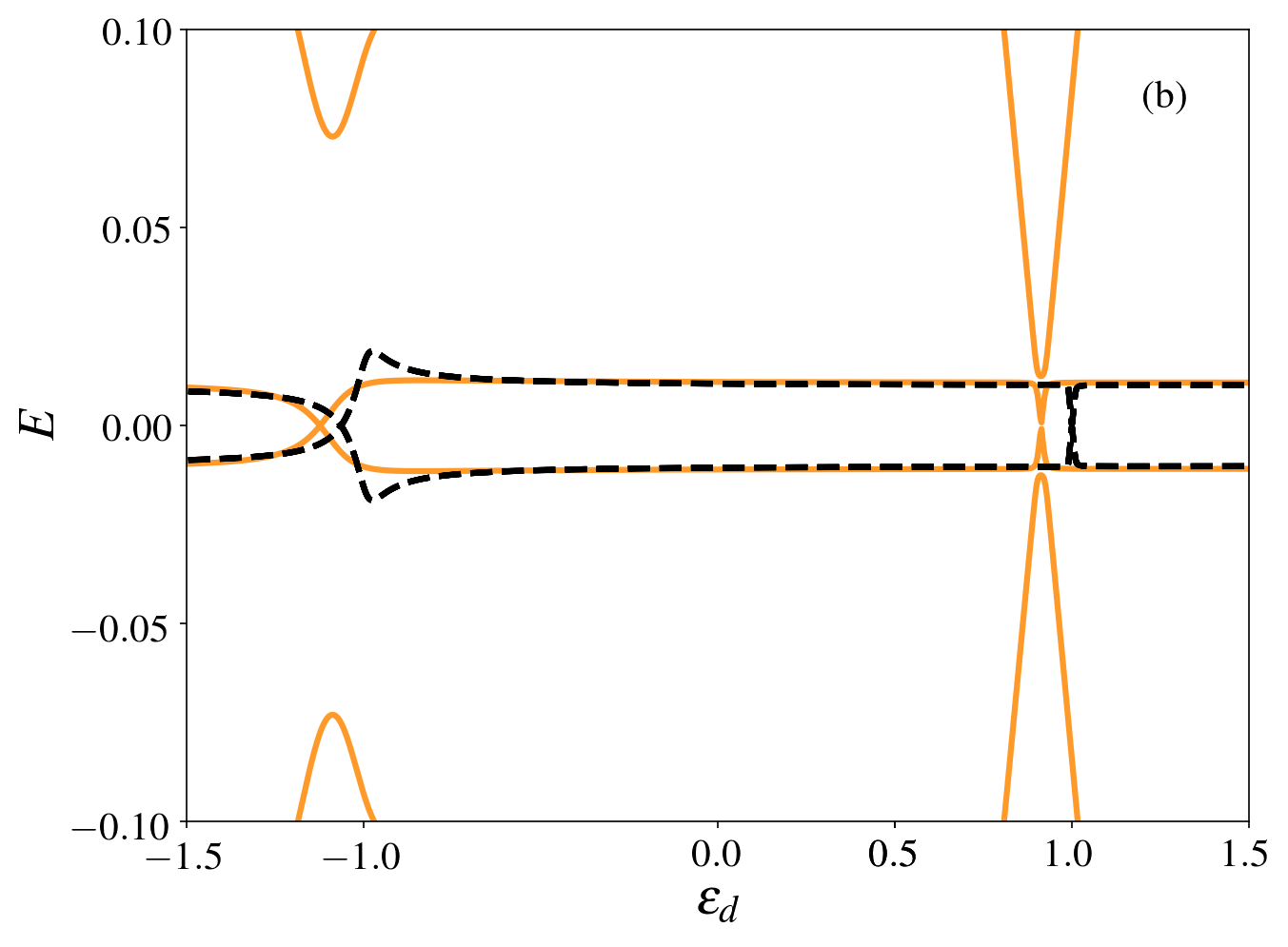}}\vspace{-0.1cm}
\subfloat{\includegraphics[width= 3.3in]{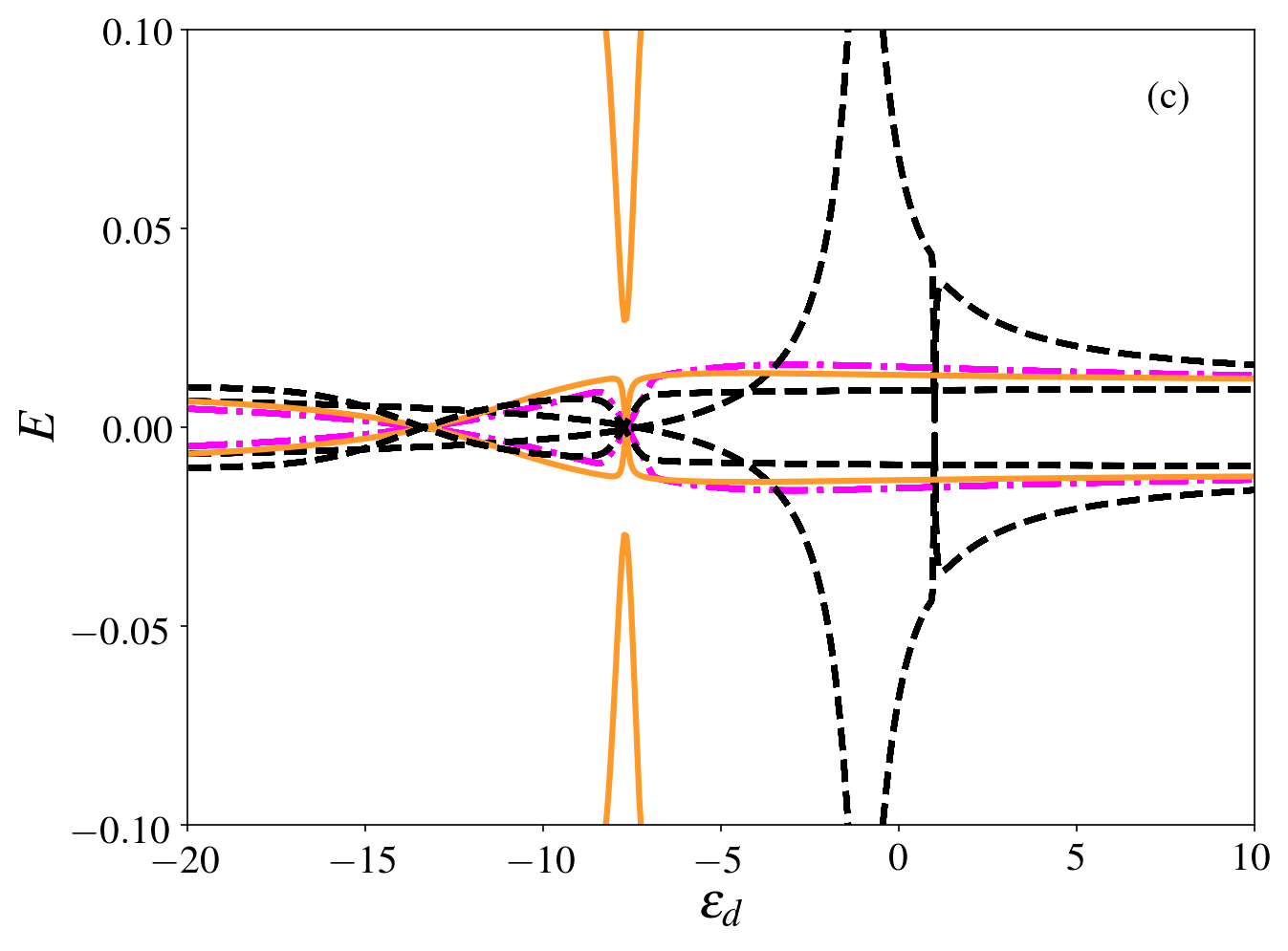}}\vspace{-0.1cm}
\caption{Sub-gap spectrum as a function of the energy $\varepsilon_d$ of the quantum dot for a system with $L=250 a$, $B=1meV$, $\lambda=0.5meV$,
$\Delta=0.2 meV$ and $\mu=1.01\mu_{\rm c}$.
Solid and dashed lines correspond, respectively, to the results obtained with the exact and the effective Hamiltonian. (a): $t_{\rm c S}=0.1meV$.  The  (non-vanishing) parameters of the effective Hamiltonian are:
$w_r=0.02$, $w_l=0.008$, $\varepsilon_M=0.01 meV$, $\theta_M=0.045\pi$. The value of the tilt was obtained from the results of  the exact diagonalization of the wire without quantum dot.
(b):   $t_{\rm c S}=1meV$.
(c): $t_{\rm c S}=10meV$. In panel (c), the results of the effective Hamiltonian including the contribution of two supragap states is shown in magenta dash-dotted lines. 
The corresponding parameters are:  $E_{1}=0.11meV$,
$E_{2}=0.14meV$, $\tilde{\alpha}_{1,\uparrow}=1.5\tilde{\beta}_{1,\uparrow}, \;\tilde{\beta}_{1,\uparrow}=0.073  $,
$\tilde{\alpha}_{2,\downarrow}=1.7\tilde{\beta}_{2,\downarrow}, \;\tilde{\beta}_{2,\downarrow}=0.08meV $, while  the rest are the same as in the other panels. The results obtained with 
$H_{\rm eff}$ considering  $\tilde{\alpha}_{n,\sigma}=\tilde{\beta}_{n,\sigma}=0$ are shown in black dotted lines.
}\label{Fig:fig4}
\end{figure}

\subsubsection{Topological phase: $\mu < \mu_{\rm c}$} 
We consider here a wire with parameters in the topological phase, $\mu<\mu_{\rm c}$ 
with a length $L=N_{\rm w} a$, with $N_{\rm w}=250$. In this scenario, 
the Majorana end modes of the topological phase are not fully decoupled but are weakly hybridized. As a result of their hybridization, 
the spectrum of the wire contains low-energy subgap 
particle and hole states at finite energies $\pm \varepsilon_M$ which can be described by Eq. (\ref{gamam}).

We start by analyzing the spectral properties that result from the hybridization of these modes with the quantum dot, assuming that it  has the same  Zeeman field as the wire
($B_{\rm d}=B_{\rm S}=1meV$). Such a quantum dot behaves as a magnetic impurity hybridized with the superconductor. 
The subgap states arising from its hybridization with the superconducting wire may exhibit different regimes, which depend on the degree of coupling between these two systems. In the weak-coupling regime (corresponding to $t_{\rm cS} \ll \Delta$)
 we simply expect bonding and antibonding-like combinations of the low-energy states of the wire with those of the quantum dot, when these two systems are in resonance. This happens when one of the Zeeman levels of the quantum dot
has energies $\varepsilon_{\rm d, \sigma}=\varepsilon_{\rm d} \pm B_{\rm d} \simeq \varepsilon_M, \; \sigma=\uparrow, \downarrow$. For strong coupling, the supra-gap states also play a role, as we will discuss.

The subgap spectrum originated in the hybridization of the topological modes with the polarized quantum dot is analyzed in Figs. \ref{Fig:fig4} for the case of a very weakly connected quantum dot (a) an intermediate hybridization (b) and a strongly coupled quantum dot (c). These figures show the exact spectra, as calculated by diagonalizing the Hamiltonian of the wire coupled to the quantum dot as well as the prediction based on diagonalyzing the effective Hamiltonian of Eq. (\ref{heff}). 
In the three cases exhibited in the figure, we can identify features of the decoupled superconducting wire (see the straight lines at energy 
$\pm \varepsilon_M$) for sufficiently large values of $|\varepsilon_{\rm d}|$.
For the weakly coupled quantum dot, these asymptotic values are reached when $\varepsilon_{\rm d}$ slightly departs from the resonant values determined by the Zeeman field, corresponding at
$\varepsilon_{\rm d} \pm B_{\rm d}$, associated to the $ \uparrow, \;  \downarrow$ spin states of the isolated quantum dot.
 Close to these values, we can also identify the lines crossing zero energy.
 The effect of the hybridization between these two systems can be identified in two features of the spectrum shown in Fig.  \ref{Fig:fig4} : (i) the opening of a small gap between the two lowest-energy levels of both particle and hole sectors at $\varepsilon_{\rm d}=\pm B_{\rm d}$, and (ii)  the shift in the crossing at zero energy. 
In the weak coupling limit (see top panel of Fig. \ref{Fig:fig4}) the low-energy sector of the spectrum can be accurately reproduced by the effective Hamiltonian of Eq. (\ref{heff})
upon neglecting the coupling of the quantum dot with the supragap states ($t^{\prime}_{\sigma,n}=\Delta^{\prime}_{\sigma,n} =0$). This is shown in 
dashed lines in Fig.  \ref{Fig:fig4}. For the present parameters the particle and hole excitations of the Majorana  modes have a small angle $\theta_M$ with respect to the orientation of the magnetic field, which generates  a
significant asymmetry in the net coupling between these modes and the $\uparrow, \; \downarrow$ states of the quantum dot. This  is explicitly accounted for the effective Hamiltonian. In fact, we see in Eqs. (\ref{param1}) that the hybridization is
$\propto \cos(\theta_M)$ for the $\uparrow$ spin orientation and $\propto \sin(\theta_M)$ for the $\downarrow$ one. From this effective model, it is easy to calculate the
 crossing with the horizontal axis, for which the low-energy modes have zero energy. These crossings take place at $ \varepsilon_{\rm d} \simeq \pm B_{\rm d}-
 \delta \varepsilon_{\sigma}$, with $\delta \varepsilon_{\sigma} =
 \left(|t_{\sigma}|^2-|\Delta_{\sigma}|^2\right)/\varepsilon_M $ (see Appendix \ref{cross}). Hence, taking into account Eq. (\ref{param1}) we see that this crossing provides valuable information about the weights $w_{1,l}, w_{1,r}$ of the Majorana modes localized at the left and right end of the wire, on the first site of the wire that is connected with the quantum dot. In particular, the crossing at zero energy takes place at a value of $\varepsilon_{\rm d}$ which is shifted from the one determined by the Zeeman splitting $\pm B_{\rm d}$ by an amount
$\delta \varepsilon_{\sigma} = \left(|\Delta_{\sigma}|^2-|t_{\sigma}|^2\right)/\varepsilon_M $. After substituting Eq. (\ref{param1}) it is found
\begin{eqnarray}\label{shift}
\delta \varepsilon_{\uparrow} &=& \pm 4 \frac{t_{\rm cS}^2}{\varepsilon_M} w_{1,l} w_{1,r} \cos^2(\theta_M/2), \nonumber \\
\delta \varepsilon_{\downarrow} &= & \mp 4 \frac{t_{\rm cS}^2}{\varepsilon_M} w_{1,l} w_{1,r} \sin^2(\theta_M/2).
\end{eqnarray}
 In Ref. \onlinecite{prada2017measuring}, the gap between the low-energy levels of the effective Hamiltonian [see Eq. (\ref{eff-levels})]
at $\varepsilon_{\rm d}=\pm B_{\rm d}$
was pointed out as a measure of the non-locality of the hybridized Majorana modes. Here, we would like to highlight that such information is also
encoded in the value of $\varepsilon_{\rm d}$ for which zero-energy crossings take place. Nevertheless, the precise description of the low-energy states of the spectrum, in particular, the precise
positions of these crossings are significantly affected by the effect of the coupling with the  supragap states and we further discuss this 
feature below. 

In Fig. \ref{Fig:fig4} (b) it is shown that, as the coupling between the quantum dot and the superconducting wire becomes stronger,
the behavior of the low-energy spectrum departs from the prediction of the simplest version of the effective Hamiltonian, based only on the hybridization of the dot with the
combination of Majorana modes. The position of the zero-energy crossing is  particularly  affected, as well as the functional dependence of the lowest-energy states with 
$\varepsilon_{\rm d}$. Such a departure becomes even stronger for higher coupling, as illustrated in  Fig. \ref{Fig:fig4} (c). Remarkably, we see that the
zero-energy crossings are strongly shifted away from the Zeeman values $\pm B_{\rm d}$. These features can be reproduced by the effective Hamiltonian upon 
including the effect of the supragap states. Results are shown in the figure with dot-dashed lines. To define the effective parameters, we have followed a phenomenological approach, by
considering two lowest-energy supragap states with energies close to the ones  of  the spectrum of the exact spectrum of the wire, while we
selected the rest of the parameters in order to fit the zero-energy crossings.
The corresponding values are specified in the caption of the figure.
For comparison, we also show the results obtained with the effective Hamiltonian for the coupling of the dot with the lowest-energy modes without the hybridization with the supragap states (see dotted lines). We can identify here the diamond-type shape of this effective spectrum, characterizing strongly hybridized Majorana modes with the quantum dot, as discussed in 
Ref. \cite{prada2017measuring}. However, we see that such an effective description is not able to reproduce important features of the exact spectrum, like the crossing at zero energy. The proper description of the lowest energy states demands the consideration of the coupling with the supragap states also, as verified with the more complete effective description leading to the results shown in the dot-dashed lines.

In what follows we discuss the correspondence between these features and those expected to be observed in experiments measuring the conductance 
in these configurations of wires and quantum dots. Here we add to the description the coupling of the quantum dot to the normal lead. In addition, we consider the other end of the  wire coupled to an ordinary superconductor [see sites $j=N_{\rm w}+1,\ldots$ in Eq. (\ref{hsreal})].
We assume the same magnetic field applied throughout  the wire, the quantum dot and the normal lead.

\begin{figure}[htb]
\centering
\includegraphics[width=\linewidth]{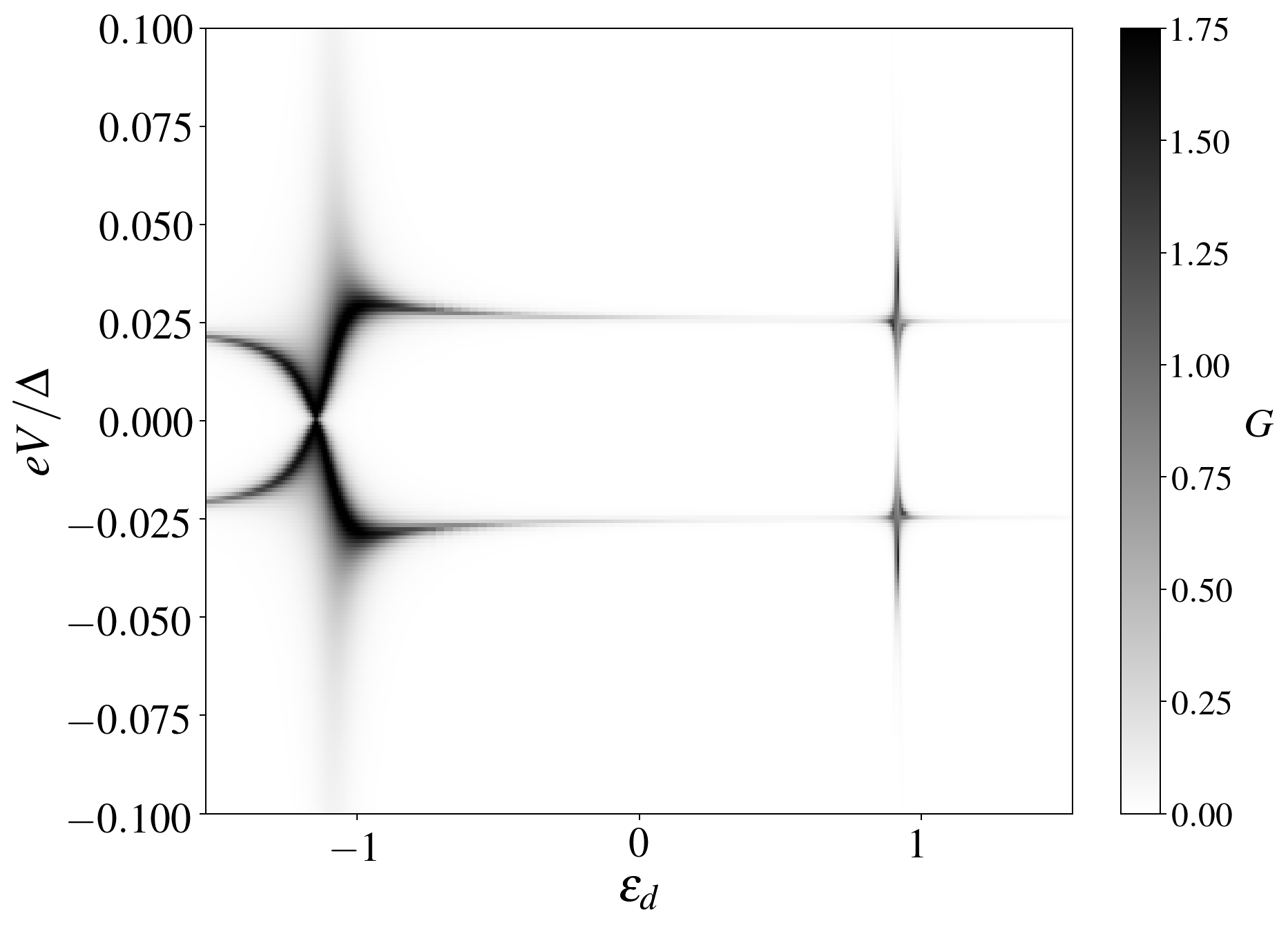}
\caption{Total conductance as a function of the energy of the quantum dot level, $\varepsilon_{\rm d}$ (in units of $meV$), and the applied bias voltage $V$ in units of $\Delta$. The quantum dot is coupled with $t_{cN}=1.0meV$ to the normal lead. The other parameters are the same as in  Fig. \ref{Fig:fig4} (b). }\label{Fig:fig5bis}
\end{figure}

Results for the conductance as a function of the dot level energy and the bias voltage are shown in Fig. \ref{Fig:fig5bis} for the same parameters of Fig. \ref{Fig:fig4} (b). We can identify in the conductance a similar behavior as in the spectrum. In particular,
high values of the conductance at zero-bias at energies $\varepsilon_{\rm d}$ close to those corresponding to the zero-energy crossings in the spectrum.

\begin{figure}[htb]
\centering
\includegraphics[width=\linewidth]{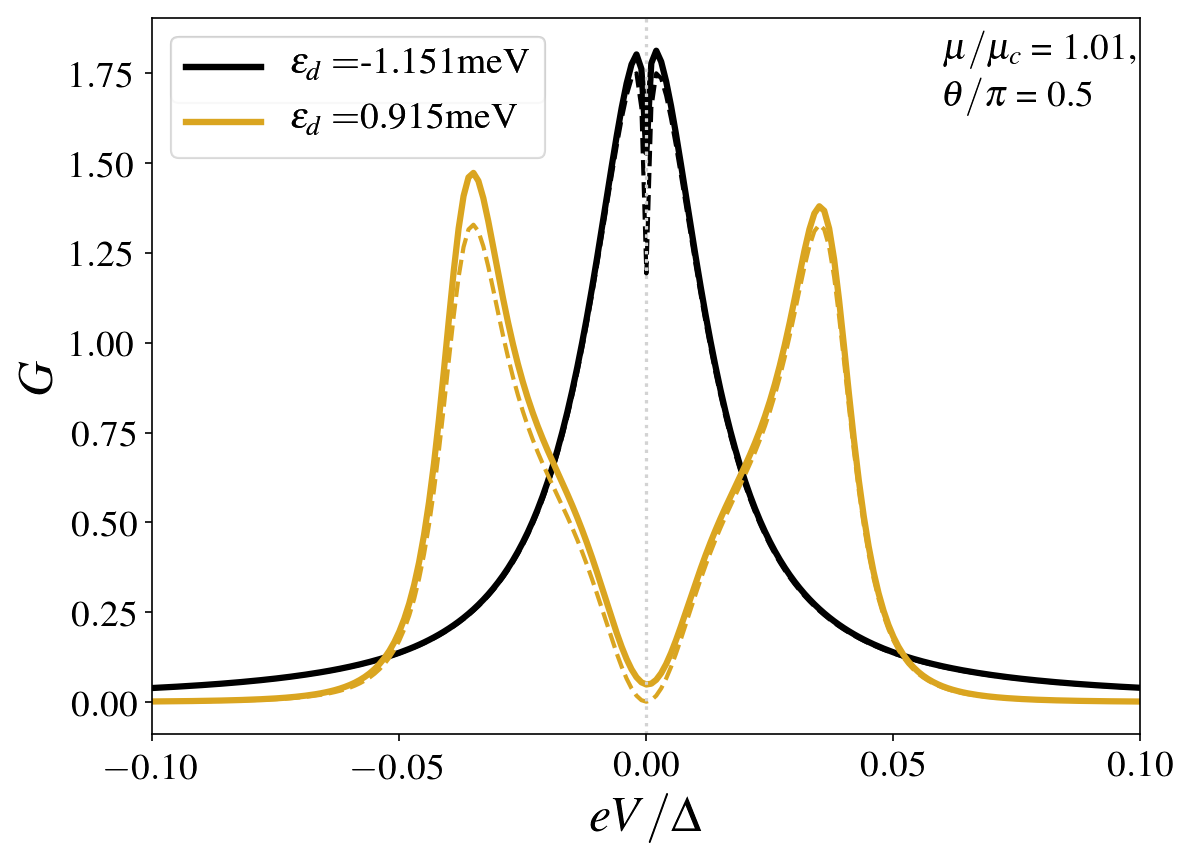}
\includegraphics[width=\linewidth]{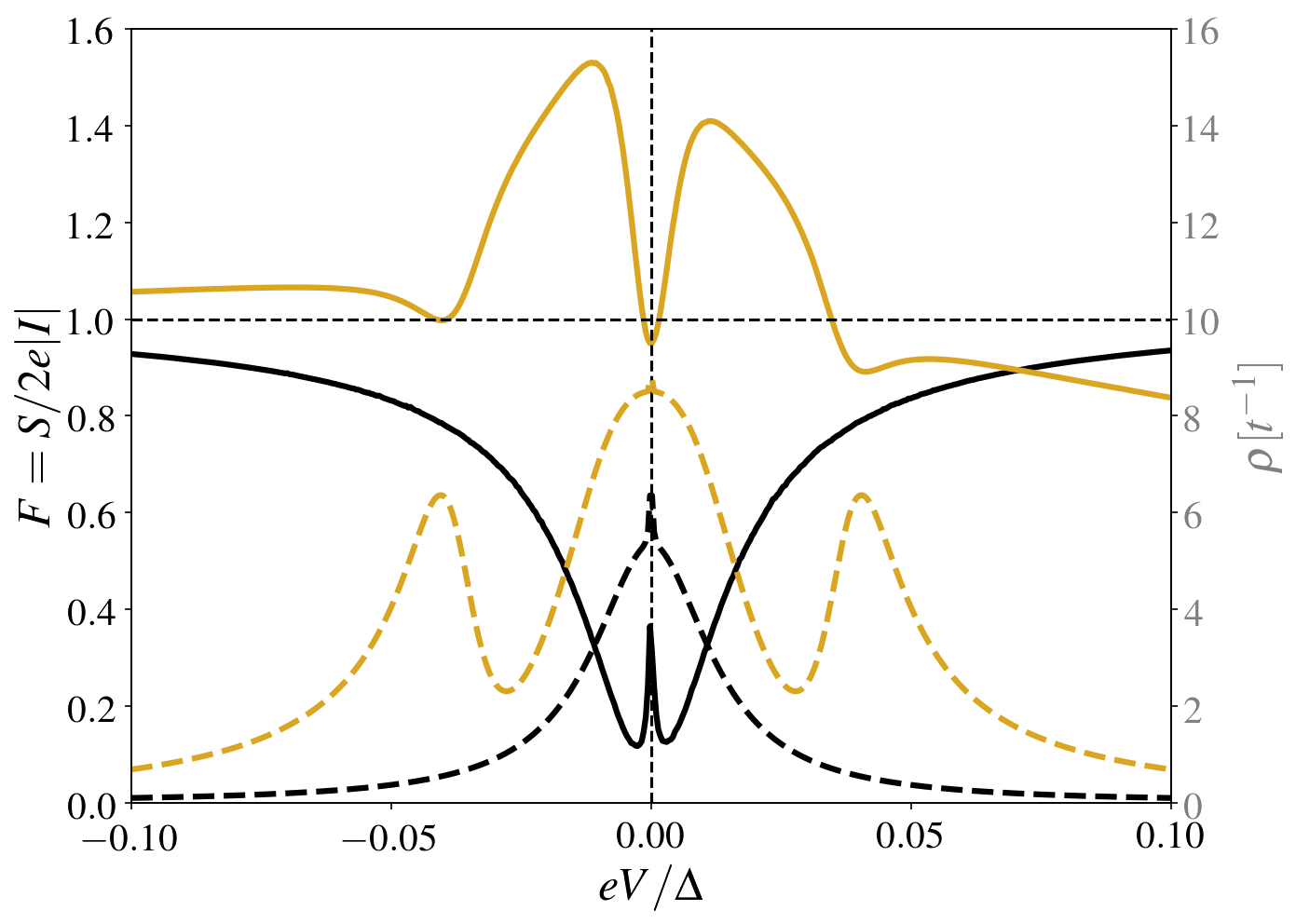}
\caption{Top: Total conductance, in units of $G_0=e^2/h$, as a function of the  bias voltage $V$ with the energies $\varepsilon_{\rm d}$ corresponding to the zero-energy
crossings. 
Bottom: Fano factor (solid lines) and LDOS (dashed lines) of the quantum coupled to both reservoirs for the same parameters of the top panel. 
The quantum dot is coupled with $t_{\rm cN}=0.5meV$ to the normal lead. The other parameters are the same as in  Fig. \ref{Fig:fig4} (b).}\label{Fig:fig_GtvseV_comparacion_eds_topo_}
\end{figure}

In order to gather more information on the zero-bias response at these points we show in the top panel of Fig. \ref{Fig:fig_GtvseV_comparacion_eds_topo_}
the behavior of the conductance as a function of the bias voltage for these particular values of $\varepsilon_{\rm d}$, along with the contribution 
of the Andreev reflection processes (see dashed lines). For the value of  $\varepsilon_{\rm d}$ associated to  the $\uparrow$-Zeeman state of the quantum dot,
the zero-bias conductance  is large albeit lower than the ideal quantized value $2 G_0$ expected for the pure Majorana zero modes in semi-infinite wires. 
 Instead, 
for the value of $\varepsilon_{\rm d}$ associated to the hybridization of the $\downarrow$-Zeeman state of the quantum dot, the zero-bias conductance is very low and there is a pronounced
splitting into two peaks departed from $V=0$. In both cases we  see that  the conductance is almost completely due to Andreev processes within this
range of low $V$.

For the same parameters, we also show the Fano factor $F$ defined in Eq. (\ref{fano}) in the bottom panel of 
Fig. \ref{Fig:fig_GtvseV_comparacion_eds_topo_} along with the local density of states (LDOS) at the quantum dot (see dashed lines). 
In the LDOS we can identify a single zero-energy feature in the black plot, which is associated to a resonant mode resulting from  the Majorana bound states of the wire hybridized with the $\uparrow$  Zeeman level of the quantum dot slightly widen because of the coupling to the normal lead. Instead, in the yellow plot we observe a wide zero-energy peak resulting from the hybridization of the $\downarrow$ 
Zeeman level of the quantum dot with the normal lead and very weakly hybridized with the wire. We also observe side peaks at higher energies, which are associated to the hybridization of the dot level with the supragap states of the wire. Notice that, as a consequence of the strong polarization of the Majorana modes along the oposite spin orientation,
the  coupling between this level of the quantum dot and the wire is very weak, hence, the conductance is very low. 
In the two cases  shown here, we observe a rather complex behavior of
$F$ as a function of $V$ but the main feature to highlight 
is  $F \leq 1$  in the neighborhood of $V=0$ for the value  of 
$\varepsilon_{\rm d}$ associated to the  $\uparrow$-Zeeman state (see plot in black solid lines). In this case, the conductance achieves a large value at zero bias and consequently,
the Fano factor is low (although it does not reach the ideal value zero). For the other value of $\varepsilon_{\rm d}$, associated to the 
the opposite Zeeman state (see plots in yellow lines), the Fano factor is significantly larger and close to $F=1$ around $V=0$
in accordance with the low conductance.

\begin{figure}[htb]
\centering
\includegraphics[width=\linewidth]{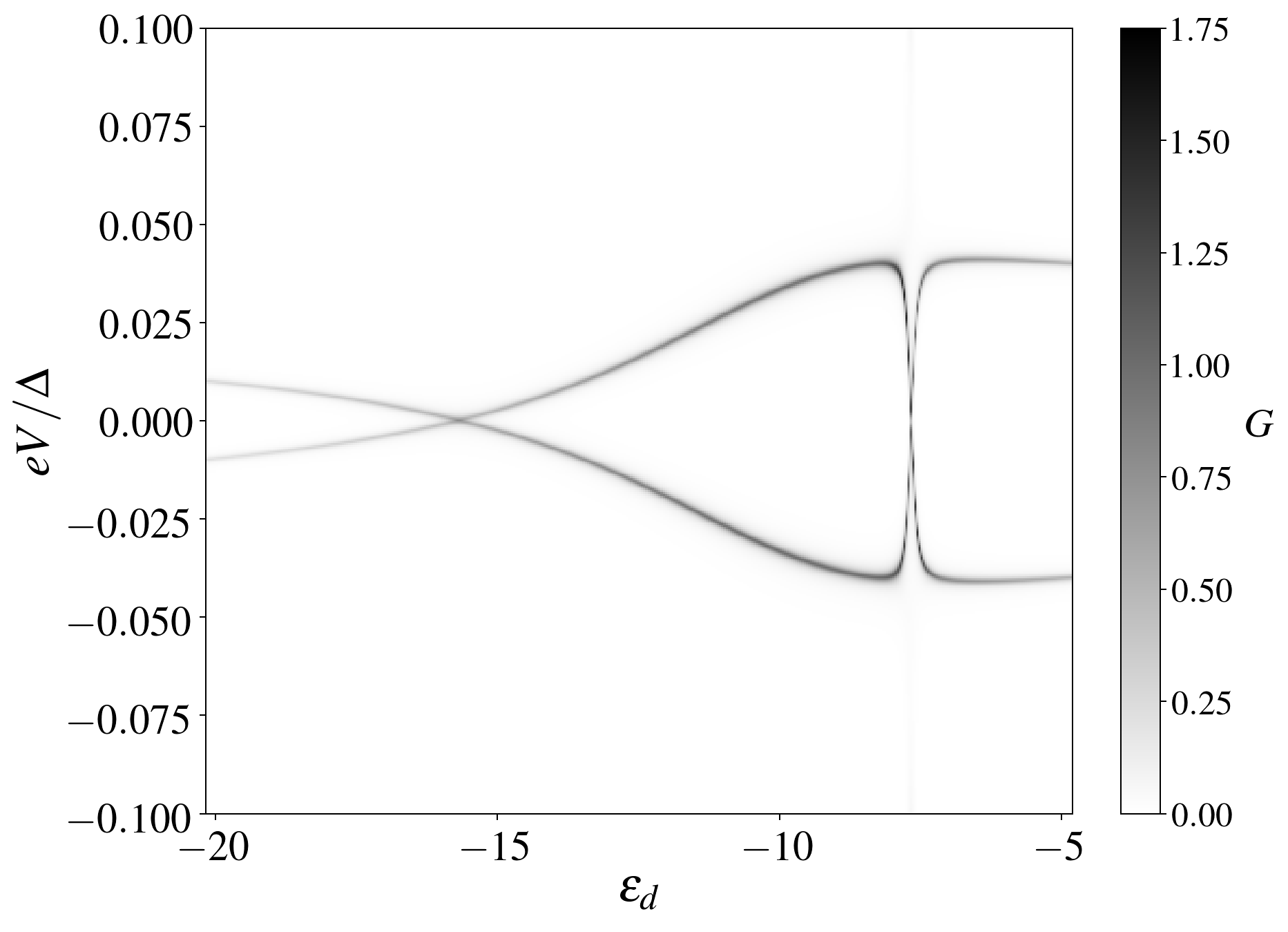}
\caption{Same as Fig. \ref{Fig:fig5bis}. The quantum dot is coupled with $t_{cS}=10meV$ to the normal lead. }
\label{Fig:fig_tcS_10meV}
\end{figure}

\begin{figure}[htb]
\centering
\includegraphics[width=\linewidth]{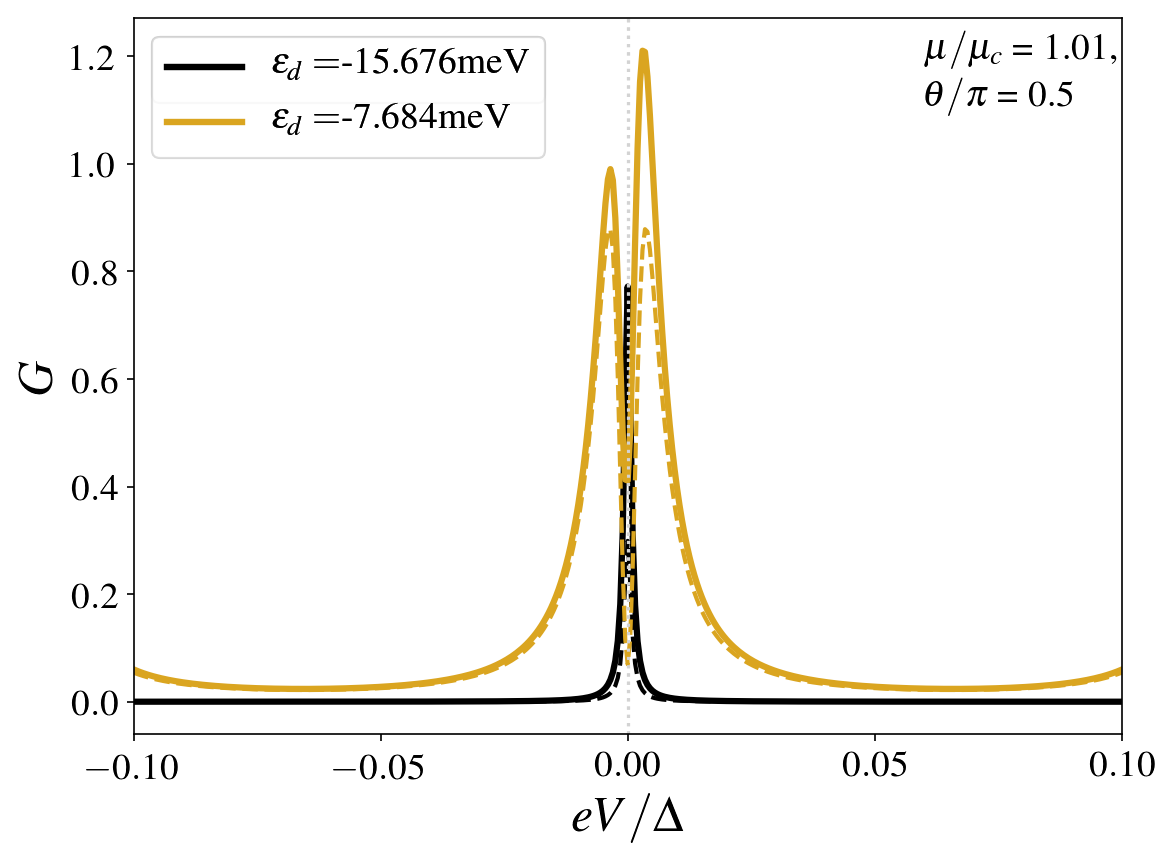}
\includegraphics[width=\linewidth]{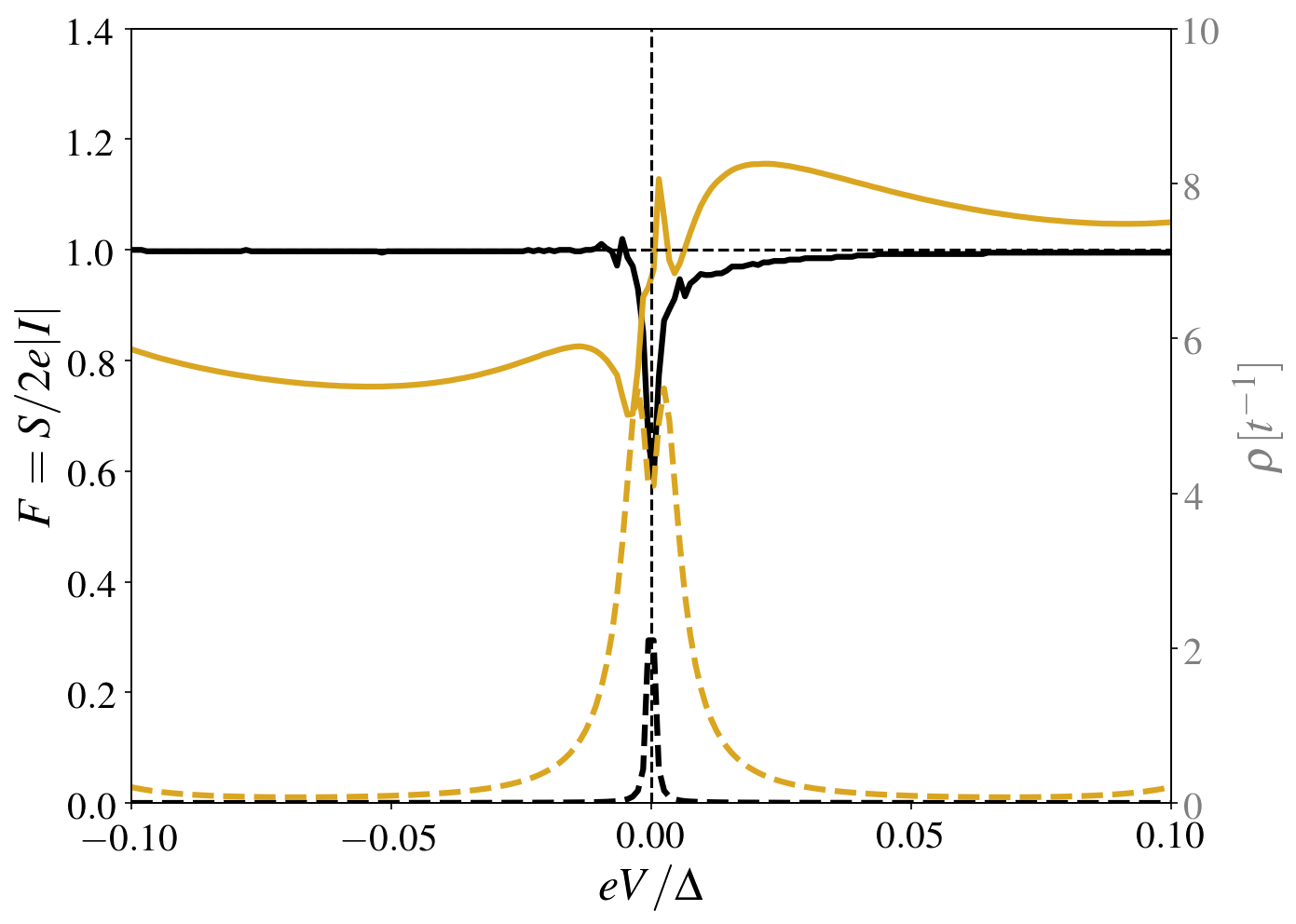}
\caption{Same as Fig. \ref{Fig:fig_GtvseV_comparacion_eds_topo_} 
for $\varepsilon_{\rm d}$ at the zero-energy crossings  of Fig. \ref{Fig:fig_tcS_10meV}.  The other parameters are the same as in Fig. \ref{Fig:fig4} (c).}
\label{Fig:fig_GtvseV_eds_max_tcS_grande_}
\end{figure}

We extend the analysis of the transport properties to the configuration where the quantum dot is strongly coupled. This corresponds to the
parameters of the bottom panel of Fig. \ref{Fig:fig4}, where we identified in the spectrum a large contribution of the supragap states. 
Results for the behavior of the total conductance as a function of $V$ and $\varepsilon_{\rm d}$ are shown in 
Fig. \ref{Fig:fig_tcS_10meV}.
As in the case of the more weakly coupled quantum dot, we see that the map for the maxima of the conductance bares a close 
resemblance with the spectrum. The corresponding plots of $G(V)$ for $\varepsilon_{\rm d}$ fixed at the values  zero-bias crossings
are shown in Fig. \ref{Fig:fig_GtvseV_eds_max_tcS_grande_} (top panel) along with the corresponding Fano factor (bottom panel).
The behavior is qualitatively similar to the weaker case analyzed in \ref{Fig:fig_GtvseV_comparacion_eds_topo_}. Namely, the zero-bias conductance has a high weight for the lowest value of
$\varepsilon_{\rm d}$, associated to the $\uparrow$-Zeeman level of the quantum dot (see plots in black lines), while it is vanishing small (in particular the Andreev component)
for the other value. The response is much lower in the present case than for the  case analyzed in Fig. \ref{Fig:fig_GtvseV_comparacion_eds_topo_}.
The behavior of the  Fano factor is much more irregular than in the weaker coupled quantum dot. However, as before, we can observe also here
a dip as $|V|\rightarrow 0$ in the case of $\varepsilon_{\rm d}$ associated to the $\uparrow$-Zeeman level (see plots in black lines).

\subsubsection{Non-topological phase: $\mu > \mu_{\rm c}$.} 
We now discuss the behavior of the low-energy spectrum of a superconducting wire with the same values of $B$, $\lambda$, $\Delta$ and 
$N_{\rm w}$ as in the cases previously analyzed, but 
with a value of the chemical potential within the non-topological phase. For these parameters, the effective superconducting gap is small but the Fermi energy crosses the two bands of the Hamiltonian of Eq. (\ref{wirekit}) and the dominant pairing 
is $\Delta^S_{k_F}$. As the wire has a finite length, the supragap spectrum consists of a sequence of discrete levels, as already mentioned. In this regime the origin of the zero-energy crossings is similar to
the ones generated by magnetic impurities in superconductors (Yu-Shiba-Rusinov states). In fact, the spectrum without impurity is gapped while the hybridization of the impurity with the supragap states results in bound states in the gap. For suitable values of the parameters,
these bound states cross at zero energy.

\begin{figure}[htb]
\centering
\includegraphics[width=\linewidth]{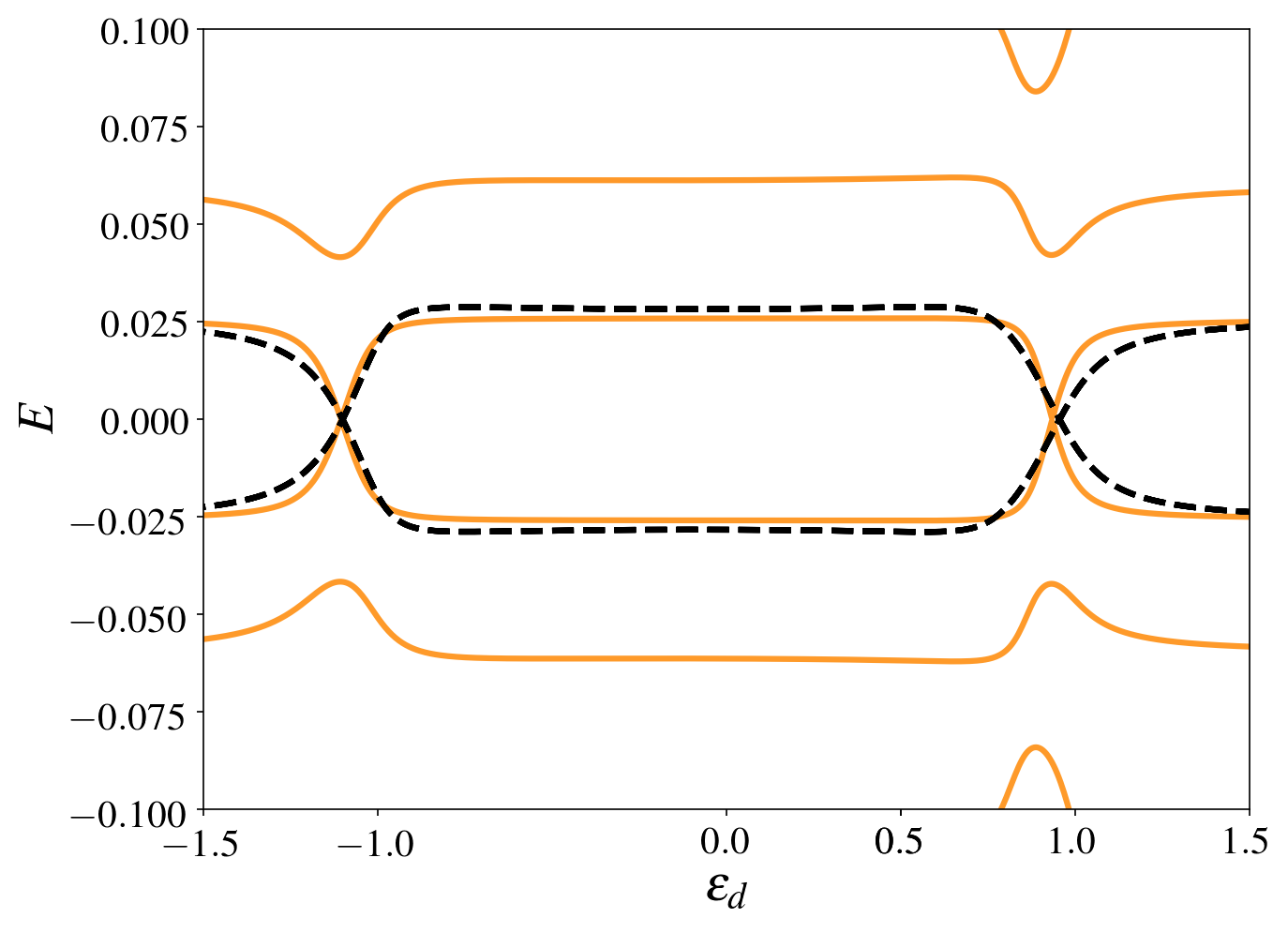}
\includegraphics[width=\linewidth]{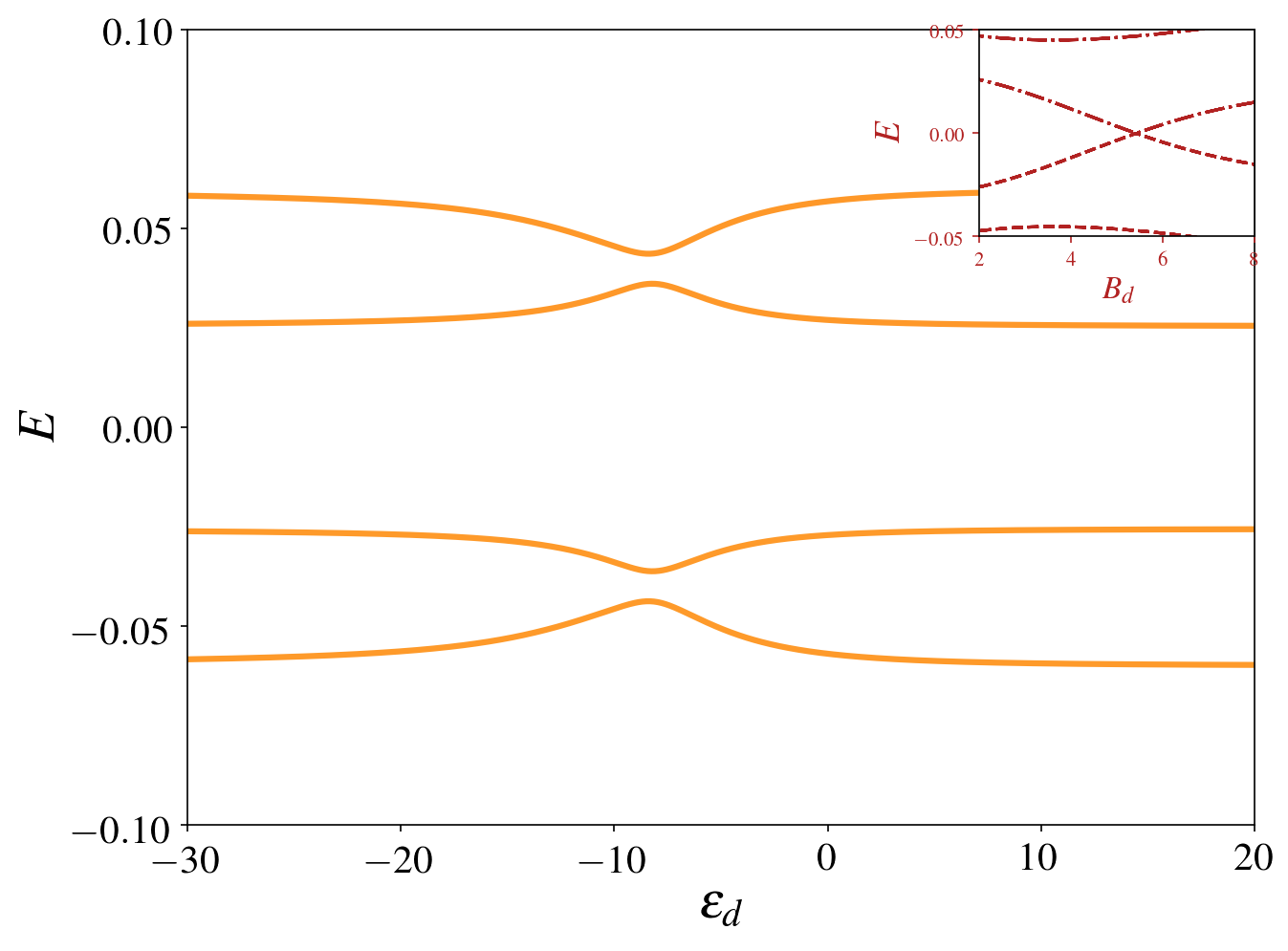}
\caption{Sub-gap spectrum as a function of the energy $\varepsilon_d$ of the quantum dot for a system with $N_{\rm w}=250 $. Solid and dashed lines correspond, respectively, to the exact and the effective Hamiltonian. The parameters are the same as in Fig. \ref{Fig:fig4}: $\lambda=0.5meV$,
$B=1meV$ but with a chemical potential above the critical value, $\mu=0.9 \mu_{\rm c}$.
Top and bottom panels correspond, respectively, to $t_{\rm c S}=1meV, \; 10meV$. 
The effective Hamiltonian corresponds to Eq. (\ref{heff}) without the terms with $\Gamma_M$ and two supragap modes with energies
$E_1=0.026meV,\; E_2=0.59meV$ and parameters $t^{\prime}_{1,\uparrow}=\Delta^{\prime}_{1,\downarrow}=0.06meV$, $t^{\prime}_{1,\downarrow}=\Delta^{\prime}_{1,\uparrow}=0.048meV$,
$t^{\prime}_{2,\sigma}=1.2 t^{\prime}_{1,\sigma}$, $\Delta^{\prime}_{2,\sigma}=\Delta^{\prime}_{1,\sigma}$.
Inset: same parameters as in the bottom panel with $\varepsilon_{\rm qd}=-13meV$ as a function of $B$. 
}\label{Fig:fig6}
\end{figure}

\begin{figure}[htb]
\centering
\includegraphics[width=\linewidth]{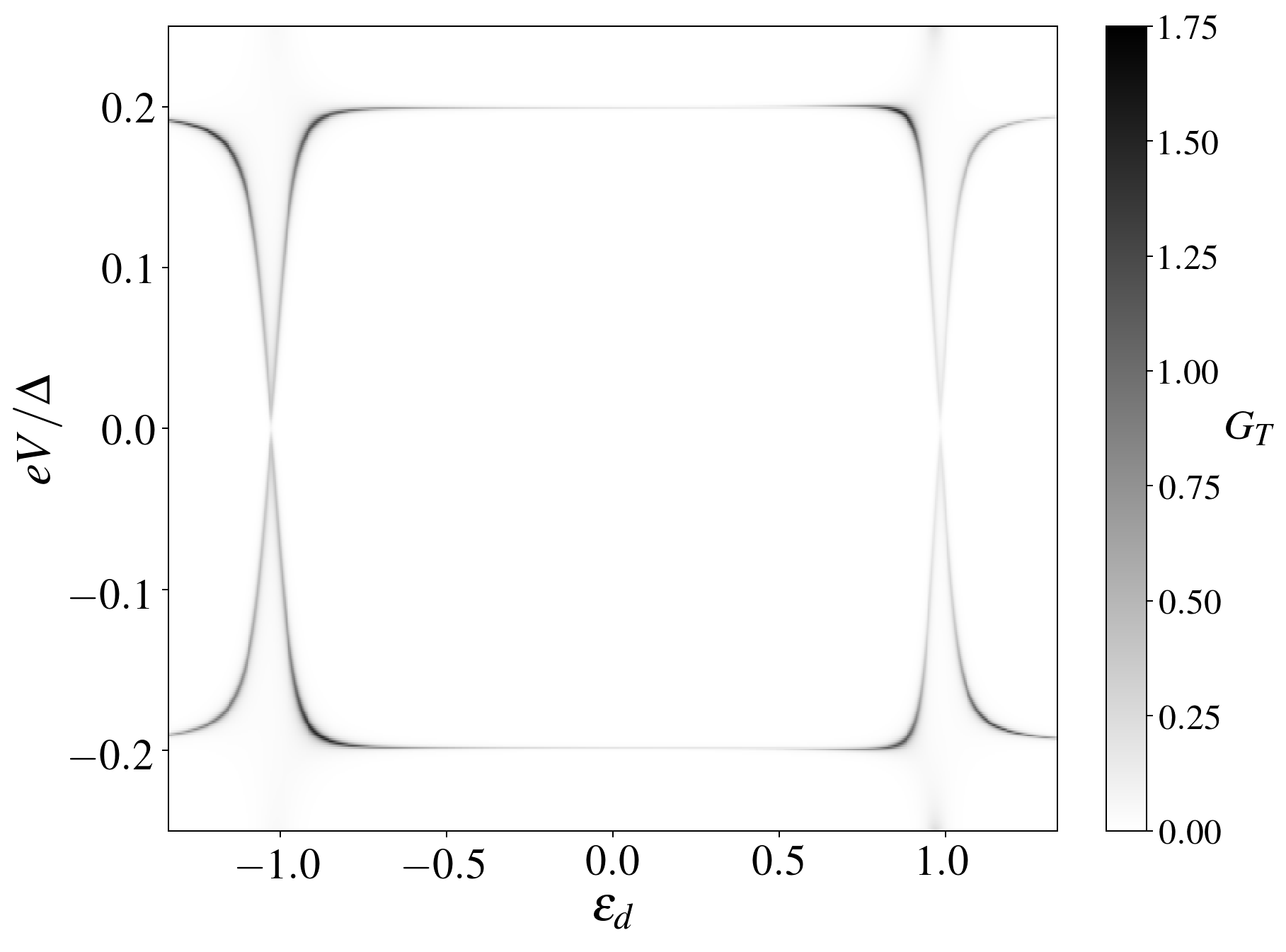}
\caption{Total conductance as a function of the energy of the quantum dot and the applied bias voltage $V$. The quantum dot is coupled with $t_{cN}=1.0meV$ to the normal lead. The other parameters are the same as in the top panel of Fig. \ref{Fig:fig6}}
\label{Fig:fig6bis}
\end{figure}

\begin{figure}[htb]
\centering
\includegraphics[width=0.5\textwidth]{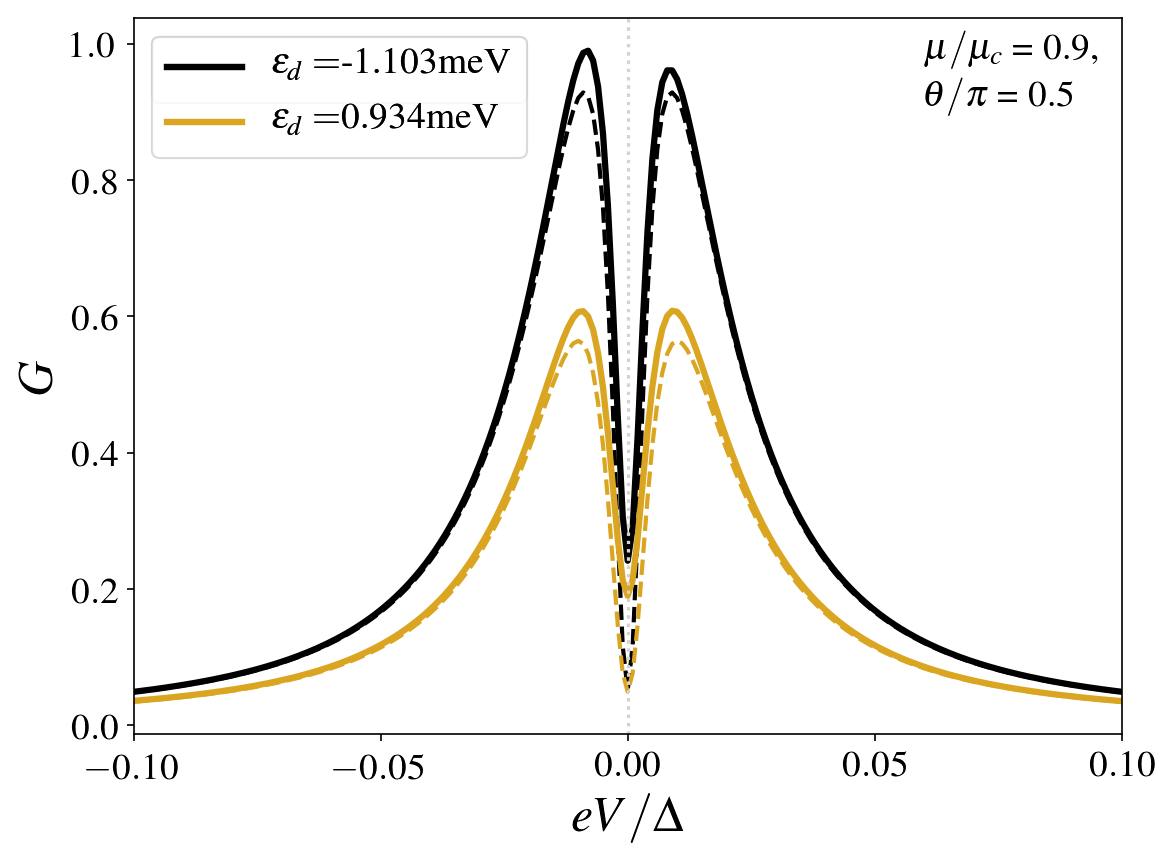}
\includegraphics[width=0.5\textwidth]{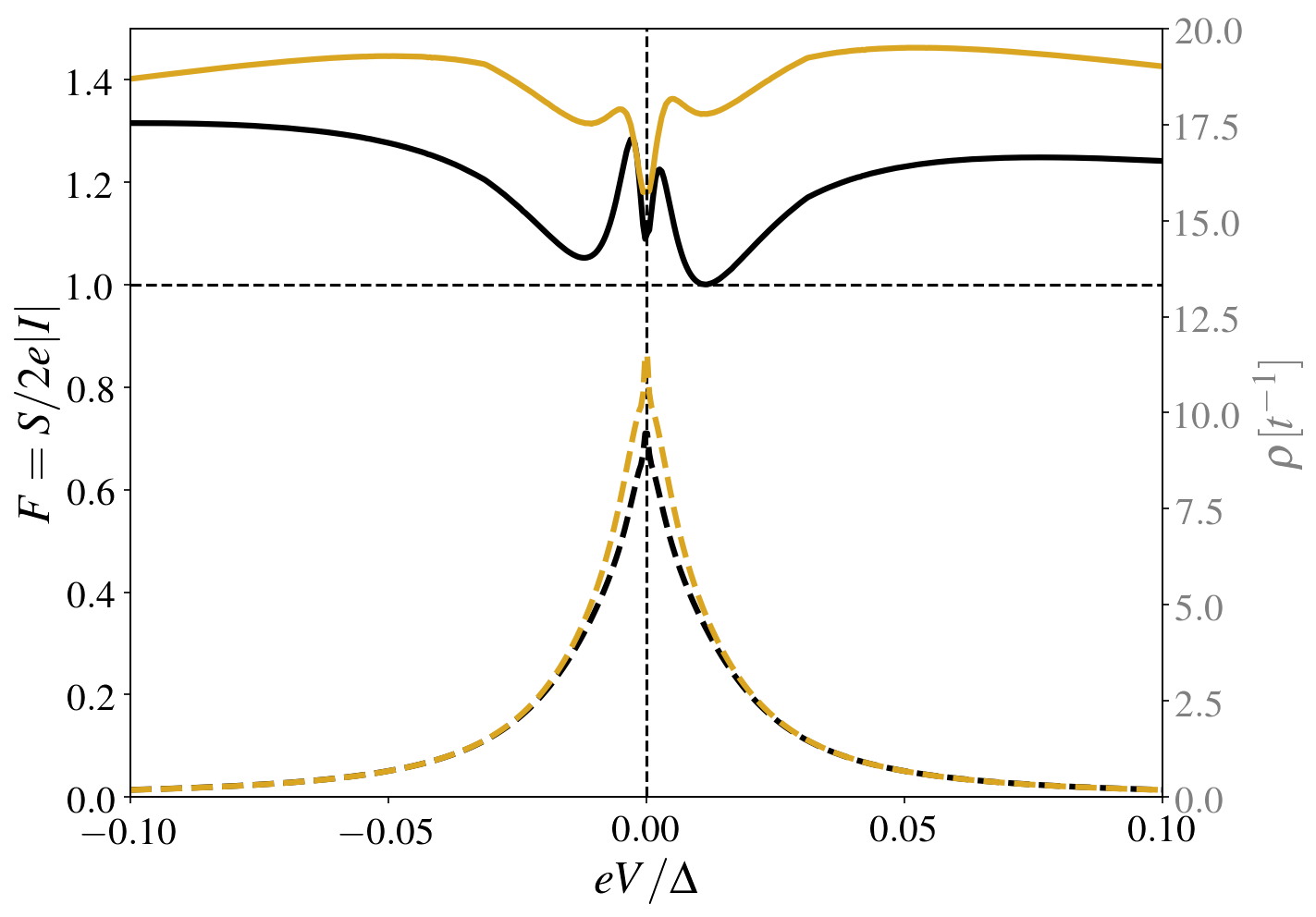}%
\caption{Same as Fig. \ref{Fig:fig_GtvseV_comparacion_eds_topo_} for the parameters of top panel of Fig. \ref{Fig:fig6}.
}
\label{Fig:fig_GvseV_both_ed_non_topo_}
\end{figure}

The low-energy spectrum for $\mu=0.9 \mu_{\rm c}$ is shown in Fig. \ref{Fig:fig6} for two different couplings with the quantum dot. The top panel
corresponds to the same strength of the coupling considered in  Fig. \ref{Fig:fig4} (b) and the main hybridization takes place between
the Zeeman levels of the quantum dot and the two lowest-energy supragap state. This is confirmed by the comparison of the results 
 of exact diagonalization with those calculated the effective Hamiltonian (see dashed lines in the upper panel of the Fig. \ref{Fig:fig6}).
We can identify   zero-energy crossings close to $\varepsilon_{\rm d}=\pm B_{\rm d}$. Unlike the topological case, the behavior is symmetric for
positive and negative values. This reflects the spin nature of these excitations, which have similar weights on the $\uparrow$ and
$\downarrow$ components. In contrast, in the topological case, we recall that the lowest energy excitations are combinations of Majorana modes defined
by fermionic operators polarized along a direction with a small tilt  $\theta_M$ with respect to the magnetic field. 

The effect of a stronger coupling between the quantum dot and the wire is analyzed in the bottom panel of the figure. As the coupling becomes larger, the quantum dot hybridizes with
a larger number of supragap states. The result is a behavior of the spectrum without crossings, which resembles  weakly bounded Yu-Shiba-Rusinov states. 
As in the case of such bound states, the crossings take place for higher values of the magnetic field in the quantum dot, $B_{\rm d}$. 
This is illustrated in the inset of the bottom panel of the figure. To analyze this limit, it is useful to recall the results expected for a magnetic impurity coupled to a singlet superconductor with a constant density of states. In that case, the crossing at zero energy of the subgap states resulting from the hybridization of the impurity with the s-wave superconductor with  density of states 
$\nu_{\rm S}\simeq \pi/4t_{\rm S}$ \cite{balatsky2006impurity}
takes place at 
\begin{equation}
\varepsilon_{\rm d}=\pm \sqrt{B_{\rm d}^2-t_{\rm cS}^2 \nu_{\rm S}},
\end{equation} 
which has real solutions for $B_{\rm d} \ge t_{\rm cS} \sqrt{\nu_{\rm S}}$. In the example shown in the bottom panel of the Fig. \ref{Fig:fig6} this corresponds to 
 $B_{\rm d} > 2.8 meV $ and this is in good agreement with the results shown in the inset.

 As in the previous section, we contrast the properties of the spectrum with the corresponding transport features. In Fig. \ref{Fig:fig6bis} we show the behavior of the conductance as a function of the energy of the quantum dot and the bias voltage. We can identify similar features as those of Fig. \ref{Fig:fig6}, in particular the 
 crossings at zero energy. 
 Fig. \ref{Fig:fig_GvseV_both_ed_non_topo_} shows the behavior of the conductance (top panel) and noise (bottom panel) for the values of $\varepsilon_{\rm d}$ corresponding to
 these crossings. Unlike the topological phase, the conductance has a very low weight at zero bias for both values of $\varepsilon_{\rm d}$. 
 We recall that in the topological phase the conductance through the Zeeman level of the quantum dot having a large spin projection with the polarization of the Majorana
 mode is much higher and robust than the one through the other Zeeman level. Instead, in the non-topological phase the orientation of the subgap states follow the orientation of the spin of the quantum dot, which behaves as a magnetic impurity. As these states do not have the robustness of the  topological zero modes, they split when the system is hybridized with the normal lead. The result is a double-peak feature with a dip at zero bias 
 in the conductance. Such a different behavior between the topological and non-topological cases could be useful to identify them in experiments.
 Consequently, the behavior of the noise 
  in the present case is clearly different from the response in the topological phase shown in Figs. \ref{Fig:fig_GtvseV_eds_max_tcS_grande_}
and \ref{Fig:fig_GtvseV_comparacion_eds_topo_}. In particular, the Fano factor $F$ exhibits values  $F\geq 1$ within the full range of voltages. 

\subsection{Non-perpendicular SOC and magnetic field}
\begin{figure}[htb]
   \centering
   \includegraphics[width=0.49\textwidth]{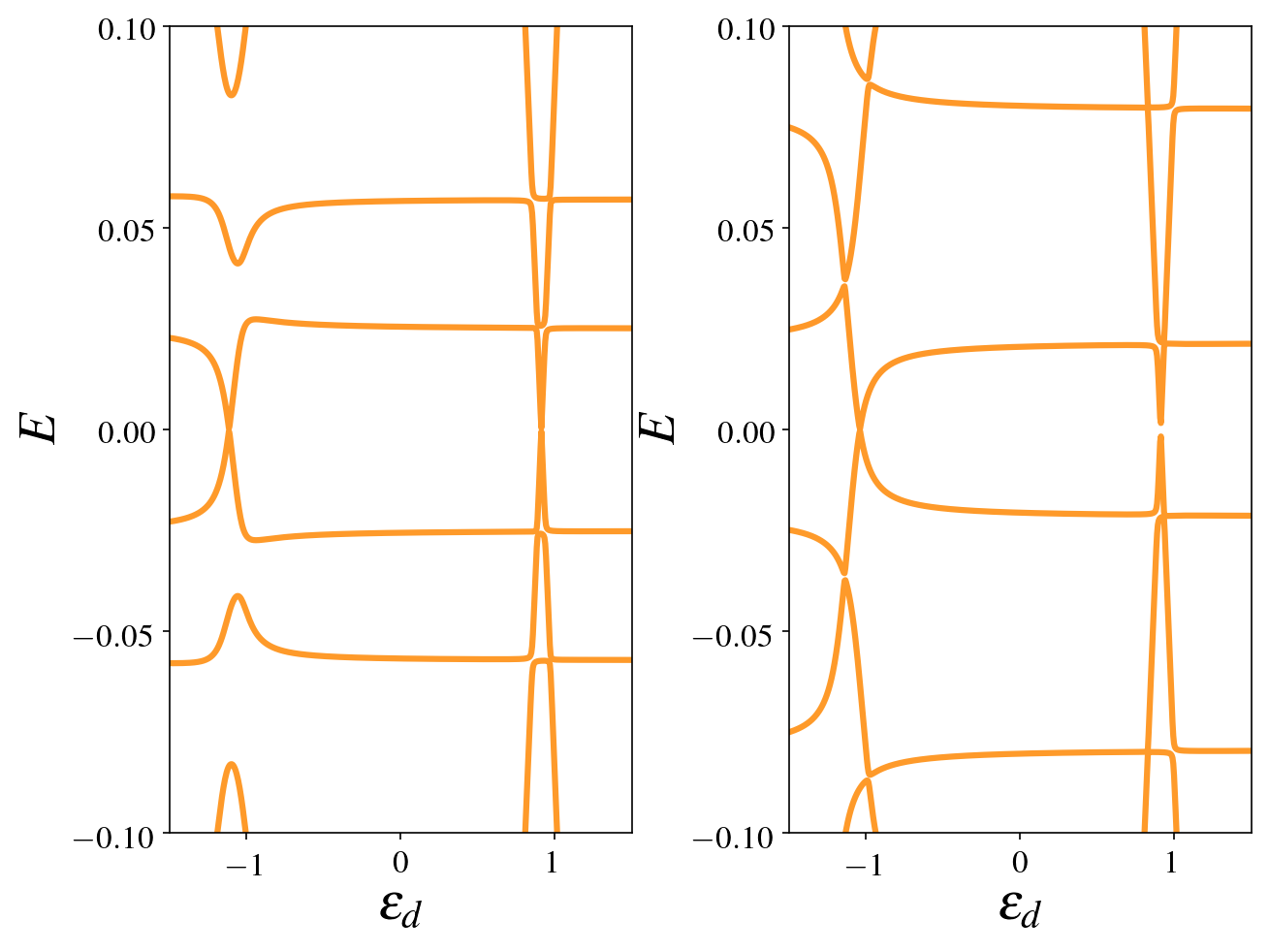}
   \includegraphics[width=0.49\textwidth]{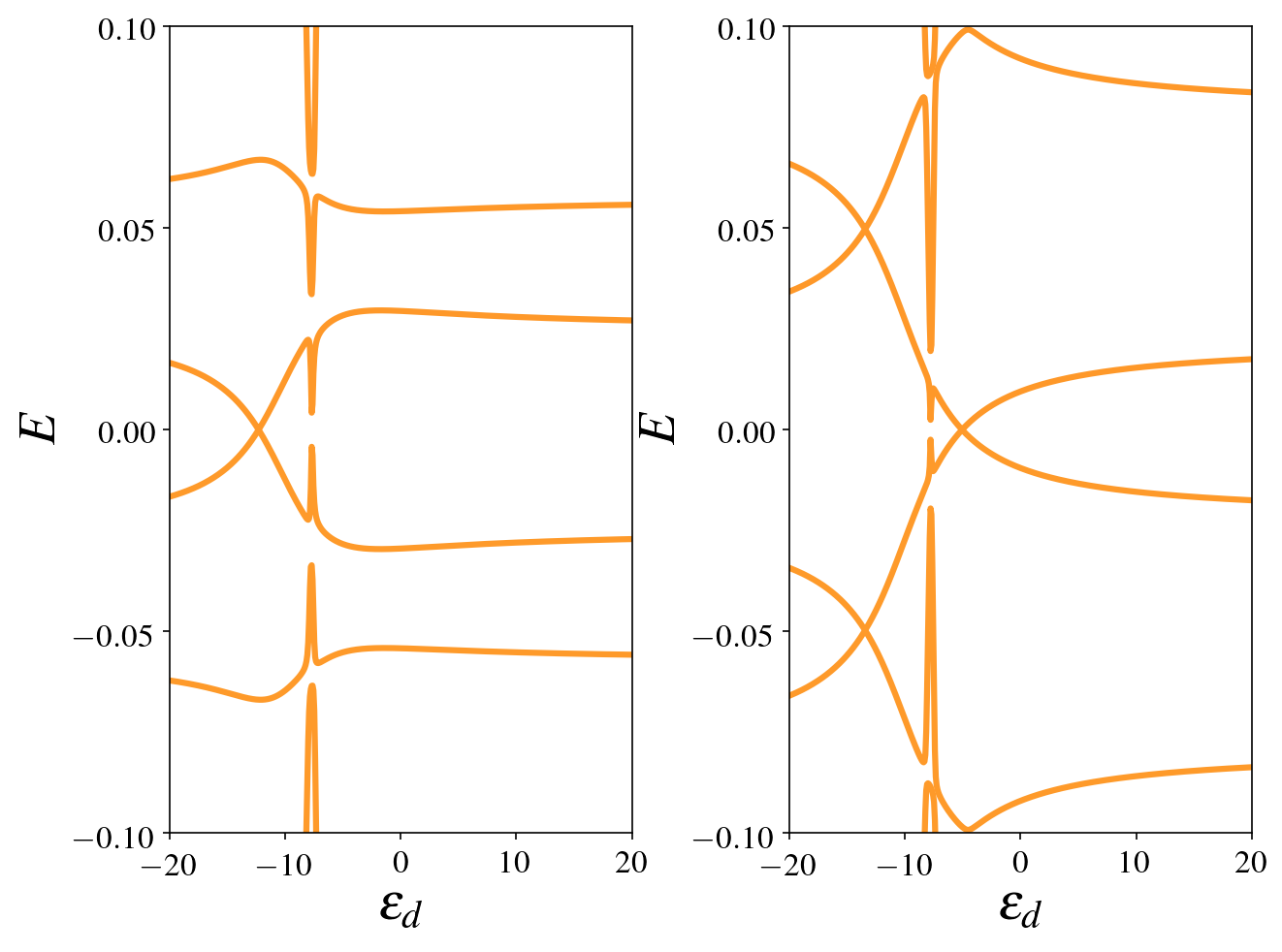} 
\caption{Sub-gap spectrum as a function of the energy $\varepsilon_d$ of the quantum dot for a system with $L=250 a$. All the cases correspond to the exact Hamiltonian. Top and bottom panels correspond, respectively to $t_{\rm c S}=1,\;10meV$.
The rest of the  parameters are the same as in Fig. \ref{Fig:fig4}: $\lambda=0.5meV$,
$B=1meV$, $\mu=1.01 \mu_{\rm c}$. The two angles between $\vec{n}_{\lambda}$ and $\vec{n}_{B}$ are $\theta=\pi/2-\delta\theta$, with
$\delta \theta=0.1 \pi,\; 0.35 \pi$ (left and right panels). 
}\label{Fig:fig7}
\end{figure}

The orientation of the magnetic field is a natural knob to experimentally explore the transition to the topological phase \cite{mourik2012signatures,wang2022observation}. We analyze here the effect of a small departure from the ideal perpendicular configuration with 
respect to the orientation of the SOC.
Results for configurations where $\vec{n}_{\lambda}$ and $\vec{n}_B$ are non-perpendicular are shown in Fig. \ref{Fig:fig7}.
For these parameters, the value of the critical angle defined in Eq. (\ref{bound}) is $\theta_{\rm c} \simeq 0.436 \pi$, meaning 
that a departure $\delta \theta_{\rm c}=0.064 \pi$ with respect to the perpendicular configuration is enough to drive the wire away from the topological phase. The two angles between $\vec{n}_{\lambda}$ and $\vec{n}_{B}$ are $\theta=\pi/2-\delta\theta$, with
$\delta \theta=0.1 \pi,\; 0.35 \pi$ and they are shown in the Fig. \ref{Fig:fig7} are in the non-topological phase. In each case, we consider  moderate  and 
strong couplings between the wire and the quantum dot. 

\begin{figure}[htb]
\centering
\includegraphics[width=\linewidth]{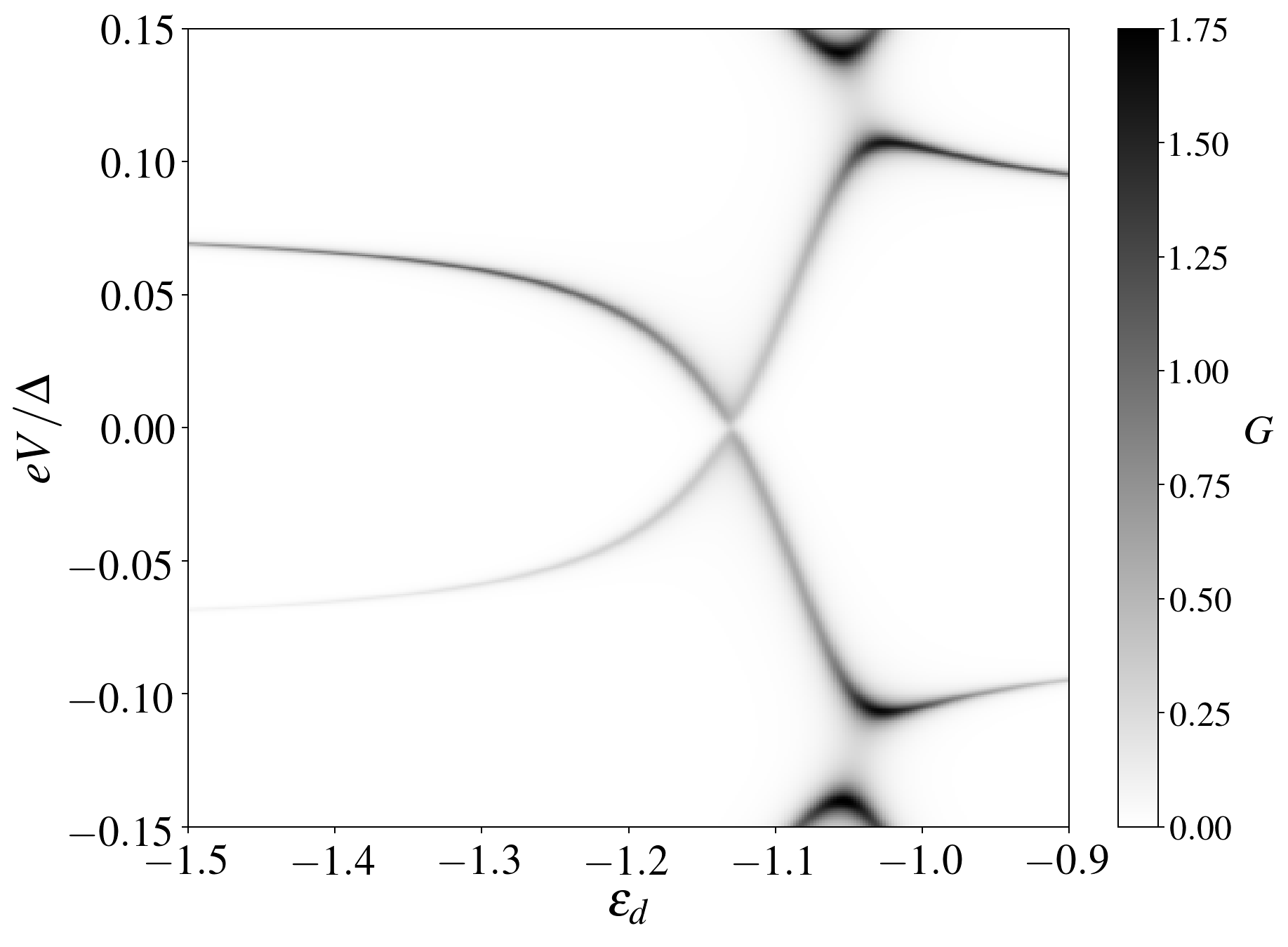}
\caption{Total conductance as a function of the energy of the quantum dot and the applied bias voltage $V$ for 
$\theta=0.5\pi + \delta \theta$ with $\delta \theta= 0.1\pi$. The QD is coupled with $t_{cN}=1.0meV$ to the normal lead. The parameters are the same as in the left column in Fig. \ref{Fig:fig7} }
\label{Fig:fig_heatmap_non_perpendicular_1}
\end{figure}

\begin{figure}[htb]
\centering
\includegraphics[width=\linewidth]{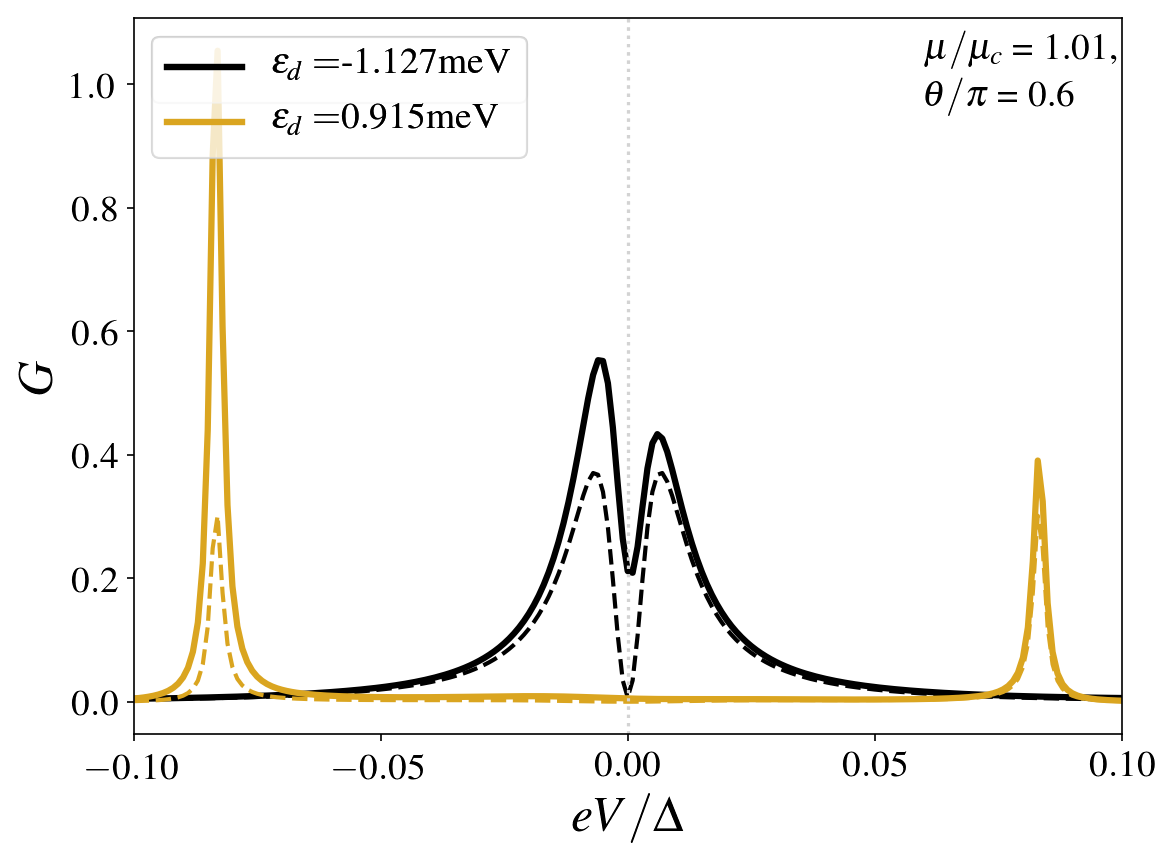}
\includegraphics[width=\linewidth]{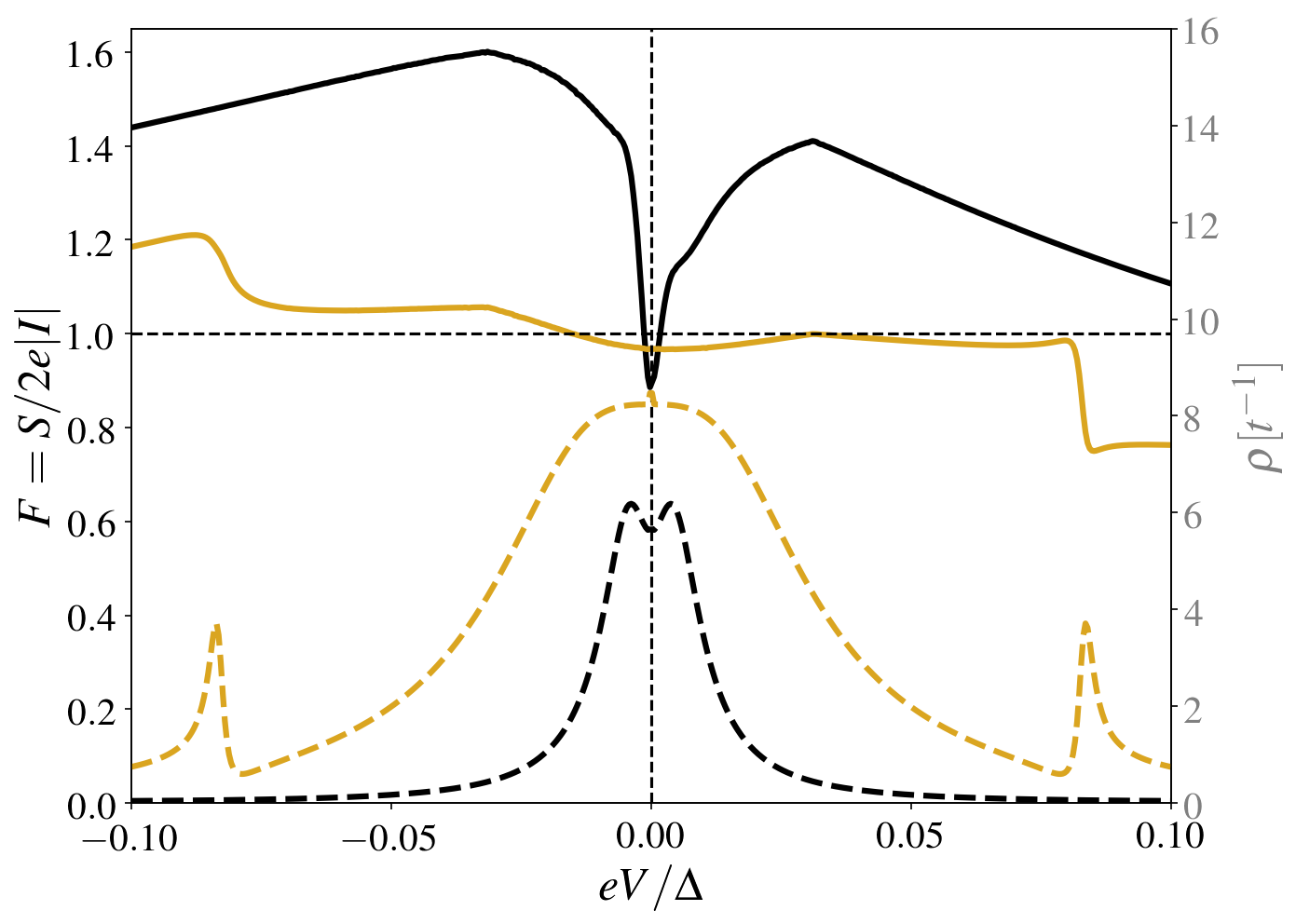}%
\caption{Same as Fig. \ref{Fig:fig_GtvseV_comparacion_eds_topo_} for the parameters of top panel of upper left panel of Fig. \ref{Fig:fig7}.
The quantum dot is coupled with $t_{cN}=0.5meV$ to the normal lead. In this case $\theta= 0.5\pi + \delta \theta$ with $\delta \theta=0.1\pi $. 
}
\label{Fig:fig_13}
\end{figure}

In all the cases we observe  similar features in these spectra as those discussed for the perpendicular case shown in Fig. \ref{Fig:fig4}. In the case of 
$\delta \theta=0.35 \pi$  we can observe some crossings in the excited states above the ones closest to zero energy. The corresponding behavior of the conductance 
as a function of $V$ and $\varepsilon_{\rm d}$ is shown in Fig. \ref{Fig:fig_heatmap_non_perpendicular_1}. At a first glance, we can identify similar features as in 
the perpendicular case shown in Fig. \ref{Fig:fig_GtvseV_comparacion_eds_topo_}. However, a closer analysis reveals a stronger asymmetry in the transport through the two Zeeman levels in the non-perpendicular case. This can be clearly observed in the behavior of the conductance as a function of $V$ for $\varepsilon_{\rm d}$ fixed at the values for which the zero-energy crossings take place. We notice  that the 
conductance at low voltages through the $\downarrow$ level of the quantum dot is basically vanishing. This is because the
superconducting wire is fully gapped and, in addition, the quasiparticles are strongly polarized. The finite density of states at zero energy that is observed in the dashed yellow plots of the low panel of the Fig. \ref{Fig:fig_13} are purely
due to the hybridization of the quantum dot with the normal wire, while the corresponding density of states with this spin orientation is vanishing low at the wire. 
In comparison with the topological case analyzed in Fig. \ref{Fig:fig_GtvseV_comparacion_eds_topo_}, we also see a lower contribution of the Andreev reflection to the total conductance. 
In the lower panel of Fig. 
\ref{Fig:fig_13} we show the corresponding behavior of the noise in solid lines. As in the non-topological case illustrated in Fig. 
\ref{Fig:fig_GvseV_both_ed_non_topo_}, we observe values $F\geq 1$ within almost all the subgap range.

\section{Summary and conclusions}\label{last}
We have analyzed the transport properties of a finite-length superconducting wire with spin-orbit coupling and magnetic field inside and outside the
topological phase, with focus on the parameters that are relevant for experiments. We have also considered in the device an embedded quantum dot
in the N-S junction, as it appears to be the case in many experimental setups \cite{deng2012majorana,deng2016majorana,deng2018}. 

We have guided the study with a previous analysis of the spectral properties of the wire contacted to the quantum dot by means of exactly diagonalyzing the Hamiltonian describing this uncoupled system as well as by deriving and solving analytically the  effective low-energy Hamiltonians. The underlying picture in the topological phase is that the Majorana end modes combine in the finite-size wire to form low non-zero-energy fermionic excitations. These hybridyze with the quantum-dot levels, generating states that cross zero energy for some parameters. For low hybridization with the quantum dot, the behavior of these states 
keep track of the spin structure of the topological modes. Under such circumstances, the zero-energy crossings contain very valuable information of the structure of the Majorana modes and their degree of localization. However,
for strong hybridization, supragap states contribute, introducing additional features. Within the non-topological phase, with the spin-orbit direction perpendicular to the magnetic field, the quantum dot behaves as a magnetic
impurity and the spectrum of the coupled system can be understood as the generation of Yu-Shiba-Rusinov bound states inside the effective gap. For non-perpendicular orientations of these two fields, the spectrum is very similar to that of the topological phase when focusing at low energies when the angle $\theta$ is slightly larger than the critical 
value $\theta_{\rm c}$.

The concomitant behavior of the conductance follows closely the features of the spectrum of the combined wire and quantum dot. Importantly, for a quantum dot with a Zeeman splitting as the case we considered here,  the behavior of the conductance is clearly different for  the two spin orientations of the quantum dot when the wire is in the topological phase. 
For parameters consistent with this phase, the conductance
 has an
important weight at zero bias --albeit lower than the limit $2G_0$ expected in a semi-infinite topological wire-- for the spin orientation having a large component 
with the spin of the particle component of the Majorana zero mode. However it is vanishing small for the other spin orientation. Consequently, in the first case the Fano factor
is $F \ll 1$ in a neighborhood of $V=0$ while in the second case $F \sim 1$. These features are very clear for weakly coupled quantum dots. However, when the coupling becomes large, supragap states partially mask the response, hence, the zero-bias peak and the corresponding Fano dip become very narrow.
Away from the topological phase in configurations where the SOC and the magnetic field are perpendicular, zero energy crossings are more likely to take place for weakly coupled quantum dots. Unlike the topological phase, the transport features (conductance and Fano factor) are very similar for the two spin orientations of the quantum dot. This approximate symmetry for the two spin orientations provides an important distinction between non-topological and topological states for the case of orthogonal SOC and Zeeman fields. 
Overall, however, it is difficult to clearly identify features that unambiguously distinguish the topological from the non-topological phase in systems with
 non-perpendicular orientations of the spin-orbit and magnetic field. 
 
 Our results highlight the role
of the degree of coupling between the quantum dot and the wire. In fact,  the response of the topological wire weakly hybridized with the quantum dot can be understood 
in terms of a simple picture based on the hybridization of the levels of the quantum dot with the end modes of the wire. Instead,  as the coupling becomes stronger,
the hybridization between the quantum dot and the supragap states play a role. This has impact in the behavior of the transport properties. In particular, the conductance
decreases and the noise becomes higher, thus hindering the identification of the topological phase. Altogether, these observations may serve as a guide for the design of future experiments.

\section{Acknowledgements}
LA and LG thank CONICET as well as FonCyT, Argentina through grants PICT 2017, PICT-2018-04536 and PICT 2020-A-03661. Support from Spanish AEI through grant PID2020-117671GB-I00 is acknowledged.

\appendix
\section{Diagonalization of Eq. (\ref{hbdg}) with $\Delta=0$}\label{sing-trip}
To identify the properties of the bulk, it is useful to express the Hamiltonian of the wire in the absence of superconductivity in a diagonal basis. Assuming periodic boundary conditions we get \cite{aligia2020tomography},
\begin{eqnarray}
f_{k,+}&=&u_k c_{k\uparrow}+ v_k c_{k\downarrow},\nonumber \\
f_{k,-}&=&-v_k c_{k\uparrow}+ v_k c_{k\downarrow},
\end{eqnarray}
being $u_k=\sqrt{1+B/r_k}, \;v_k=\mbox{sgn}(\lambda_k)\sqrt{1-B/r_k}$, with $r_k=\sqrt{\lambda_k^2+B^2}$.
In this basis, the Hamiltonian of the wire, including the superconducting terms reads 
\begin{eqnarray}\label{wirekit}
    H_{\rm w}&=& \sum_{k,s=\pm}\left(\varepsilon_{k,s} f^{\dagger}_{k,s} f_{k,s}+ \Delta^{\rm T}_k f_{k,s} f_{-k,s}+ \text{H.c.}\right)\nonumber \\
    & & + \sum_{k} \left(\Delta^{\rm S}_k f_{k,+} f_{-k,-}+\text{H.c.}\right),
\end{eqnarray}
with
\begin{equation}
    \varepsilon_{k,\pm} = - 2 t_{\rm S} \cos(ka)  \pm r_k -\mu,
\end{equation}
which define two energy bands separated by a gap defined by $2 r_k$.

\section{Calculation of the conductance}\label{apa}
In terms of the Green's functions the normal transmission probability and the Andreev reflection functions read
\begin{eqnarray}\label{tna}
{\cal T}_N(\varepsilon) &=& \sum_{j=1,2}\left[ \Gamma^{\rm N}(\varepsilon) G_d^r(\varepsilon)
\Gamma^{\rm S}(\varepsilon) G_d^a(\varepsilon) \right]_{j,j},\nonumber \\
{\cal R}_A(\varepsilon) &=& \sum_{j=1,2}\left\{\Gamma^{\rm N}_{j,j}(\varepsilon) [G_d^r(\varepsilon)]_{j,\overline{j}}
\Gamma^{\rm S}_{\overline{j},\overline{j}}((\varepsilon) [G_d^a(\varepsilon)]_{\overline{j},j} \right\},
\end{eqnarray}
where  we denote the Nambu indices with $\overline{1}\equiv 4$ and $\overline{2}\equiv 3$. We have introduced the retarded and advanced Green's functions for the quantum dot $G_d^r(\varepsilon)$ and $G_d^a(\varepsilon)$, respectively.
 They read 
\begin{equation}\label{grdot}
    G_{\rm d}^{r/a}(\varepsilon)= \left[\varepsilon \hat{1} - h_{\rm d,0} -\Sigma^{r/a}_{\rm S}(\varepsilon) -\Sigma^{r/a}_{\rm N}(\varepsilon)\right]^{-1},
\end{equation}
where $\hat{1}$ is the $4\times 4$ unitary matrix and $h_{\rm d,0}$ is the Hamiltonian matrix of $H_{\rm d,0}$ expressed in the Nambu basis.
The self-energy matrices 
\begin{equation}
    \Sigma^{r/a}_{\alpha}(\varepsilon) = |t_{\rm c \alpha}|^2  g_{\alpha}^{r/a}(\varepsilon),
\end{equation}
with $\alpha={\rm S, N}$, are obtained after integrating-out the degrees of freedom of the S and N wires by solving the Dyson equation for the coupling between the quantum dot and these systems. They are defined from the Green's functions of the uncoupled systems $g_{\alpha}^{r/a}(\varepsilon)$.
In Eqs. (\ref{tna}) we have also introduced the definition
\begin{equation} 
\Gamma^{\alpha}(\varepsilon) = -i \left[  \Sigma^{r}_{\alpha}(\varepsilon)-  \Sigma^{a}_{\alpha}(\varepsilon)\right].
\end{equation}

\section{Spectral of the effective Hamiltonian of Eq. (\ref{heff}) }\label{cross}
We start by analyzing a
weakly connected quantum dot, so that the effect of the hybridization with the supragap states can be neglected.
This corresponds to the quantum dot  hybridized only with the combination of the Majorana modes. We also focus on a Zeeman energy larger than the hybridization of the quantum dot  with the wire.

The spectrum of the effective  Hamiltonian formulated in Eq. (\ref{heff}) with $t^{\prime}_{\sigma}=\Delta^{\prime}_{\sigma}=0$, assuming that the subspaces associated to $\sigma=\uparrow, \downarrow$ are blocked, is
\begin{eqnarray}\label{eff-levels}
E_{1,\sigma}^{\pm} &=& \pm\frac{1}{2} \left|
\sqrt{\left(\varepsilon_M - \varepsilon_{\rm d,\sigma}\right)^2 + 4 \Delta_{\sigma}^2}
+ \sqrt{\left(\varepsilon_M + \varepsilon_{\rm d,\sigma}\right)^2 + 4 t_{\sigma}^2} \right|, \nonumber \\
E_{2,\sigma}^{\pm} &=& \pm\frac{1}{2} \left|
\sqrt{\left(\varepsilon_M - \varepsilon_{\rm d,\sigma}\right)^2 + 4 \Delta_{\sigma}^2}
- \sqrt{\left(\varepsilon_M + \varepsilon_{\rm d,\sigma}\right)^2 + 4 t_{\sigma}^2} \right|,\nonumber
\end{eqnarray}
with $\varepsilon_{\rm d, \sigma}=\varepsilon_{\rm d} \pm B_{\rm d}$.

Alternatively, we can rely on a description based on Green's functions and calculate the
the retarded Green's function of the quantum dot. For weak coupling we can neglect the effective coupling between the two spin states of the quantum dot mediated by the hybridization with the superconducting wires. Hence, the inverse of the retarded Green's function associated to one of the spin orientation reads
\begin{equation}
G^{-1}_{\sigma}(\varepsilon)=\varepsilon\tau^0+ (\varepsilon_{\rm qd} \mp B)\tau^z -\Sigma_{\sigma}(\varepsilon),
\end{equation}
where $\tau^0$ is the $2\times 2$ unit matrix and  $\tau^z$ is the Pauli matrix operating in the particle-hole degree of freedom. $\Sigma_{\sigma}(\varepsilon)$ is a 
$2\times 2$ matrix with elements
\begin{eqnarray}\label{self}
\Sigma_{11}(\varepsilon)&=&t_{\sigma}^2 g_{M}(\varepsilon) +\Delta_{\sigma}^2 \overline{g}_{M}(\varepsilon)=-\Sigma_{22}(-\varepsilon),\nonumber \\
\Sigma_{12}(\varepsilon)&=&-t_{\sigma} \Delta_{\sigma} \left[ g_{M}(\varepsilon) + \overline{g}_{M}(\varepsilon) \right] =
\Sigma_{21}(\varepsilon),
\end{eqnarray}
being
\begin{equation}
g_{M}(\varepsilon) = \frac{1}{\varepsilon-\varepsilon_{M}+i0^+}, \;\;\;\;\;\;\overline{g}_{M}(\varepsilon) = 
\frac{1}{\varepsilon+\varepsilon_{M}+i 0^+}.
\end{equation}0
The spectrum of low-energy levels with weight on the quantum dot is defined by the poles of $G_{\sigma}(\varepsilon)$, which are calculated from
\begin{equation}
\mbox{Det}\left[G^{-1}_{\sigma}(\varepsilon)\right]=0.
\end{equation}
In turn, the crossings at zero energy are defined from
\begin{equation}
\mbox{Det}\left[G^{-1}_{\sigma}(\varepsilon=0)\right]=0.
\end{equation}
This equation leads to 
\begin{equation}
\varepsilon_{\rm qd} = \mp B + \frac{\left(\Delta_{\sigma}^2- t_{\sigma}^2\right)}{\varepsilon_{M}},
\end{equation}
where the upper/lower sign corresponds to $\sigma=\uparrow, \downarrow$.

This procedure to include in the description the effect of the supragap states introduced in Eq. (\ref{heff}) is basically the same. The crucial step is the
addition to extra terms in the self-energy of Eq. (\ref{self}) as follows
\begin{eqnarray}\label{self}
\Sigma_{11,n}(\varepsilon)&=&(t^{\prime}_{\sigma})^2 g_{n}(\varepsilon) +(\Delta^{\prime}_{\sigma})^2 \overline{g}_{n}(\varepsilon)=-\Sigma_{22,n}(-\varepsilon),\nonumber \\
\Sigma_{12,n}(\varepsilon)&=&-t^{\prime}_{\sigma} \Delta^{\prime}_{\sigma} \left[ g_{n}(\varepsilon) + \overline{g}_{n}(\varepsilon) \right] =
\Sigma_{21,n}(\varepsilon),
\end{eqnarray}
with
\begin{equation}
g_{n}(\varepsilon) = \frac{1}{\varepsilon-E_{n}+i0^+}, \;\;\;\;\;\;\overline{g}_{n}(\varepsilon) = 
\frac{1}{\varepsilon+E_{n}+i 0^+}.
\end{equation}

\bibliography{topsn.bib}

\begin{thebibliography}{78}%
\makeatletter
\providecommand \@ifxundefined [1]{%
 \@ifx{#1\undefined}
}%
\providecommand \@ifnum [1]{%
 \ifnum #1\expandafter \@firstoftwo
 \else \expandafter \@secondoftwo
 \fi
}%
\providecommand \@ifx [1]{%
 \ifx #1\expandafter \@firstoftwo
 \else \expandafter \@secondoftwo
 \fi
}%
\providecommand \natexlab [1]{#1}%
\providecommand \enquote  [1]{``#1''}%
\providecommand \bibnamefont  [1]{#1}%
\providecommand \bibfnamefont [1]{#1}%
\providecommand \citenamefont [1]{#1}%
\providecommand \href@noop [0]{\@secondoftwo}%
\providecommand \href [0]{\begingroup \@sanitize@url \@href}%
\providecommand \@href[1]{\@@startlink{#1}\@@href}%
\providecommand \@@href[1]{\endgroup#1\@@endlink}%
\providecommand \@sanitize@url [0]{\catcode `\\12\catcode `\$12\catcode
  `\&12\catcode `\#12\catcode `\^12\catcode `\_12\catcode `\%12\relax}%
\providecommand \@@startlink[1]{}%
\providecommand \@@endlink[0]{}%
\providecommand \url  [0]{\begingroup\@sanitize@url \@url }%
\providecommand \@url [1]{\endgroup\@href {#1}{\urlprefix }}%
\providecommand \urlprefix  [0]{URL }%
\providecommand \Eprint [0]{\href }%
\providecommand \doibase [0]{https://doi.org/}%
\providecommand \selectlanguage [0]{\@gobble}%
\providecommand \bibinfo  [0]{\@secondoftwo}%
\providecommand \bibfield  [0]{\@secondoftwo}%
\providecommand \translation [1]{[#1]}%
\providecommand \BibitemOpen [0]{}%
\providecommand \bibitemStop [0]{}%
\providecommand \bibitemNoStop [0]{.\EOS\space}%
\providecommand \EOS [0]{\spacefactor3000\relax}%
\providecommand \BibitemShut  [1]{\csname bibitem#1\endcsname}%
\let\auto@bib@innerbib\@empty
\bibitem [{\citenamefont {Kitaev}(2001)}]{kitaev2001unpaired}%
  \BibitemOpen
  \bibfield  {author} {\bibinfo {author} {\bibfnamefont {A.~Y.}\ \bibnamefont
  {Kitaev}},\ }\bibfield  {title} {\bibinfo {title} {Unpaired majorana fermions
  in quantum wires},\ }\href@noop {} {\bibfield  {journal} {\bibinfo  {journal}
  {Physics-Uspekhi}\ }\textbf {\bibinfo {volume} {44}},\ \bibinfo {pages} {131}
  (\bibinfo {year} {2001})}\BibitemShut {NoStop}%
\bibitem [{\citenamefont {Kitaev}(2003)}]{kitaev2003fault}%
  \BibitemOpen
  \bibfield  {author} {\bibinfo {author} {\bibfnamefont {A.~Y.}\ \bibnamefont
  {Kitaev}},\ }\bibfield  {title} {\bibinfo {title} {Fault-tolerant quantum
  computation by anyons},\ }\href@noop {} {\bibfield  {journal} {\bibinfo
  {journal} {Annals of Physics}\ }\textbf {\bibinfo {volume} {303}},\ \bibinfo
  {pages} {2} (\bibinfo {year} {2003})}\BibitemShut {NoStop}%
\bibitem [{\citenamefont {Nayak}\ \emph {et~al.}(2008)\citenamefont {Nayak},
  \citenamefont {Simon}, \citenamefont {Stern}, \citenamefont {Freedman},\ and\
  \citenamefont {Sarma}}]{nayak2008non}%
  \BibitemOpen
  \bibfield  {author} {\bibinfo {author} {\bibfnamefont {C.}~\bibnamefont
  {Nayak}}, \bibinfo {author} {\bibfnamefont {S.~H.}\ \bibnamefont {Simon}},
  \bibinfo {author} {\bibfnamefont {A.}~\bibnamefont {Stern}}, \bibinfo
  {author} {\bibfnamefont {M.}~\bibnamefont {Freedman}},\ and\ \bibinfo
  {author} {\bibfnamefont {S.~D.}\ \bibnamefont {Sarma}},\ }\bibfield  {title}
  {\bibinfo {title} {Non-abelian anyons and topological quantum computation},\
  }\href@noop {} {\bibfield  {journal} {\bibinfo  {journal} {Reviews of Modern
  Physics}\ }\textbf {\bibinfo {volume} {80}},\ \bibinfo {pages} {1083}
  (\bibinfo {year} {2008})}\BibitemShut {NoStop}%
\bibitem [{\citenamefont {Alicea}(2012)}]{alicea2012new}%
  \BibitemOpen
  \bibfield  {author} {\bibinfo {author} {\bibfnamefont {J.}~\bibnamefont
  {Alicea}},\ }\bibfield  {title} {\bibinfo {title} {New directions in the
  pursuit of majorana fermions in solid state systems},\ }\href@noop {}
  {\bibfield  {journal} {\bibinfo  {journal} {Reports on progress in physics}\
  }\textbf {\bibinfo {volume} {75}},\ \bibinfo {pages} {076501} (\bibinfo
  {year} {2012})}\BibitemShut {NoStop}%
\bibitem [{\citenamefont {Lutchyn}\ \emph {et~al.}(2010)\citenamefont
  {Lutchyn}, \citenamefont {Sau},\ and\ \citenamefont {Sarma}}]{wires1}%
  \BibitemOpen
  \bibfield  {author} {\bibinfo {author} {\bibfnamefont {R.~M.}\ \bibnamefont
  {Lutchyn}}, \bibinfo {author} {\bibfnamefont {J.~D.}\ \bibnamefont {Sau}},\
  and\ \bibinfo {author} {\bibfnamefont {S.~D.}\ \bibnamefont {Sarma}},\
  }\bibfield  {title} {\bibinfo {title} {Majorana fermions and a topological
  phase transition in semiconductor-superconductor heterostructures},\
  }\href@noop {} {\bibfield  {journal} {\bibinfo  {journal} {Physical review
  letters}\ }\textbf {\bibinfo {volume} {105}},\ \bibinfo {pages} {077001}
  (\bibinfo {year} {2010})}\BibitemShut {NoStop}%
\bibitem [{\citenamefont {Oreg}\ \emph {et~al.}(2010)\citenamefont {Oreg},
  \citenamefont {Refael},\ and\ \citenamefont {von Oppen}}]{wires2}%
  \BibitemOpen
  \bibfield  {author} {\bibinfo {author} {\bibfnamefont {Y.}~\bibnamefont
  {Oreg}}, \bibinfo {author} {\bibfnamefont {G.}~\bibnamefont {Refael}},\ and\
  \bibinfo {author} {\bibfnamefont {F.}~\bibnamefont {von Oppen}},\ }\bibfield
  {title} {\bibinfo {title} {Helical liquids and majorana bound states in
  quantum wires},\ }\href {https://doi.org/10.1103/PhysRevLett.105.177002}
  {\bibfield  {journal} {\bibinfo  {journal} {Phys. Rev. Lett.}\ }\textbf
  {\bibinfo {volume} {105}},\ \bibinfo {pages} {177002} (\bibinfo {year}
  {2010})}\BibitemShut {NoStop}%
\bibitem [{\citenamefont {Rex}\ and\ \citenamefont
  {Sudb{\o}}(2014)}]{rex2014tilting}%
  \BibitemOpen
  \bibfield  {author} {\bibinfo {author} {\bibfnamefont {S.}~\bibnamefont
  {Rex}}\ and\ \bibinfo {author} {\bibfnamefont {A.}~\bibnamefont {Sudb{\o}}},\
  }\bibfield  {title} {\bibinfo {title} {Tilting of the magnetic field in
  majorana nanowires: Critical angle and zero-energy differential
  conductance},\ }\href@noop {} {\bibfield  {journal} {\bibinfo  {journal}
  {Physical Review B}\ }\textbf {\bibinfo {volume} {90}},\ \bibinfo {pages}
  {115429} (\bibinfo {year} {2014})}\BibitemShut {NoStop}%
\bibitem [{\citenamefont {Osca}\ \emph {et~al.}(2014)\citenamefont {Osca},
  \citenamefont {Ruiz},\ and\ \citenamefont {Serra}}]{osca2014effects}%
  \BibitemOpen
  \bibfield  {author} {\bibinfo {author} {\bibfnamefont {J.}~\bibnamefont
  {Osca}}, \bibinfo {author} {\bibfnamefont {D.}~\bibnamefont {Ruiz}},\ and\
  \bibinfo {author} {\bibfnamefont {L.}~\bibnamefont {Serra}},\ }\bibfield
  {title} {\bibinfo {title} {Effects of tilting the magnetic field in
  one-dimensional majorana nanowires},\ }\href@noop {} {\bibfield  {journal}
  {\bibinfo  {journal} {Physical Review B}\ }\textbf {\bibinfo {volume} {89}},\
  \bibinfo {pages} {245405} (\bibinfo {year} {2014})}\BibitemShut {NoStop}%
\bibitem [{\citenamefont {Klinovaja}\ and\ \citenamefont
  {Loss}(2015)}]{klinovaja2015fermionic}%
  \BibitemOpen
  \bibfield  {author} {\bibinfo {author} {\bibfnamefont {J.}~\bibnamefont
  {Klinovaja}}\ and\ \bibinfo {author} {\bibfnamefont {D.}~\bibnamefont
  {Loss}},\ }\bibfield  {title} {\bibinfo {title} {Fermionic and majorana bound
  states in hybrid nanowires with non-uniform spin-orbit interaction},\
  }\href@noop {} {\bibfield  {journal} {\bibinfo  {journal} {The European
  Physical Journal B}\ }\textbf {\bibinfo {volume} {88}},\ \bibinfo {pages} {1}
  (\bibinfo {year} {2015})}\BibitemShut {NoStop}%
\bibitem [{\citenamefont {Aligia}\ \emph {et~al.}(2020)\citenamefont {Aligia},
  \citenamefont {Daroca},\ and\ \citenamefont
  {Arrachea}}]{aligia2020tomography}%
  \BibitemOpen
  \bibfield  {author} {\bibinfo {author} {\bibfnamefont {A.~A.}\ \bibnamefont
  {Aligia}}, \bibinfo {author} {\bibfnamefont {D.~P.}\ \bibnamefont {Daroca}},\
  and\ \bibinfo {author} {\bibfnamefont {L.}~\bibnamefont {Arrachea}},\
  }\bibfield  {title} {\bibinfo {title} {Tomography of zero-energy end modes in
  topological superconducting wires},\ }\href@noop {} {\bibfield  {journal}
  {\bibinfo  {journal} {Physical Review Letters}\ }\textbf {\bibinfo {volume}
  {125}},\ \bibinfo {pages} {256801} (\bibinfo {year} {2020})}\BibitemShut
  {NoStop}%
\bibitem [{\citenamefont {Daroca}\ and\ \citenamefont
  {Aligia}(2021)}]{daroca2021phase}%
  \BibitemOpen
  \bibfield  {author} {\bibinfo {author} {\bibfnamefont {D.~P.}\ \bibnamefont
  {Daroca}}\ and\ \bibinfo {author} {\bibfnamefont {A.~A.}\ \bibnamefont
  {Aligia}},\ }\bibfield  {title} {\bibinfo {title} {Phase diagram of a model
  for topological superconducting wires},\ }\href@noop {} {\bibfield  {journal}
  {\bibinfo  {journal} {Physical Review B}\ }\textbf {\bibinfo {volume}
  {104}},\ \bibinfo {pages} {115125} (\bibinfo {year} {2021})}\BibitemShut
  {NoStop}%
\bibitem [{\citenamefont {Mourik}\ \emph {et~al.}(2012)\citenamefont {Mourik},
  \citenamefont {Zuo}, \citenamefont {Frolov}, \citenamefont {Plissard},
  \citenamefont {Bakkers},\ and\ \citenamefont
  {Kouwenhoven}}]{mourik2012signatures}%
  \BibitemOpen
  \bibfield  {author} {\bibinfo {author} {\bibfnamefont {V.}~\bibnamefont
  {Mourik}}, \bibinfo {author} {\bibfnamefont {K.}~\bibnamefont {Zuo}},
  \bibinfo {author} {\bibfnamefont {S.~M.}\ \bibnamefont {Frolov}}, \bibinfo
  {author} {\bibfnamefont {S.}~\bibnamefont {Plissard}}, \bibinfo {author}
  {\bibfnamefont {E.~P.}\ \bibnamefont {Bakkers}},\ and\ \bibinfo {author}
  {\bibfnamefont {L.~P.}\ \bibnamefont {Kouwenhoven}},\ }\bibfield  {title}
  {\bibinfo {title} {Signatures of majorana fermions in hybrid
  superconductor-semiconductor nanowire devices},\ }\href@noop {} {\bibfield
  {journal} {\bibinfo  {journal} {Science}\ }\textbf {\bibinfo {volume}
  {336}},\ \bibinfo {pages} {1003} (\bibinfo {year} {2012})}\BibitemShut
  {NoStop}%
\bibitem [{\citenamefont {Deng}\ \emph {et~al.}(2016)\citenamefont {Deng},
  \citenamefont {Vaitiek{\.e}nas}, \citenamefont {Hansen}, \citenamefont
  {Danon}, \citenamefont {Leijnse}, \citenamefont {Flensberg}, \citenamefont
  {Nyg{\aa}rd}, \citenamefont {Krogstrup},\ and\ \citenamefont
  {Marcus}}]{deng2016majorana}%
  \BibitemOpen
  \bibfield  {author} {\bibinfo {author} {\bibfnamefont {M.}~\bibnamefont
  {Deng}}, \bibinfo {author} {\bibfnamefont {S.}~\bibnamefont
  {Vaitiek{\.e}nas}}, \bibinfo {author} {\bibfnamefont {E.~B.}\ \bibnamefont
  {Hansen}}, \bibinfo {author} {\bibfnamefont {J.}~\bibnamefont {Danon}},
  \bibinfo {author} {\bibfnamefont {M.}~\bibnamefont {Leijnse}}, \bibinfo
  {author} {\bibfnamefont {K.}~\bibnamefont {Flensberg}}, \bibinfo {author}
  {\bibfnamefont {J.}~\bibnamefont {Nyg{\aa}rd}}, \bibinfo {author}
  {\bibfnamefont {P.}~\bibnamefont {Krogstrup}},\ and\ \bibinfo {author}
  {\bibfnamefont {C.~M.}\ \bibnamefont {Marcus}},\ }\bibfield  {title}
  {\bibinfo {title} {Majorana bound state in a coupled quantum-dot
  hybrid-nanowire system},\ }\href@noop {} {\bibfield  {journal} {\bibinfo
  {journal} {Science}\ }\textbf {\bibinfo {volume} {354}},\ \bibinfo {pages}
  {1557} (\bibinfo {year} {2016})}\BibitemShut {NoStop}%
\bibitem [{\citenamefont {Chen}\ \emph {et~al.}(2017)\citenamefont {Chen},
  \citenamefont {Yu}, \citenamefont {Stenger}, \citenamefont {Hocevar},
  \citenamefont {Car}, \citenamefont {Plissard}, \citenamefont {Bakkers},
  \citenamefont {Stanescu},\ and\ \citenamefont
  {Frolov}}]{chen2017experimental}%
  \BibitemOpen
  \bibfield  {author} {\bibinfo {author} {\bibfnamefont {J.}~\bibnamefont
  {Chen}}, \bibinfo {author} {\bibfnamefont {P.}~\bibnamefont {Yu}}, \bibinfo
  {author} {\bibfnamefont {J.}~\bibnamefont {Stenger}}, \bibinfo {author}
  {\bibfnamefont {M.}~\bibnamefont {Hocevar}}, \bibinfo {author} {\bibfnamefont
  {D.}~\bibnamefont {Car}}, \bibinfo {author} {\bibfnamefont {S.~R.}\
  \bibnamefont {Plissard}}, \bibinfo {author} {\bibfnamefont {E.~P.}\
  \bibnamefont {Bakkers}}, \bibinfo {author} {\bibfnamefont {T.~D.}\
  \bibnamefont {Stanescu}},\ and\ \bibinfo {author} {\bibfnamefont {S.~M.}\
  \bibnamefont {Frolov}},\ }\bibfield  {title} {\bibinfo {title} {Experimental
  phase diagram of zero-bias conductance peaks in superconductor/semiconductor
  nanowire devices},\ }\href@noop {} {\bibfield  {journal} {\bibinfo  {journal}
  {Science advances}\ }\textbf {\bibinfo {volume} {3}},\ \bibinfo {pages}
  {e1701476} (\bibinfo {year} {2017})}\BibitemShut {NoStop}%
\bibitem [{\citenamefont {Nichele}\ \emph {et~al.}(2017)\citenamefont
  {Nichele}, \citenamefont {Drachmann}, \citenamefont {Whiticar}, \citenamefont
  {O’Farrell}, \citenamefont {Suominen}, \citenamefont {Fornieri},
  \citenamefont {Wang}, \citenamefont {Gardner}, \citenamefont {Thomas},
  \citenamefont {Hatke} \emph {et~al.}}]{nichele2017scaling}%
  \BibitemOpen
  \bibfield  {author} {\bibinfo {author} {\bibfnamefont {F.}~\bibnamefont
  {Nichele}}, \bibinfo {author} {\bibfnamefont {A.~C.}\ \bibnamefont
  {Drachmann}}, \bibinfo {author} {\bibfnamefont {A.~M.}\ \bibnamefont
  {Whiticar}}, \bibinfo {author} {\bibfnamefont {E.~C.}\ \bibnamefont
  {O’Farrell}}, \bibinfo {author} {\bibfnamefont {H.~J.}\ \bibnamefont
  {Suominen}}, \bibinfo {author} {\bibfnamefont {A.}~\bibnamefont {Fornieri}},
  \bibinfo {author} {\bibfnamefont {T.}~\bibnamefont {Wang}}, \bibinfo {author}
  {\bibfnamefont {G.~C.}\ \bibnamefont {Gardner}}, \bibinfo {author}
  {\bibfnamefont {C.}~\bibnamefont {Thomas}}, \bibinfo {author} {\bibfnamefont
  {A.~T.}\ \bibnamefont {Hatke}}, \emph {et~al.},\ }\bibfield  {title}
  {\bibinfo {title} {Scaling of majorana zero-bias conductance peaks},\
  }\href@noop {} {\bibfield  {journal} {\bibinfo  {journal} {Physical review
  letters}\ }\textbf {\bibinfo {volume} {119}},\ \bibinfo {pages} {136803}
  (\bibinfo {year} {2017})}\BibitemShut {NoStop}%
\bibitem [{\citenamefont {Vaitiekėnas}\ \emph {et~al.}(2020)\citenamefont
  {Vaitiekėnas}, \citenamefont {Winkler}, \citenamefont {van Heck},
  \citenamefont {Karzig}, \citenamefont {Deng}, \citenamefont {Flensberg},
  \citenamefont {Glazman}, \citenamefont {Nayak}, \citenamefont {Krogstrup},
  \citenamefont {Lutchyn},\ and\ \citenamefont
  {Marcus}}]{vaitiekenas2020fullshell}%
  \BibitemOpen
  \bibfield  {author} {\bibinfo {author} {\bibfnamefont {S.}~\bibnamefont
  {Vaitiekėnas}}, \bibinfo {author} {\bibfnamefont {G.~W.}\ \bibnamefont
  {Winkler}}, \bibinfo {author} {\bibfnamefont {B.}~\bibnamefont {van Heck}},
  \bibinfo {author} {\bibfnamefont {T.}~\bibnamefont {Karzig}}, \bibinfo
  {author} {\bibfnamefont {M.-T.}\ \bibnamefont {Deng}}, \bibinfo {author}
  {\bibfnamefont {K.}~\bibnamefont {Flensberg}}, \bibinfo {author}
  {\bibfnamefont {L.~I.}\ \bibnamefont {Glazman}}, \bibinfo {author}
  {\bibfnamefont {C.}~\bibnamefont {Nayak}}, \bibinfo {author} {\bibfnamefont
  {P.}~\bibnamefont {Krogstrup}}, \bibinfo {author} {\bibfnamefont {R.~M.}\
  \bibnamefont {Lutchyn}},\ and\ \bibinfo {author} {\bibfnamefont {C.~M.}\
  \bibnamefont {Marcus}},\ }\bibfield  {title} {\bibinfo {title} {Flux-induced
  topological superconductivity in full-shell nanowires},\ }\href
  {https://doi.org/10.1126/science.aav3392} {\bibfield  {journal} {\bibinfo
  {journal} {Science}\ }\textbf {\bibinfo {volume} {367}},\ \bibinfo {pages}
  {eaav3392} (\bibinfo {year} {2020})},\ \Eprint
  {https://arxiv.org/abs/https://www.science.org/doi/pdf/10.1126/science.aav3392}
  {https://www.science.org/doi/pdf/10.1126/science.aav3392} \BibitemShut
  {NoStop}%
\bibitem [{\citenamefont {Tanaka}\ and\ \citenamefont
  {Kashiwaya}(1995)}]{tanaka95}%
  \BibitemOpen
  \bibfield  {author} {\bibinfo {author} {\bibfnamefont {Y.}~\bibnamefont
  {Tanaka}}\ and\ \bibinfo {author} {\bibfnamefont {S.}~\bibnamefont
  {Kashiwaya}},\ }\bibfield  {title} {\bibinfo {title} {Theory of tunneling
  spectroscopy of $\mathit{d}$-wave superconductors},\ }\href
  {https://doi.org/10.1103/PhysRevLett.74.3451} {\bibfield  {journal} {\bibinfo
   {journal} {Phys. Rev. Lett.}\ }\textbf {\bibinfo {volume} {74}},\ \bibinfo
  {pages} {3451} (\bibinfo {year} {1995})}\BibitemShut {NoStop}%
\bibitem [{\citenamefont {Tanaka}\ and\ \citenamefont
  {Kashiwaya}(2004)}]{tanaka04}%
  \BibitemOpen
  \bibfield  {author} {\bibinfo {author} {\bibfnamefont {Y.}~\bibnamefont
  {Tanaka}}\ and\ \bibinfo {author} {\bibfnamefont {S.}~\bibnamefont
  {Kashiwaya}},\ }\bibfield  {title} {\bibinfo {title} {Anomalous charge
  transport in triplet superconductor junctions},\ }\href
  {https://doi.org/10.1103/PhysRevB.70.012507} {\bibfield  {journal} {\bibinfo
  {journal} {Phys. Rev. B}\ }\textbf {\bibinfo {volume} {70}},\ \bibinfo
  {pages} {012507} (\bibinfo {year} {2004})}\BibitemShut {NoStop}%
\bibitem [{\citenamefont {Kells}\ \emph {et~al.}(2012)\citenamefont {Kells},
  \citenamefont {Meidan},\ and\ \citenamefont {Brouwer}}]{kells2012near}%
  \BibitemOpen
  \bibfield  {author} {\bibinfo {author} {\bibfnamefont {G.}~\bibnamefont
  {Kells}}, \bibinfo {author} {\bibfnamefont {D.}~\bibnamefont {Meidan}},\ and\
  \bibinfo {author} {\bibfnamefont {P.}~\bibnamefont {Brouwer}},\ }\bibfield
  {title} {\bibinfo {title} {Near-zero-energy end states in topologically
  trivial spin-orbit coupled superconducting nanowires with a smooth
  confinement},\ }\href@noop {} {\bibfield  {journal} {\bibinfo  {journal}
  {Physical Review B}\ }\textbf {\bibinfo {volume} {86}},\ \bibinfo {pages}
  {100503} (\bibinfo {year} {2012})}\BibitemShut {NoStop}%
\bibitem [{\citenamefont {Prada}\ \emph {et~al.}(2012)\citenamefont {Prada},
  \citenamefont {San-Jose},\ and\ \citenamefont {Aguado}}]{prada2012transport}%
  \BibitemOpen
  \bibfield  {author} {\bibinfo {author} {\bibfnamefont {E.}~\bibnamefont
  {Prada}}, \bibinfo {author} {\bibfnamefont {P.}~\bibnamefont {San-Jose}},\
  and\ \bibinfo {author} {\bibfnamefont {R.}~\bibnamefont {Aguado}},\
  }\bibfield  {title} {\bibinfo {title} {Transport spectroscopy of n s nanowire
  junctions with majorana fermions},\ }\href@noop {} {\bibfield  {journal}
  {\bibinfo  {journal} {Physical Review B}\ }\textbf {\bibinfo {volume} {86}},\
  \bibinfo {pages} {180503} (\bibinfo {year} {2012})}\BibitemShut {NoStop}%
\bibitem [{\citenamefont {Roy}\ \emph {et~al.}(2013)\citenamefont {Roy},
  \citenamefont {Bondyopadhaya},\ and\ \citenamefont {Tewari}}]{roy2013}%
  \BibitemOpen
  \bibfield  {author} {\bibinfo {author} {\bibfnamefont {D.}~\bibnamefont
  {Roy}}, \bibinfo {author} {\bibfnamefont {N.}~\bibnamefont {Bondyopadhaya}},\
  and\ \bibinfo {author} {\bibfnamefont {S.}~\bibnamefont {Tewari}},\
  }\bibfield  {title} {\bibinfo {title} {Topologically trivial zero-bias
  conductance peak in semiconductor majorana wires from boundary effects},\
  }\href {https://doi.org/10.1103/PhysRevB.88.020502} {\bibfield  {journal}
  {\bibinfo  {journal} {Phys. Rev. B}\ }\textbf {\bibinfo {volume} {88}},\
  \bibinfo {pages} {020502} (\bibinfo {year} {2013})}\BibitemShut {NoStop}%
\bibitem [{\citenamefont {Liu}\ \emph {et~al.}(2017)\citenamefont {Liu},
  \citenamefont {Sau}, \citenamefont {Stanescu},\ and\ \citenamefont
  {Sarma}}]{liu2017andreev}%
  \BibitemOpen
  \bibfield  {author} {\bibinfo {author} {\bibfnamefont {C.-X.}\ \bibnamefont
  {Liu}}, \bibinfo {author} {\bibfnamefont {J.~D.}\ \bibnamefont {Sau}},
  \bibinfo {author} {\bibfnamefont {T.~D.}\ \bibnamefont {Stanescu}},\ and\
  \bibinfo {author} {\bibfnamefont {S.~D.}\ \bibnamefont {Sarma}},\ }\bibfield
  {title} {\bibinfo {title} {Andreev bound states versus majorana bound states
  in quantum dot-nanowire-superconductor hybrid structures: Trivial versus
  topological zero-bias conductance peaks},\ }\href@noop {} {\bibfield
  {journal} {\bibinfo  {journal} {Physical Review B}\ }\textbf {\bibinfo
  {volume} {96}},\ \bibinfo {pages} {075161} (\bibinfo {year}
  {2017})}\BibitemShut {NoStop}%
\bibitem [{\citenamefont {Moore}\ \emph
  {et~al.}(2018{\natexlab{a}})\citenamefont {Moore}, \citenamefont {Stanescu},\
  and\ \citenamefont {Tewari}}]{moore2018two}%
  \BibitemOpen
  \bibfield  {author} {\bibinfo {author} {\bibfnamefont {C.}~\bibnamefont
  {Moore}}, \bibinfo {author} {\bibfnamefont {T.~D.}\ \bibnamefont
  {Stanescu}},\ and\ \bibinfo {author} {\bibfnamefont {S.}~\bibnamefont
  {Tewari}},\ }\bibfield  {title} {\bibinfo {title} {Two-terminal charge
  tunneling: Disentangling majorana zero modes from partially separated andreev
  bound states in semiconductor-superconductor heterostructures},\ }\href@noop
  {} {\bibfield  {journal} {\bibinfo  {journal} {Physical Review B}\ }\textbf
  {\bibinfo {volume} {97}},\ \bibinfo {pages} {165302} (\bibinfo {year}
  {2018}{\natexlab{a}})}\BibitemShut {NoStop}%
\bibitem [{\citenamefont {Moore}\ \emph
  {et~al.}(2018{\natexlab{b}})\citenamefont {Moore}, \citenamefont {Zeng},
  \citenamefont {Stanescu},\ and\ \citenamefont {Tewari}}]{moore2018quantized}%
  \BibitemOpen
  \bibfield  {author} {\bibinfo {author} {\bibfnamefont {C.}~\bibnamefont
  {Moore}}, \bibinfo {author} {\bibfnamefont {C.}~\bibnamefont {Zeng}},
  \bibinfo {author} {\bibfnamefont {T.~D.}\ \bibnamefont {Stanescu}},\ and\
  \bibinfo {author} {\bibfnamefont {S.}~\bibnamefont {Tewari}},\ }\bibfield
  {title} {\bibinfo {title} {Quantized zero-bias conductance plateau in
  semiconductor-superconductor heterostructures without topological majorana
  zero modes},\ }\href@noop {} {\bibfield  {journal} {\bibinfo  {journal}
  {Physical Review B}\ }\textbf {\bibinfo {volume} {98}},\ \bibinfo {pages}
  {155314} (\bibinfo {year} {2018}{\natexlab{b}})}\BibitemShut {NoStop}%
\bibitem [{\citenamefont {Fleckenstein}\ \emph {et~al.}(2018)\citenamefont
  {Fleckenstein}, \citenamefont {Dom{\'\i}nguez}, \citenamefont {Ziani},\ and\
  \citenamefont {Trauzettel}}]{fleckenstein2018decaying}%
  \BibitemOpen
  \bibfield  {author} {\bibinfo {author} {\bibfnamefont {C.}~\bibnamefont
  {Fleckenstein}}, \bibinfo {author} {\bibfnamefont {F.}~\bibnamefont
  {Dom{\'\i}nguez}}, \bibinfo {author} {\bibfnamefont {N.~T.}\ \bibnamefont
  {Ziani}},\ and\ \bibinfo {author} {\bibfnamefont {B.}~\bibnamefont
  {Trauzettel}},\ }\bibfield  {title} {\bibinfo {title} {Decaying spectral
  oscillations in a majorana wire with finite coherence length},\ }\href@noop
  {} {\bibfield  {journal} {\bibinfo  {journal} {Physical Review B}\ }\textbf
  {\bibinfo {volume} {97}},\ \bibinfo {pages} {155425} (\bibinfo {year}
  {2018})}\BibitemShut {NoStop}%
\bibitem [{\citenamefont {Prada}\ \emph {et~al.}(2020)\citenamefont {Prada},
  \citenamefont {San-Jose}, \citenamefont {de~Moor}, \citenamefont {Geresdi},
  \citenamefont {Lee}, \citenamefont {Klinovaja}, \citenamefont {Loss},
  \citenamefont {Nyg{\aa}rd}, \citenamefont {Aguado},\ and\ \citenamefont
  {Kouwenhoven}}]{prada2020andreev}%
  \BibitemOpen
  \bibfield  {author} {\bibinfo {author} {\bibfnamefont {E.}~\bibnamefont
  {Prada}}, \bibinfo {author} {\bibfnamefont {P.}~\bibnamefont {San-Jose}},
  \bibinfo {author} {\bibfnamefont {M.~W.}\ \bibnamefont {de~Moor}}, \bibinfo
  {author} {\bibfnamefont {A.}~\bibnamefont {Geresdi}}, \bibinfo {author}
  {\bibfnamefont {E.~J.}\ \bibnamefont {Lee}}, \bibinfo {author} {\bibfnamefont
  {J.}~\bibnamefont {Klinovaja}}, \bibinfo {author} {\bibfnamefont
  {D.}~\bibnamefont {Loss}}, \bibinfo {author} {\bibfnamefont {J.}~\bibnamefont
  {Nyg{\aa}rd}}, \bibinfo {author} {\bibfnamefont {R.}~\bibnamefont {Aguado}},\
  and\ \bibinfo {author} {\bibfnamefont {L.~P.}\ \bibnamefont {Kouwenhoven}},\
  }\bibfield  {title} {\bibinfo {title} {From andreev to majorana bound states
  in hybrid superconductor--semiconductor nanowires},\ }\href@noop {}
  {\bibfield  {journal} {\bibinfo  {journal} {Nature Reviews Physics}\ }\textbf
  {\bibinfo {volume} {2}},\ \bibinfo {pages} {575} (\bibinfo {year}
  {2020})}\BibitemShut {NoStop}%
\bibitem [{\citenamefont {Vuik}\ \emph {et~al.}(2019)\citenamefont {Vuik},
  \citenamefont {Nijholt}, \citenamefont {Akhmerov},\ and\ \citenamefont
  {Wimmer}}]{vuik2019reproducing}%
  \BibitemOpen
  \bibfield  {author} {\bibinfo {author} {\bibfnamefont {A.}~\bibnamefont
  {Vuik}}, \bibinfo {author} {\bibfnamefont {B.}~\bibnamefont {Nijholt}},
  \bibinfo {author} {\bibfnamefont {A.}~\bibnamefont {Akhmerov}},\ and\
  \bibinfo {author} {\bibfnamefont {M.}~\bibnamefont {Wimmer}},\ }\bibfield
  {title} {\bibinfo {title} {Reproducing topological properties with
  quasi-majorana states},\ }\href@noop {} {\bibfield  {journal} {\bibinfo
  {journal} {SciPost Physics}\ }\textbf {\bibinfo {volume} {7}},\ \bibinfo
  {pages} {061} (\bibinfo {year} {2019})}\BibitemShut {NoStop}%
\bibitem [{\citenamefont {Zhang}\ \emph {et~al.}(2022)\citenamefont {Zhang},
  \citenamefont {Wang}, \citenamefont {Pan}, \citenamefont {Li}, \citenamefont
  {Lu}, \citenamefont {Li}, \citenamefont {Zhang}, \citenamefont {Liu},
  \citenamefont {Cao}, \citenamefont {Liu} \emph
  {et~al.}}]{zhang2022suppressing}%
  \BibitemOpen
  \bibfield  {author} {\bibinfo {author} {\bibfnamefont {S.}~\bibnamefont
  {Zhang}}, \bibinfo {author} {\bibfnamefont {Z.}~\bibnamefont {Wang}},
  \bibinfo {author} {\bibfnamefont {D.}~\bibnamefont {Pan}}, \bibinfo {author}
  {\bibfnamefont {H.}~\bibnamefont {Li}}, \bibinfo {author} {\bibfnamefont
  {S.}~\bibnamefont {Lu}}, \bibinfo {author} {\bibfnamefont {Z.}~\bibnamefont
  {Li}}, \bibinfo {author} {\bibfnamefont {G.}~\bibnamefont {Zhang}}, \bibinfo
  {author} {\bibfnamefont {D.}~\bibnamefont {Liu}}, \bibinfo {author}
  {\bibfnamefont {Z.}~\bibnamefont {Cao}}, \bibinfo {author} {\bibfnamefont
  {L.}~\bibnamefont {Liu}}, \emph {et~al.},\ }\bibfield  {title} {\bibinfo
  {title} {Suppressing andreev bound state zero bias peaks using a strongly
  dissipative lead},\ }\href@noop {} {\bibfield  {journal} {\bibinfo  {journal}
  {Physical Review Letters}\ }\textbf {\bibinfo {volume} {128}},\ \bibinfo
  {pages} {076803} (\bibinfo {year} {2022})}\BibitemShut {NoStop}%
\bibitem [{\citenamefont {Frolov}(2021)}]{frolov2021quantum}%
  \BibitemOpen
  \bibfield  {author} {\bibinfo {author} {\bibfnamefont {S.}~\bibnamefont
  {Frolov}},\ }\href@noop {} {\bibinfo {title} {Quantum computing’s
  reproducibility crisis: Majorana fermions}} (\bibinfo {year}
  {2021})\BibitemShut {NoStop}%
\bibitem [{\citenamefont {Asano}\ and\ \citenamefont
  {Tanaka}(2013)}]{tanaka13}%
  \BibitemOpen
  \bibfield  {author} {\bibinfo {author} {\bibfnamefont {Y.}~\bibnamefont
  {Asano}}\ and\ \bibinfo {author} {\bibfnamefont {Y.}~\bibnamefont {Tanaka}},\
  }\bibfield  {title} {\bibinfo {title} {Majorana fermions and odd-frequency
  cooper pairs in a normal-metal nanowire proximity-coupled to a topological
  superconductor},\ }\href {https://doi.org/10.1103/PhysRevB.87.104513}
  {\bibfield  {journal} {\bibinfo  {journal} {Phys. Rev. B}\ }\textbf {\bibinfo
  {volume} {87}},\ \bibinfo {pages} {104513} (\bibinfo {year}
  {2013})}\BibitemShut {NoStop}%
\bibitem [{\citenamefont {Bondyopadhaya}\ and\ \citenamefont
  {Roy}(2019)}]{roy2019}%
  \BibitemOpen
  \bibfield  {author} {\bibinfo {author} {\bibfnamefont {N.}~\bibnamefont
  {Bondyopadhaya}}\ and\ \bibinfo {author} {\bibfnamefont {D.}~\bibnamefont
  {Roy}},\ }\bibfield  {title} {\bibinfo {title} {Dynamics of hybrid junctions
  of majorana wires},\ }\href {https://doi.org/10.1103/PhysRevB.99.214514}
  {\bibfield  {journal} {\bibinfo  {journal} {Phys. Rev. B}\ }\textbf {\bibinfo
  {volume} {99}},\ \bibinfo {pages} {214514} (\bibinfo {year}
  {2019})}\BibitemShut {NoStop}%
\bibitem [{\citenamefont {Lai}\ \emph {et~al.}(2021)\citenamefont {Lai},
  \citenamefont {Sarma},\ and\ \citenamefont {Sau}}]{lai2021quality}%
  \BibitemOpen
  \bibfield  {author} {\bibinfo {author} {\bibfnamefont {Y.-H.}\ \bibnamefont
  {Lai}}, \bibinfo {author} {\bibfnamefont {S.~D.}\ \bibnamefont {Sarma}},\
  and\ \bibinfo {author} {\bibfnamefont {J.~D.}\ \bibnamefont {Sau}},\
  }\bibfield  {title} {\bibinfo {title} {Quality factor for zero-bias
  conductance peaks in majorana nanowire},\ }\href@noop {} {\bibfield
  {journal} {\bibinfo  {journal} {arXiv preprint arXiv:2111.01178}\ } (\bibinfo
  {year} {2021})}\BibitemShut {NoStop}%
\bibitem [{\citenamefont {Lobos}\ and\ \citenamefont
  {Sarma}(2015)}]{lobos2015tunneling}%
  \BibitemOpen
  \bibfield  {author} {\bibinfo {author} {\bibfnamefont {A.~M.}\ \bibnamefont
  {Lobos}}\ and\ \bibinfo {author} {\bibfnamefont {S.~D.}\ \bibnamefont
  {Sarma}},\ }\bibfield  {title} {\bibinfo {title} {Tunneling transport in nsn
  majorana junctions across the topological quantum phase transition},\
  }\href@noop {} {\bibfield  {journal} {\bibinfo  {journal} {New Journal of
  Physics}\ }\textbf {\bibinfo {volume} {17}},\ \bibinfo {pages} {065010}
  (\bibinfo {year} {2015})}\BibitemShut {NoStop}%
\bibitem [{\citenamefont {Gramich}\ \emph {et~al.}(2017)\citenamefont
  {Gramich}, \citenamefont {Baumgartner},\ and\ \citenamefont
  {Sch{\"o}nenberger}}]{gramich2017andreev}%
  \BibitemOpen
  \bibfield  {author} {\bibinfo {author} {\bibfnamefont {J.}~\bibnamefont
  {Gramich}}, \bibinfo {author} {\bibfnamefont {A.}~\bibnamefont
  {Baumgartner}},\ and\ \bibinfo {author} {\bibfnamefont {C.}~\bibnamefont
  {Sch{\"o}nenberger}},\ }\bibfield  {title} {\bibinfo {title} {Andreev bound
  states probed in three-terminal quantum dots},\ }\href@noop {} {\bibfield
  {journal} {\bibinfo  {journal} {Physical Review B}\ }\textbf {\bibinfo
  {volume} {96}},\ \bibinfo {pages} {195418} (\bibinfo {year}
  {2017})}\BibitemShut {NoStop}%
\bibitem [{\citenamefont {Zazunov}\ \emph {et~al.}(2017)\citenamefont
  {Zazunov}, \citenamefont {Egger}, \citenamefont {Alvarado},\ and\
  \citenamefont {Yeyati}}]{zazunov2017multiterminal}%
  \BibitemOpen
  \bibfield  {author} {\bibinfo {author} {\bibfnamefont {A.}~\bibnamefont
  {Zazunov}}, \bibinfo {author} {\bibfnamefont {R.}~\bibnamefont {Egger}},
  \bibinfo {author} {\bibfnamefont {M.}~\bibnamefont {Alvarado}},\ and\
  \bibinfo {author} {\bibfnamefont {A.~L.}\ \bibnamefont {Yeyati}},\ }\bibfield
   {title} {\bibinfo {title} {Josephson effect in multiterminal topological
  junctions},\ }\href {https://doi.org/10.1103/PhysRevB.96.024516} {\bibfield
  {journal} {\bibinfo  {journal} {Phys. Rev. B}\ }\textbf {\bibinfo {volume}
  {96}},\ \bibinfo {pages} {024516} (\bibinfo {year} {2017})}\BibitemShut
  {NoStop}%
\bibitem [{\citenamefont {Jonckheere}\ \emph {et~al.}(2019)\citenamefont
  {Jonckheere}, \citenamefont {Rech}, \citenamefont {Zazunov}, \citenamefont
  {Egger}, \citenamefont {Yeyati},\ and\ \citenamefont
  {Martin}}]{jonckheere2019trijunction}%
  \BibitemOpen
  \bibfield  {author} {\bibinfo {author} {\bibfnamefont {T.}~\bibnamefont
  {Jonckheere}}, \bibinfo {author} {\bibfnamefont {J.}~\bibnamefont {Rech}},
  \bibinfo {author} {\bibfnamefont {A.}~\bibnamefont {Zazunov}}, \bibinfo
  {author} {\bibfnamefont {R.}~\bibnamefont {Egger}}, \bibinfo {author}
  {\bibfnamefont {A.~L.}\ \bibnamefont {Yeyati}},\ and\ \bibinfo {author}
  {\bibfnamefont {T.}~\bibnamefont {Martin}},\ }\bibfield  {title} {\bibinfo
  {title} {Giant shot noise from majorana zero modes in topological
  trijunctions},\ }\href {https://doi.org/10.1103/PhysRevLett.122.097003}
  {\bibfield  {journal} {\bibinfo  {journal} {Phys. Rev. Lett.}\ }\textbf
  {\bibinfo {volume} {122}},\ \bibinfo {pages} {097003} (\bibinfo {year}
  {2019})}\BibitemShut {NoStop}%
\bibitem [{\citenamefont {Danon}\ \emph {et~al.}(2020)\citenamefont {Danon},
  \citenamefont {Hellenes}, \citenamefont {Hansen}, \citenamefont {Casparis},
  \citenamefont {Higginbotham},\ and\ \citenamefont
  {Flensberg}}]{danon2020nonlocal}%
  \BibitemOpen
  \bibfield  {author} {\bibinfo {author} {\bibfnamefont {J.}~\bibnamefont
  {Danon}}, \bibinfo {author} {\bibfnamefont {A.~B.}\ \bibnamefont {Hellenes}},
  \bibinfo {author} {\bibfnamefont {E.~B.}\ \bibnamefont {Hansen}}, \bibinfo
  {author} {\bibfnamefont {L.}~\bibnamefont {Casparis}}, \bibinfo {author}
  {\bibfnamefont {A.~P.}\ \bibnamefont {Higginbotham}},\ and\ \bibinfo {author}
  {\bibfnamefont {K.}~\bibnamefont {Flensberg}},\ }\bibfield  {title} {\bibinfo
  {title} {Nonlocal conductance spectroscopy of andreev bound states: Symmetry
  relations and bcs charges},\ }\href@noop {} {\bibfield  {journal} {\bibinfo
  {journal} {Physical Review Letters}\ }\textbf {\bibinfo {volume} {124}},\
  \bibinfo {pages} {036801} (\bibinfo {year} {2020})}\BibitemShut {NoStop}%
\bibitem [{\citenamefont {Melo}\ \emph {et~al.}(2021)\citenamefont {Melo},
  \citenamefont {Liu}, \citenamefont {Ro{\.z}ek}, \citenamefont {Rosdahl},\
  and\ \citenamefont {Wimmer}}]{melo2021conductance}%
  \BibitemOpen
  \bibfield  {author} {\bibinfo {author} {\bibfnamefont {A.}~\bibnamefont
  {Melo}}, \bibinfo {author} {\bibfnamefont {C.-X.}\ \bibnamefont {Liu}},
  \bibinfo {author} {\bibfnamefont {P.}~\bibnamefont {Ro{\.z}ek}}, \bibinfo
  {author} {\bibfnamefont {T.~{\"O}.}\ \bibnamefont {Rosdahl}},\ and\ \bibinfo
  {author} {\bibfnamefont {M.}~\bibnamefont {Wimmer}},\ }\bibfield  {title}
  {\bibinfo {title} {Conductance asymmetries in mesoscopic superconducting
  devices due to finite bias},\ }\href@noop {} {\bibfield  {journal} {\bibinfo
  {journal} {SciPost Physics}\ }\textbf {\bibinfo {volume} {10}},\ \bibinfo
  {pages} {037} (\bibinfo {year} {2021})}\BibitemShut {NoStop}%
\bibitem [{\citenamefont {Pan}\ \emph {et~al.}(2021)\citenamefont {Pan},
  \citenamefont {Sau},\ and\ \citenamefont {Sarma}}]{pan2021three}%
  \BibitemOpen
  \bibfield  {author} {\bibinfo {author} {\bibfnamefont {H.}~\bibnamefont
  {Pan}}, \bibinfo {author} {\bibfnamefont {J.~D.}\ \bibnamefont {Sau}},\ and\
  \bibinfo {author} {\bibfnamefont {S.~D.}\ \bibnamefont {Sarma}},\ }\bibfield
  {title} {\bibinfo {title} {Three-terminal nonlocal conductance in majorana
  nanowires: Distinguishing topological and trivial in realistic systems with
  disorder and inhomogeneous potential},\ }\href@noop {} {\bibfield  {journal}
  {\bibinfo  {journal} {Physical Review B}\ }\textbf {\bibinfo {volume}
  {103}},\ \bibinfo {pages} {014513} (\bibinfo {year} {2021})}\BibitemShut
  {NoStop}%
\bibitem [{\citenamefont {Banerjee}\ \emph {et~al.}(2023)\citenamefont
  {Banerjee}, \citenamefont {Lesser}, \citenamefont {Rahman}, \citenamefont
  {Thomas}, \citenamefont {Wang}, \citenamefont {Manfra}, \citenamefont {Berg},
  \citenamefont {Oreg}, \citenamefont {Stern},\ and\ \citenamefont
  {Marcus}}]{banerjee2023local}%
  \BibitemOpen
  \bibfield  {author} {\bibinfo {author} {\bibfnamefont {A.}~\bibnamefont
  {Banerjee}}, \bibinfo {author} {\bibfnamefont {O.}~\bibnamefont {Lesser}},
  \bibinfo {author} {\bibfnamefont {M.}~\bibnamefont {Rahman}}, \bibinfo
  {author} {\bibfnamefont {C.}~\bibnamefont {Thomas}}, \bibinfo {author}
  {\bibfnamefont {T.}~\bibnamefont {Wang}}, \bibinfo {author} {\bibfnamefont
  {M.}~\bibnamefont {Manfra}}, \bibinfo {author} {\bibfnamefont
  {E.}~\bibnamefont {Berg}}, \bibinfo {author} {\bibfnamefont {Y.}~\bibnamefont
  {Oreg}}, \bibinfo {author} {\bibfnamefont {A.}~\bibnamefont {Stern}},\ and\
  \bibinfo {author} {\bibfnamefont {C.}~\bibnamefont {Marcus}},\ }\bibfield
  {title} {\bibinfo {title} {Local and nonlocal transport spectroscopy in
  planar josephson junctions},\ }\href@noop {} {\bibfield  {journal} {\bibinfo
  {journal} {Physical Review Letters}\ }\textbf {\bibinfo {volume} {130}},\
  \bibinfo {pages} {096202} (\bibinfo {year} {2023})}\BibitemShut {NoStop}%
\bibitem [{\citenamefont {Hess}\ \emph {et~al.}(2022)\citenamefont {Hess},
  \citenamefont {Legg}, \citenamefont {Loss},\ and\ \citenamefont
  {Klinovaja}}]{hess-2022}%
  \BibitemOpen
  \bibfield  {author} {\bibinfo {author} {\bibfnamefont {R.}~\bibnamefont
  {Hess}}, \bibinfo {author} {\bibfnamefont {H.}~\bibnamefont {Legg}}, \bibinfo
  {author} {\bibfnamefont {D.}~\bibnamefont {Loss}},\ and\ \bibinfo {author}
  {\bibfnamefont {J.}~\bibnamefont {Klinovaja}},\ }\bibfield  {title} {\bibinfo
  {title} {Trivial andreev band mimicking topological bulk gap reopening in the
  non-local conductance of long rashba nanowires},\ }\href@noop {} {\bibfield
  {journal} {\bibinfo  {journal} {arXiv:2210.03507}\ } (\bibinfo {year}
  {2022})}\BibitemShut {NoStop}%
\bibitem [{\citenamefont {Yu}\ \emph {et~al.}(2021)\citenamefont {Yu},
  \citenamefont {Chen}, \citenamefont {Gomanko}, \citenamefont {Badawy},
  \citenamefont {Bakkers}, \citenamefont {Zuo}, \citenamefont {Mourik},\ and\
  \citenamefont {Frolov}}]{yu2021non}%
  \BibitemOpen
  \bibfield  {author} {\bibinfo {author} {\bibfnamefont {P.}~\bibnamefont
  {Yu}}, \bibinfo {author} {\bibfnamefont {J.}~\bibnamefont {Chen}}, \bibinfo
  {author} {\bibfnamefont {M.}~\bibnamefont {Gomanko}}, \bibinfo {author}
  {\bibfnamefont {G.}~\bibnamefont {Badawy}}, \bibinfo {author} {\bibfnamefont
  {E.}~\bibnamefont {Bakkers}}, \bibinfo {author} {\bibfnamefont
  {K.}~\bibnamefont {Zuo}}, \bibinfo {author} {\bibfnamefont {V.}~\bibnamefont
  {Mourik}},\ and\ \bibinfo {author} {\bibfnamefont {S.}~\bibnamefont
  {Frolov}},\ }\bibfield  {title} {\bibinfo {title} {Non-majorana states yield
  nearly quantized conductance in proximatized nanowires},\ }\href@noop {}
  {\bibfield  {journal} {\bibinfo  {journal} {Nature Physics}\ }\textbf
  {\bibinfo {volume} {17}},\ \bibinfo {pages} {482} (\bibinfo {year}
  {2021})}\BibitemShut {NoStop}%
\bibitem [{\citenamefont {Wang}\ \emph
  {et~al.}(2022{\natexlab{a}})\citenamefont {Wang}, \citenamefont {Song},
  \citenamefont {Pan}, \citenamefont {Zhang}, \citenamefont {Miao},
  \citenamefont {Li}, \citenamefont {Cao}, \citenamefont {Zhang}, \citenamefont
  {Liu}, \citenamefont {Wen} \emph {et~al.}}]{wang2022observation}%
  \BibitemOpen
  \bibfield  {author} {\bibinfo {author} {\bibfnamefont {Z.}~\bibnamefont
  {Wang}}, \bibinfo {author} {\bibfnamefont {H.}~\bibnamefont {Song}}, \bibinfo
  {author} {\bibfnamefont {D.}~\bibnamefont {Pan}}, \bibinfo {author}
  {\bibfnamefont {Z.}~\bibnamefont {Zhang}}, \bibinfo {author} {\bibfnamefont
  {W.}~\bibnamefont {Miao}}, \bibinfo {author} {\bibfnamefont {R.}~\bibnamefont
  {Li}}, \bibinfo {author} {\bibfnamefont {Z.}~\bibnamefont {Cao}}, \bibinfo
  {author} {\bibfnamefont {G.}~\bibnamefont {Zhang}}, \bibinfo {author}
  {\bibfnamefont {L.}~\bibnamefont {Liu}}, \bibinfo {author} {\bibfnamefont
  {L.}~\bibnamefont {Wen}}, \emph {et~al.},\ }\bibfield  {title} {\bibinfo
  {title} {Observation of plateau regions for zero bias peaks within 5\% of the
  quantized conductance value $2e^{2}/h$},\ }\href@noop {} {\bibfield
  {journal} {\bibinfo  {journal} {arXiv preprint arXiv:2205.06736}\ } (\bibinfo
  {year} {2022}{\natexlab{a}})}\BibitemShut {NoStop}%
\bibitem [{\citenamefont {Wang}\ \emph
  {et~al.}(2022{\natexlab{b}})\citenamefont {Wang}, \citenamefont {Song},
  \citenamefont {Pan}, \citenamefont {Zhang}, \citenamefont {Miao},
  \citenamefont {Li}, \citenamefont {Cao}, \citenamefont {Zhang}, \citenamefont
  {Liu}, \citenamefont {Wen} \emph {et~al.}}]{wang2022plateau}%
  \BibitemOpen
  \bibfield  {author} {\bibinfo {author} {\bibfnamefont {Z.}~\bibnamefont
  {Wang}}, \bibinfo {author} {\bibfnamefont {H.}~\bibnamefont {Song}}, \bibinfo
  {author} {\bibfnamefont {D.}~\bibnamefont {Pan}}, \bibinfo {author}
  {\bibfnamefont {Z.}~\bibnamefont {Zhang}}, \bibinfo {author} {\bibfnamefont
  {W.}~\bibnamefont {Miao}}, \bibinfo {author} {\bibfnamefont {R.}~\bibnamefont
  {Li}}, \bibinfo {author} {\bibfnamefont {Z.}~\bibnamefont {Cao}}, \bibinfo
  {author} {\bibfnamefont {G.}~\bibnamefont {Zhang}}, \bibinfo {author}
  {\bibfnamefont {L.}~\bibnamefont {Liu}}, \bibinfo {author} {\bibfnamefont
  {L.}~\bibnamefont {Wen}}, \emph {et~al.},\ }\bibfield  {title} {\bibinfo
  {title} {Plateau regions for zero-bias peaks within 5\% of the quantized
  conductance value 2 e 2/h},\ }\href@noop {} {\bibfield  {journal} {\bibinfo
  {journal} {Physical Review Letters}\ }\textbf {\bibinfo {volume} {129}},\
  \bibinfo {pages} {167702} (\bibinfo {year} {2022}{\natexlab{b}})}\BibitemShut
  {NoStop}%
\bibitem [{\citenamefont {Pikulin}\ \emph {et~al.}(2021)\citenamefont
  {Pikulin}, \citenamefont {van Heck}, \citenamefont {Karzig}, \citenamefont
  {Martinez}, \citenamefont {Nijholt}, \citenamefont {Laeven}, \citenamefont
  {Winkler}, \citenamefont {Watson}, \citenamefont {Heedt}, \citenamefont
  {Temurhan} \emph {et~al.}}]{pikulin2021protocol}%
  \BibitemOpen
  \bibfield  {author} {\bibinfo {author} {\bibfnamefont {D.~I.}\ \bibnamefont
  {Pikulin}}, \bibinfo {author} {\bibfnamefont {B.}~\bibnamefont {van Heck}},
  \bibinfo {author} {\bibfnamefont {T.}~\bibnamefont {Karzig}}, \bibinfo
  {author} {\bibfnamefont {E.~A.}\ \bibnamefont {Martinez}}, \bibinfo {author}
  {\bibfnamefont {B.}~\bibnamefont {Nijholt}}, \bibinfo {author} {\bibfnamefont
  {T.}~\bibnamefont {Laeven}}, \bibinfo {author} {\bibfnamefont {G.~W.}\
  \bibnamefont {Winkler}}, \bibinfo {author} {\bibfnamefont {J.~D.}\
  \bibnamefont {Watson}}, \bibinfo {author} {\bibfnamefont {S.}~\bibnamefont
  {Heedt}}, \bibinfo {author} {\bibfnamefont {M.}~\bibnamefont {Temurhan}},
  \emph {et~al.},\ }\bibfield  {title} {\bibinfo {title} {Protocol to identify
  a topological superconducting phase in a three-terminal device},\ }\href@noop
  {} {\bibfield  {journal} {\bibinfo  {journal} {arXiv preprint
  arXiv:2103.12217}\ } (\bibinfo {year} {2021})}\BibitemShut {NoStop}%
\bibitem [{\citenamefont {Stanescu}\ \emph {et~al.}(2012)\citenamefont
  {Stanescu}, \citenamefont {Tewari}, \citenamefont {Sau},\ and\ \citenamefont
  {Sarma}}]{stanescu2012close}%
  \BibitemOpen
  \bibfield  {author} {\bibinfo {author} {\bibfnamefont {T.~D.}\ \bibnamefont
  {Stanescu}}, \bibinfo {author} {\bibfnamefont {S.}~\bibnamefont {Tewari}},
  \bibinfo {author} {\bibfnamefont {J.~D.}\ \bibnamefont {Sau}},\ and\ \bibinfo
  {author} {\bibfnamefont {S.~D.}\ \bibnamefont {Sarma}},\ }\bibfield  {title}
  {\bibinfo {title} {To close or not to close: the fate of the superconducting
  gap across the topological quantum phase transition in majorana-carrying
  semiconductor nanowires},\ }\href@noop {} {\bibfield  {journal} {\bibinfo
  {journal} {Physical review letters}\ }\textbf {\bibinfo {volume} {109}},\
  \bibinfo {pages} {266402} (\bibinfo {year} {2012})}\BibitemShut {NoStop}%
\bibitem [{\citenamefont {Sarma}\ \emph {et~al.}(2012)\citenamefont {Sarma},
  \citenamefont {Sau},\ and\ \citenamefont {Stanescu}}]{sarma2012splitting}%
  \BibitemOpen
  \bibfield  {author} {\bibinfo {author} {\bibfnamefont {S.~D.}\ \bibnamefont
  {Sarma}}, \bibinfo {author} {\bibfnamefont {J.~D.}\ \bibnamefont {Sau}},\
  and\ \bibinfo {author} {\bibfnamefont {T.~D.}\ \bibnamefont {Stanescu}},\
  }\bibfield  {title} {\bibinfo {title} {Splitting of the zero-bias conductance
  peak as smoking gun evidence for the existence of the majorana mode in a
  superconductor-semiconductor nanowire},\ }\href@noop {} {\bibfield  {journal}
  {\bibinfo  {journal} {Physical Review B}\ }\textbf {\bibinfo {volume} {86}},\
  \bibinfo {pages} {220506} (\bibinfo {year} {2012})}\BibitemShut {NoStop}%
\bibitem [{\citenamefont {Chevallier}\ \emph {et~al.}(2013)\citenamefont
  {Chevallier}, \citenamefont {Simon},\ and\ \citenamefont
  {Bena}}]{chevallier2013andreev}%
  \BibitemOpen
  \bibfield  {author} {\bibinfo {author} {\bibfnamefont {D.}~\bibnamefont
  {Chevallier}}, \bibinfo {author} {\bibfnamefont {P.}~\bibnamefont {Simon}},\
  and\ \bibinfo {author} {\bibfnamefont {C.}~\bibnamefont {Bena}},\ }\bibfield
  {title} {\bibinfo {title} {From andreev bound states to majorana fermions in
  topological wires on superconducting substrates: A story of mutation},\
  }\href@noop {} {\bibfield  {journal} {\bibinfo  {journal} {Physical Review
  B}\ }\textbf {\bibinfo {volume} {88}},\ \bibinfo {pages} {165401} (\bibinfo
  {year} {2013})}\BibitemShut {NoStop}%
\bibitem [{\citenamefont {Dmytruk}\ and\ \citenamefont
  {Klinovaja}(2018)}]{dmytruk2018suppression}%
  \BibitemOpen
  \bibfield  {author} {\bibinfo {author} {\bibfnamefont {O.}~\bibnamefont
  {Dmytruk}}\ and\ \bibinfo {author} {\bibfnamefont {J.}~\bibnamefont
  {Klinovaja}},\ }\bibfield  {title} {\bibinfo {title} {Suppression of the
  overlap between majorana fermions by orbital magnetic effects in
  semiconducting-superconducting nanowires},\ }\href@noop {} {\bibfield
  {journal} {\bibinfo  {journal} {Physical Review B}\ }\textbf {\bibinfo
  {volume} {97}},\ \bibinfo {pages} {155409} (\bibinfo {year}
  {2018})}\BibitemShut {NoStop}%
\bibitem [{\citenamefont {Rainis}\ \emph {et~al.}(2013)\citenamefont {Rainis},
  \citenamefont {Trifunovic}, \citenamefont {Klinovaja},\ and\ \citenamefont
  {Loss}}]{rainis2013towards}%
  \BibitemOpen
  \bibfield  {author} {\bibinfo {author} {\bibfnamefont {D.}~\bibnamefont
  {Rainis}}, \bibinfo {author} {\bibfnamefont {L.}~\bibnamefont {Trifunovic}},
  \bibinfo {author} {\bibfnamefont {J.}~\bibnamefont {Klinovaja}},\ and\
  \bibinfo {author} {\bibfnamefont {D.}~\bibnamefont {Loss}},\ }\bibfield
  {title} {\bibinfo {title} {Towards a realistic transport modeling in a
  superconducting nanowire with majorana fermions},\ }\href@noop {} {\bibfield
  {journal} {\bibinfo  {journal} {Physical Review B}\ }\textbf {\bibinfo
  {volume} {87}},\ \bibinfo {pages} {024515} (\bibinfo {year}
  {2013})}\BibitemShut {NoStop}%
\bibitem [{\citenamefont {Danon}\ \emph {et~al.}(2017)\citenamefont {Danon},
  \citenamefont {Hansen},\ and\ \citenamefont
  {Flensberg}}]{danon2017conductance}%
  \BibitemOpen
  \bibfield  {author} {\bibinfo {author} {\bibfnamefont {J.}~\bibnamefont
  {Danon}}, \bibinfo {author} {\bibfnamefont {E.~B.}\ \bibnamefont {Hansen}},\
  and\ \bibinfo {author} {\bibfnamefont {K.}~\bibnamefont {Flensberg}},\
  }\bibfield  {title} {\bibinfo {title} {Conductance spectroscopy on majorana
  wires and the inverse proximity effect},\ }\href@noop {} {\bibfield
  {journal} {\bibinfo  {journal} {Physical Review B}\ }\textbf {\bibinfo
  {volume} {96}},\ \bibinfo {pages} {125420} (\bibinfo {year}
  {2017})}\BibitemShut {NoStop}%
\bibitem [{\citenamefont {Ricco}\ \emph {et~al.}(2018)\citenamefont {Ricco},
  \citenamefont {Campo~Jr}, \citenamefont {Shelykh},\ and\ \citenamefont
  {Seridonio}}]{ricco2018majorana}%
  \BibitemOpen
  \bibfield  {author} {\bibinfo {author} {\bibfnamefont {L.}~\bibnamefont
  {Ricco}}, \bibinfo {author} {\bibfnamefont {V.}~\bibnamefont {Campo~Jr}},
  \bibinfo {author} {\bibfnamefont {I.}~\bibnamefont {Shelykh}},\ and\ \bibinfo
  {author} {\bibfnamefont {A.}~\bibnamefont {Seridonio}},\ }\bibfield  {title}
  {\bibinfo {title} {Majorana oscillations modulated by fano interference and
  degree of nonlocality in a topological superconducting-nanowire--quantum-dot
  system},\ }\href@noop {} {\bibfield  {journal} {\bibinfo  {journal} {Physical
  Review B}\ }\textbf {\bibinfo {volume} {98}},\ \bibinfo {pages} {075142}
  (\bibinfo {year} {2018})}\BibitemShut {NoStop}%
\bibitem [{\citenamefont {Prada}\ \emph {et~al.}(2017)\citenamefont {Prada},
  \citenamefont {Aguado},\ and\ \citenamefont {San-Jose}}]{prada2017measuring}%
  \BibitemOpen
  \bibfield  {author} {\bibinfo {author} {\bibfnamefont {E.}~\bibnamefont
  {Prada}}, \bibinfo {author} {\bibfnamefont {R.}~\bibnamefont {Aguado}},\ and\
  \bibinfo {author} {\bibfnamefont {P.}~\bibnamefont {San-Jose}},\ }\bibfield
  {title} {\bibinfo {title} {Measuring majorana nonlocality and spin structure
  with a quantum dot},\ }\href@noop {} {\bibfield  {journal} {\bibinfo
  {journal} {Physical Review B}\ }\textbf {\bibinfo {volume} {96}},\ \bibinfo
  {pages} {085418} (\bibinfo {year} {2017})}\BibitemShut {NoStop}%
\bibitem [{\citenamefont {Deng}\ \emph {et~al.}(2018)\citenamefont {Deng},
  \citenamefont {Vaitiek\ifmmode~\dot{e}\else \.{e}\fi{}nas}, \citenamefont
  {Prada}, \citenamefont {San-Jose}, \citenamefont {Nyg\aa{}rd}, \citenamefont
  {Krogstrup}, \citenamefont {Aguado},\ and\ \citenamefont
  {Marcus}}]{deng2018}%
  \BibitemOpen
  \bibfield  {author} {\bibinfo {author} {\bibfnamefont {M.-T.}\ \bibnamefont
  {Deng}}, \bibinfo {author} {\bibfnamefont {S.}~\bibnamefont
  {Vaitiek\ifmmode~\dot{e}\else \.{e}\fi{}nas}}, \bibinfo {author}
  {\bibfnamefont {E.}~\bibnamefont {Prada}}, \bibinfo {author} {\bibfnamefont
  {P.}~\bibnamefont {San-Jose}}, \bibinfo {author} {\bibfnamefont
  {J.}~\bibnamefont {Nyg\aa{}rd}}, \bibinfo {author} {\bibfnamefont
  {P.}~\bibnamefont {Krogstrup}}, \bibinfo {author} {\bibfnamefont
  {R.}~\bibnamefont {Aguado}},\ and\ \bibinfo {author} {\bibfnamefont {C.~M.}\
  \bibnamefont {Marcus}},\ }\bibfield  {title} {\bibinfo {title} {Nonlocality
  of majorana modes in hybrid nanowires},\ }\href
  {https://doi.org/10.1103/PhysRevB.98.085125} {\bibfield  {journal} {\bibinfo
  {journal} {Phys. Rev. B}\ }\textbf {\bibinfo {volume} {98}},\ \bibinfo
  {pages} {085125} (\bibinfo {year} {2018})}\BibitemShut {NoStop}%
\bibitem [{\citenamefont {Schuray}\ \emph {et~al.}(2020)\citenamefont
  {Schuray}, \citenamefont {Rammler},\ and\ \citenamefont
  {Recher}}]{schuray2020signatures}%
  \BibitemOpen
  \bibfield  {author} {\bibinfo {author} {\bibfnamefont {A.}~\bibnamefont
  {Schuray}}, \bibinfo {author} {\bibfnamefont {M.}~\bibnamefont {Rammler}},\
  and\ \bibinfo {author} {\bibfnamefont {P.}~\bibnamefont {Recher}},\
  }\bibfield  {title} {\bibinfo {title} {Signatures of the majorana spin in
  electrical transport through a majorana nanowire},\ }\href@noop {} {\bibfield
   {journal} {\bibinfo  {journal} {Physical Review B}\ }\textbf {\bibinfo
  {volume} {102}},\ \bibinfo {pages} {045303} (\bibinfo {year}
  {2020})}\BibitemShut {NoStop}%
\bibitem [{\citenamefont {Ricco}\ \emph {et~al.}(2021)\citenamefont {Ricco},
  \citenamefont {Sanches}, \citenamefont {Marques}, \citenamefont {de~Souza},
  \citenamefont {Figueira}, \citenamefont {Shelykh},\ and\ \citenamefont
  {Seridonio}}]{ricco2021topological}%
  \BibitemOpen
  \bibfield  {author} {\bibinfo {author} {\bibfnamefont {L.}~\bibnamefont
  {Ricco}}, \bibinfo {author} {\bibfnamefont {J.}~\bibnamefont {Sanches}},
  \bibinfo {author} {\bibfnamefont {Y.}~\bibnamefont {Marques}}, \bibinfo
  {author} {\bibfnamefont {M.}~\bibnamefont {de~Souza}}, \bibinfo {author}
  {\bibfnamefont {M.}~\bibnamefont {Figueira}}, \bibinfo {author}
  {\bibfnamefont {I.}~\bibnamefont {Shelykh}},\ and\ \bibinfo {author}
  {\bibfnamefont {A.}~\bibnamefont {Seridonio}},\ }\bibfield  {title} {\bibinfo
  {title} {Topological isoconductance signatures in majorana nanowires},\
  }\href@noop {} {\bibfield  {journal} {\bibinfo  {journal} {Scientific
  Reports}\ }\textbf {\bibinfo {volume} {11}},\ \bibinfo {pages} {17310}
  (\bibinfo {year} {2021})}\BibitemShut {NoStop}%
\bibitem [{\citenamefont {Vaitiek{\.e}nas}\ \emph {et~al.}(2020)\citenamefont
  {Vaitiek{\.e}nas}, \citenamefont {Winkler}, \citenamefont {Van~Heck},
  \citenamefont {Karzig}, \citenamefont {Deng}, \citenamefont {Flensberg},
  \citenamefont {Glazman}, \citenamefont {Nayak}, \citenamefont {Krogstrup},
  \citenamefont {Lutchyn} \emph {et~al.}}]{vaitiekenas2020flux}%
  \BibitemOpen
  \bibfield  {author} {\bibinfo {author} {\bibfnamefont {S.}~\bibnamefont
  {Vaitiek{\.e}nas}}, \bibinfo {author} {\bibfnamefont {G.}~\bibnamefont
  {Winkler}}, \bibinfo {author} {\bibfnamefont {B.}~\bibnamefont {Van~Heck}},
  \bibinfo {author} {\bibfnamefont {T.}~\bibnamefont {Karzig}}, \bibinfo
  {author} {\bibfnamefont {M.-T.}\ \bibnamefont {Deng}}, \bibinfo {author}
  {\bibfnamefont {K.}~\bibnamefont {Flensberg}}, \bibinfo {author}
  {\bibfnamefont {L.}~\bibnamefont {Glazman}}, \bibinfo {author} {\bibfnamefont
  {C.}~\bibnamefont {Nayak}}, \bibinfo {author} {\bibfnamefont
  {P.}~\bibnamefont {Krogstrup}}, \bibinfo {author} {\bibfnamefont
  {R.}~\bibnamefont {Lutchyn}}, \emph {et~al.},\ }\bibfield  {title} {\bibinfo
  {title} {Flux-induced topological superconductivity in full-shell
  nanowires},\ }\href@noop {} {\bibfield  {journal} {\bibinfo  {journal}
  {Science}\ }\textbf {\bibinfo {volume} {367}},\ \bibinfo {pages} {eaav3392}
  (\bibinfo {year} {2020})}\BibitemShut {NoStop}%
\bibitem [{\citenamefont {Valentini}\ \emph {et~al.}(2021)\citenamefont
  {Valentini}, \citenamefont {Pe{\~n}aranda}, \citenamefont {Hofmann},
  \citenamefont {Brauns}, \citenamefont {Hauschild}, \citenamefont {Krogstrup},
  \citenamefont {San-Jose}, \citenamefont {Prada}, \citenamefont {Aguado},\
  and\ \citenamefont {Katsaros}}]{valentini2021nontopological}%
  \BibitemOpen
  \bibfield  {author} {\bibinfo {author} {\bibfnamefont {M.}~\bibnamefont
  {Valentini}}, \bibinfo {author} {\bibfnamefont {F.}~\bibnamefont
  {Pe{\~n}aranda}}, \bibinfo {author} {\bibfnamefont {A.}~\bibnamefont
  {Hofmann}}, \bibinfo {author} {\bibfnamefont {M.}~\bibnamefont {Brauns}},
  \bibinfo {author} {\bibfnamefont {R.}~\bibnamefont {Hauschild}}, \bibinfo
  {author} {\bibfnamefont {P.}~\bibnamefont {Krogstrup}}, \bibinfo {author}
  {\bibfnamefont {P.}~\bibnamefont {San-Jose}}, \bibinfo {author}
  {\bibfnamefont {E.}~\bibnamefont {Prada}}, \bibinfo {author} {\bibfnamefont
  {R.}~\bibnamefont {Aguado}},\ and\ \bibinfo {author} {\bibfnamefont
  {G.}~\bibnamefont {Katsaros}},\ }\bibfield  {title} {\bibinfo {title}
  {Nontopological zero-bias peaks in full-shell nanowires induced by
  flux-tunable andreev states},\ }\href@noop {} {\bibfield  {journal} {\bibinfo
   {journal} {Science}\ }\textbf {\bibinfo {volume} {373}},\ \bibinfo {pages}
  {82} (\bibinfo {year} {2021})}\BibitemShut {NoStop}%
\bibitem [{\citenamefont {Escribano}\ \emph {et~al.}(2018)\citenamefont
  {Escribano}, \citenamefont {Yeyati},\ and\ \citenamefont
  {Prada}}]{escribano2018dotformation}%
  \BibitemOpen
  \bibfield  {author} {\bibinfo {author} {\bibfnamefont {S.~D.}\ \bibnamefont
  {Escribano}}, \bibinfo {author} {\bibfnamefont {A.~L.}\ \bibnamefont
  {Yeyati}},\ and\ \bibinfo {author} {\bibfnamefont {E.}~\bibnamefont
  {Prada}},\ }\bibfield  {title} {\bibinfo {title} {Interaction-induced
  zero-energy pinning and quantum dot formation in majorana nanowires},\ }\href
  {https://doi.org/10.3762/bjnano.9.203} {\bibfield  {journal} {\bibinfo
  {journal} {Beilstein Journal of Nanotechnology}\ }\textbf {\bibinfo {volume}
  {9}},\ \bibinfo {pages} {2171} (\bibinfo {year} {2018})}\BibitemShut
  {NoStop}%
\bibitem [{\citenamefont {Pan}\ and\ \citenamefont
  {Sarma}(2020)}]{pan2020physical}%
  \BibitemOpen
  \bibfield  {author} {\bibinfo {author} {\bibfnamefont {H.}~\bibnamefont
  {Pan}}\ and\ \bibinfo {author} {\bibfnamefont {S.~D.}\ \bibnamefont
  {Sarma}},\ }\bibfield  {title} {\bibinfo {title} {Physical mechanisms for
  zero-bias conductance peaks in majorana nanowires},\ }\href@noop {}
  {\bibfield  {journal} {\bibinfo  {journal} {Physical Review Research}\
  }\textbf {\bibinfo {volume} {2}},\ \bibinfo {pages} {013377} (\bibinfo {year}
  {2020})}\BibitemShut {NoStop}%
\bibitem [{\citenamefont {Bolech}\ and\ \citenamefont
  {Demler}(2007)}]{bolech2007observing}%
  \BibitemOpen
  \bibfield  {author} {\bibinfo {author} {\bibfnamefont {C.}~\bibnamefont
  {Bolech}}\ and\ \bibinfo {author} {\bibfnamefont {E.}~\bibnamefont
  {Demler}},\ }\bibfield  {title} {\bibinfo {title} {Observing majorana bound
  states in p-wave superconductors using noise measurements in tunneling
  experiments},\ }\href@noop {} {\bibfield  {journal} {\bibinfo  {journal}
  {Physical review letters}\ }\textbf {\bibinfo {volume} {98}},\ \bibinfo
  {pages} {237002} (\bibinfo {year} {2007})}\BibitemShut {NoStop}%
\bibitem [{\citenamefont {Nilsson}\ \emph {et~al.}(2008)\citenamefont
  {Nilsson}, \citenamefont {Akhmerov},\ and\ \citenamefont
  {Beenakker}}]{nilsson2008splitting}%
  \BibitemOpen
  \bibfield  {author} {\bibinfo {author} {\bibfnamefont {J.}~\bibnamefont
  {Nilsson}}, \bibinfo {author} {\bibfnamefont {A.}~\bibnamefont {Akhmerov}},\
  and\ \bibinfo {author} {\bibfnamefont {C.}~\bibnamefont {Beenakker}},\
  }\bibfield  {title} {\bibinfo {title} {Splitting of a cooper pair by a pair
  of majorana bound states},\ }\href@noop {} {\bibfield  {journal} {\bibinfo
  {journal} {Physical review letters}\ }\textbf {\bibinfo {volume} {101}},\
  \bibinfo {pages} {120403} (\bibinfo {year} {2008})}\BibitemShut {NoStop}%
\bibitem [{\citenamefont {Golub}\ and\ \citenamefont
  {Horovitz}(2011)}]{golub2011shot}%
  \BibitemOpen
  \bibfield  {author} {\bibinfo {author} {\bibfnamefont {A.}~\bibnamefont
  {Golub}}\ and\ \bibinfo {author} {\bibfnamefont {B.}~\bibnamefont
  {Horovitz}},\ }\bibfield  {title} {\bibinfo {title} {Shot noise in a majorana
  fermion chain},\ }\href@noop {} {\bibfield  {journal} {\bibinfo  {journal}
  {Physical Review B}\ }\textbf {\bibinfo {volume} {83}},\ \bibinfo {pages}
  {153415} (\bibinfo {year} {2011})}\BibitemShut {NoStop}%
\bibitem [{\citenamefont {Zazunov}\ \emph {et~al.}(2016)\citenamefont
  {Zazunov}, \citenamefont {Egger},\ and\ \citenamefont
  {Yeyati}}]{zazunov2016low}%
  \BibitemOpen
  \bibfield  {author} {\bibinfo {author} {\bibfnamefont {A.}~\bibnamefont
  {Zazunov}}, \bibinfo {author} {\bibfnamefont {R.}~\bibnamefont {Egger}},\
  and\ \bibinfo {author} {\bibfnamefont {A.~L.}\ \bibnamefont {Yeyati}},\
  }\bibfield  {title} {\bibinfo {title} {Low-energy theory of transport in
  majorana wire junctions},\ }\href@noop {} {\bibfield  {journal} {\bibinfo
  {journal} {Physical Review B}\ }\textbf {\bibinfo {volume} {94}},\ \bibinfo
  {pages} {014502} (\bibinfo {year} {2016})}\BibitemShut {NoStop}%
\bibitem [{\citenamefont {Perrin}\ \emph {et~al.}(2021)\citenamefont {Perrin},
  \citenamefont {Civelli},\ and\ \citenamefont {Simon}}]{perrin2021noise}%
  \BibitemOpen
  \bibfield  {author} {\bibinfo {author} {\bibfnamefont {V.}~\bibnamefont
  {Perrin}}, \bibinfo {author} {\bibfnamefont {M.}~\bibnamefont {Civelli}},\
  and\ \bibinfo {author} {\bibfnamefont {P.}~\bibnamefont {Simon}},\ }\bibfield
   {title} {\bibinfo {title} {Identifying majorana bound states by tunneling
  shot-noise tomography},\ }\href
  {https://doi.org/10.1103/PhysRevB.104.L121406} {\bibfield  {journal}
  {\bibinfo  {journal} {Phys. Rev. B}\ }\textbf {\bibinfo {volume} {104}},\
  \bibinfo {pages} {L121406} (\bibinfo {year} {2021})}\BibitemShut {NoStop}%
\bibitem [{\citenamefont {Tewari}\ and\ \citenamefont {Sau}(2012)}]{tewa}%
  \BibitemOpen
  \bibfield  {author} {\bibinfo {author} {\bibfnamefont {S.}~\bibnamefont
  {Tewari}}\ and\ \bibinfo {author} {\bibfnamefont {J.~D.}\ \bibnamefont
  {Sau}},\ }\bibfield  {title} {\bibinfo {title} {Topological invariants for
  spin-orbit coupled superconductor nanowires},\ }\href
  {https://doi.org/10.1103/PhysRevLett.109.150408} {\bibfield  {journal}
  {\bibinfo  {journal} {Phys. Rev. Lett.}\ }\textbf {\bibinfo {volume} {109}},\
  \bibinfo {pages} {150408} (\bibinfo {year} {2012})}\BibitemShut {NoStop}%
\bibitem [{\citenamefont {Budich}\ and\ \citenamefont
  {Ardonne}(2013)}]{budich}%
  \BibitemOpen
  \bibfield  {author} {\bibinfo {author} {\bibfnamefont {J.~C.}\ \bibnamefont
  {Budich}}\ and\ \bibinfo {author} {\bibfnamefont {E.}~\bibnamefont
  {Ardonne}},\ }\bibfield  {title} {\bibinfo {title} {Equivalent topological
  invariants for one-dimensional majorana wires in symmetry class $d$},\ }\href
  {https://doi.org/10.1103/PhysRevB.88.075419} {\bibfield  {journal} {\bibinfo
  {journal} {Phys. Rev. B}\ }\textbf {\bibinfo {volume} {88}},\ \bibinfo
  {pages} {075419} (\bibinfo {year} {2013})}\BibitemShut {NoStop}%
\bibitem [{\citenamefont {Yu}(2005)}]{yu2005bound}%
  \BibitemOpen
  \bibfield  {author} {\bibinfo {author} {\bibfnamefont {L.}~\bibnamefont
  {Yu}},\ }\bibfield  {title} {\bibinfo {title} {Bound state in superconductors
  with paramagnetic impurities},\ }\href@noop {} {\  (\bibinfo {year}
  {2005})}\BibitemShut {NoStop}%
\bibitem [{\citenamefont {Shiba}(1968)}]{shiba1968classical}%
  \BibitemOpen
  \bibfield  {author} {\bibinfo {author} {\bibfnamefont {H.}~\bibnamefont
  {Shiba}},\ }\bibfield  {title} {\bibinfo {title} {Classical spins in
  superconductors},\ }\href@noop {} {\bibfield  {journal} {\bibinfo  {journal}
  {Progress of theoretical Physics}\ }\textbf {\bibinfo {volume} {40}},\
  \bibinfo {pages} {435} (\bibinfo {year} {1968})}\BibitemShut {NoStop}%
\bibitem [{\citenamefont {Rusinov}(1969)}]{rusinov1969theory}%
  \BibitemOpen
  \bibfield  {author} {\bibinfo {author} {\bibfnamefont {A.}~\bibnamefont
  {Rusinov}},\ }\bibfield  {title} {\bibinfo {title} {Theory of gapless
  superconductivity in alloys containing paramagnetic impurities},\ }\href@noop
  {} {\bibfield  {journal} {\bibinfo  {journal} {Sov. Phys. JETP}\ }\textbf
  {\bibinfo {volume} {29}},\ \bibinfo {pages} {1101} (\bibinfo {year}
  {1969})}\BibitemShut {NoStop}%
\bibitem [{\citenamefont {Balatsky}\ \emph {et~al.}(2006)\citenamefont
  {Balatsky}, \citenamefont {Vekhter},\ and\ \citenamefont
  {Zhu}}]{balatsky2006impurity}%
  \BibitemOpen
  \bibfield  {author} {\bibinfo {author} {\bibfnamefont {A.~V.}\ \bibnamefont
  {Balatsky}}, \bibinfo {author} {\bibfnamefont {I.}~\bibnamefont {Vekhter}},\
  and\ \bibinfo {author} {\bibfnamefont {J.-X.}\ \bibnamefont {Zhu}},\
  }\bibfield  {title} {\bibinfo {title} {Impurity-induced states in
  conventional and unconventional superconductors},\ }\href@noop {} {\bibfield
  {journal} {\bibinfo  {journal} {Reviews of Modern Physics}\ }\textbf
  {\bibinfo {volume} {78}},\ \bibinfo {pages} {373} (\bibinfo {year}
  {2006})}\BibitemShut {NoStop}%
\bibitem [{\citenamefont {Rammer}(2011)}]{rammer2011quantum}%
  \BibitemOpen
  \bibfield  {author} {\bibinfo {author} {\bibfnamefont {J.}~\bibnamefont
  {Rammer}},\ }\bibfield  {title} {\bibinfo {title} {Quantum field theory of
  non-equilibrium states},\ }\href@noop {} {\bibfield  {journal} {\bibinfo
  {journal} {Quantum Field Theory of Non-equilibrium States}\ } (\bibinfo
  {year} {2011})}\BibitemShut {NoStop}%
\bibitem [{\citenamefont {Cuevas}\ \emph {et~al.}(1996)\citenamefont {Cuevas},
  \citenamefont {Mart{\'\i}n-Rodero},\ and\ \citenamefont
  {Yeyati}}]{cuevas1996hamiltonian}%
  \BibitemOpen
  \bibfield  {author} {\bibinfo {author} {\bibfnamefont {J.}~\bibnamefont
  {Cuevas}}, \bibinfo {author} {\bibfnamefont {A.}~\bibnamefont
  {Mart{\'\i}n-Rodero}},\ and\ \bibinfo {author} {\bibfnamefont {A.~L.}\
  \bibnamefont {Yeyati}},\ }\bibfield  {title} {\bibinfo {title} {Hamiltonian
  approach to the transport properties of superconducting quantum point
  contacts},\ }\href@noop {} {\bibfield  {journal} {\bibinfo  {journal}
  {Physical Review B}\ }\textbf {\bibinfo {volume} {54}},\ \bibinfo {pages}
  {7366} (\bibinfo {year} {1996})}\BibitemShut {NoStop}%
\bibitem [{\citenamefont {Blonder}\ \emph {et~al.}(1982)\citenamefont
  {Blonder}, \citenamefont {Tinkham},\ and\ \citenamefont
  {Klapwijk}}]{blonder1982transition}%
  \BibitemOpen
  \bibfield  {author} {\bibinfo {author} {\bibfnamefont {G.}~\bibnamefont
  {Blonder}}, \bibinfo {author} {\bibfnamefont {m.~M.}\ \bibnamefont
  {Tinkham}},\ and\ \bibinfo {author} {\bibfnamefont {k.~T.}\ \bibnamefont
  {Klapwijk}},\ }\bibfield  {title} {\bibinfo {title} {Transition from metallic
  to tunneling regimes in superconducting microconstrictions: Excess current,
  charge imbalance, and supercurrent conversion},\ }\href@noop {} {\bibfield
  {journal} {\bibinfo  {journal} {Physical Review B}\ }\textbf {\bibinfo
  {volume} {25}},\ \bibinfo {pages} {4515} (\bibinfo {year}
  {1982})}\BibitemShut {NoStop}%
\bibitem [{\citenamefont {M.~P. López~Sancho}\ and\ \citenamefont
  {Rubio}(1985)}]{lopezsancho}%
  \BibitemOpen
  \bibfield  {author} {\bibinfo {author} {\bibfnamefont {J.~M. L.~S.}\
  \bibnamefont {M.~P. López~Sancho}}\ and\ \bibinfo {author} {\bibfnamefont
  {J.}~\bibnamefont {Rubio}},\ }\bibfield  {title} {\bibinfo {title} {Highly
  convergent schemes for the calculation of bulk and surface green functions},\
  }\href@noop {} {\bibfield  {journal} {\bibinfo  {journal} {J. Phys. F: Met
  Phys.}\ }\textbf {\bibinfo {volume} {15}},\ \bibinfo {pages} {851} (\bibinfo
  {year} {1985})}\BibitemShut {NoStop}%
\bibitem [{\citenamefont {Alvarado}\ \emph {et~al.}(2020)\citenamefont
  {Alvarado}, \citenamefont {Iks}, \citenamefont {Zazunov}, \citenamefont
  {Egger},\ and\ \citenamefont {Yeyati}}]{alvarado2020boundary}%
  \BibitemOpen
  \bibfield  {author} {\bibinfo {author} {\bibfnamefont {M.}~\bibnamefont
  {Alvarado}}, \bibinfo {author} {\bibfnamefont {A.}~\bibnamefont {Iks}},
  \bibinfo {author} {\bibfnamefont {A.}~\bibnamefont {Zazunov}}, \bibinfo
  {author} {\bibfnamefont {R.}~\bibnamefont {Egger}},\ and\ \bibinfo {author}
  {\bibfnamefont {A.~L.}\ \bibnamefont {Yeyati}},\ }\bibfield  {title}
  {\bibinfo {title} {Boundary green's function approach for spinful
  single-channel and multichannel majorana nanowires},\ }\href
  {https://doi.org/10.1103/PhysRevB.101.094511} {\bibfield  {journal} {\bibinfo
   {journal} {Phys. Rev. B}\ }\textbf {\bibinfo {volume} {101}},\ \bibinfo
  {pages} {094511} (\bibinfo {year} {2020})}\BibitemShut {NoStop}%
\bibitem [{\citenamefont {Klinovaja}\ and\ \citenamefont
  {Loss}(2012)}]{klinovaja2012composite}%
  \BibitemOpen
  \bibfield  {author} {\bibinfo {author} {\bibfnamefont {J.}~\bibnamefont
  {Klinovaja}}\ and\ \bibinfo {author} {\bibfnamefont {D.}~\bibnamefont
  {Loss}},\ }\bibfield  {title} {\bibinfo {title} {Composite majorana fermion
  wave functions in nanowires},\ }\href@noop {} {\bibfield  {journal} {\bibinfo
   {journal} {Physical Review B}\ }\textbf {\bibinfo {volume} {86}},\ \bibinfo
  {pages} {085408} (\bibinfo {year} {2012})}\BibitemShut {NoStop}%
\bibitem [{\citenamefont {Deng}\ \emph {et~al.}(2012)\citenamefont {Deng},
  \citenamefont {Viola},\ and\ \citenamefont {Ortiz}}]{deng2012majorana}%
  \BibitemOpen
  \bibfield  {author} {\bibinfo {author} {\bibfnamefont {S.}~\bibnamefont
  {Deng}}, \bibinfo {author} {\bibfnamefont {L.}~\bibnamefont {Viola}},\ and\
  \bibinfo {author} {\bibfnamefont {G.}~\bibnamefont {Ortiz}},\ }\bibfield
  {title} {\bibinfo {title} {Majorana modes in time-reversal invariant s-wave
  topological superconductors},\ }\href@noop {} {\bibfield  {journal} {\bibinfo
   {journal} {Physical review letters}\ }\textbf {\bibinfo {volume} {108}},\
  \bibinfo {pages} {036803} (\bibinfo {year} {2012})}\BibitemShut {NoStop}%
\end{thebibliography}%

\end{document}